\documentclass[useAMS,usenatbib]{mn2e}
\usepackage{graphicx,subfigure,floatflt,amsmath,wasysym, amssymb}
\usepackage{here}
\usepackage{setspace}
\usepackage{lscape}
\usepackage{rotating}
\title[Ionized gas in E/S0 galaxies with dust lanes]{Ionized gas in E/S0 galaxies with dust lanes{\thanks{Full Tables 3, 6 and 7 are only available at the CDS via anonymous ftp to \tt{cdsarc.u-strasbg.fr}}}} 

\author[Ido Finkelman et al.]{Ido Finkelman$^{1}$\thanks{E-mail:
ido@wise.tau.ac.il (IF); noah@wise.tau.ac.il (NB);  jfunes@specola.va (JGB); akniazev@saao.ac.za (AYK); petri@saao.ac.za (PV)}, Noah Brosch$^{1}$, Jos\'{e} G. Funes S.J.$^{2}$, Alexei Y. Kniazev$^{3,4}$,
\newauthor Petri V\"{a}is\"{a}nen$^{3,4}$\\
$^{1}$The Wise Observatory and the School of Physics and
Astronomy, the Raymond and  Beverly Sackler Faculty of Exact
Sciences, \\ Tel Aviv University, Tel Aviv 69978, Israel\\
$^{2}$Vatican Observatory, V-00120 Vatican City State, Italy\\ 
$^{3}$South African Astronomical Observatory, PO Box 9, 7935 Observatory, Cape Town, South Africa \\
$^{4}$Southern African Large Telescope Foundation, PO Box 9, 7935 Observatory, Cape Town, South Africa \\}

\begin{document}

\date{Accepted 2010 May 24. Received 2010 May 23; in original form 2010 February 24}

\pagerange{\pageref{firstpage}--\pageref{lastpage}} \pubyear{2002}

\maketitle

\label{firstpage}

\begin{abstract}
We report the results of multicolour observations of 30 E/S0 galaxies with dust lanes. For each galaxy we obtained broad-band images and narrow-band images using interference filters isolating the H$\alpha$+[NII] emission lines to derive the amount and morphology of dust and ionized gas. 
To improve the wavelength coverage we retrieved data from the SDSS and 2MASS and combined these with our data.
Ionized gas is detected in 25 galaxies and shows in most cases a smooth morphology, although knots and filamentary structure are also observed in some objects. The extended gas distribution closely follows the dust structure, with a clear correlation between the mass of both components.  
An extinction law by the extragalactic dust in the dark lanes is derived and is used to estimate the dust content of the galaxies. The derived extinction law is used to correct the measured colours for intrinsic dust extinction and the data are fitted with a stellar population synthesis model. 
We find that the H$\alpha$ emission and colours of most objects are consistent with the presence of an ``old'' stellar population ($\sim10$ Gyr) and a small fraction of a ``young'' population ($\sim10-100$ Myr). 
To check this we closely examine NGC5363, for which archival Spitzer/IRAC and GALEX data are available, as a representative dust-lane E/S0 galaxy of the sample. 
\end{abstract}

\begin{keywords}
galaxies: elliptical and lenticular, cD; galaxies: ISM; dust, extinction.
\end{keywords}

\section{Introduction}
For many years elliptical galaxies (Es) were considered as gas-poor objects (see Schweizer 1987 for a review), which formed in a single burst and evolved with no further star formation (Eggen et al.\ 1962). However, improved detector sensitivity along with new discoveries of gas and dust of low column density in E/S0s brought fresh interest to this field. We now know that E/S0s host a multi-phase gas component with hot, X-ray emitting gaseous halos (Forman, Jones \& Tucker 1985; Canizares Fabbiano \& Trinchieri 1987; O'Sullivan, Forbes \& Ponman 2001), warm ionized gas (IG) (Demoulin-Ulrich, Butcher \& Boksenberg 1984; Phillips et al.\ 1986), cold neutral-hydrogen (e.g., Knapp, Turner \& Cunniffe 1985) and molecular gas  (e.g., Lees et al.\ 1991). 
 
The gas component in E/S0s is often morphologically associated with the presence of dust.
Evidence for the presence of dust includes optical observations of obscured regions (Van Dokkum \& Franx 1995; Ferrari et al.\ 1999; Tomita et al.\ 2000; Tran et al.\ 2001) and the detection of far-IR emission from cold dust in a large fraction of E/S0s (Knapp et al.\ 1989; Roberts et al.\ 1991). The galaxies where dust obscuration produces well-defined dark lanes are known in the literature as dust-lane E/S0s.

The growing number of E/S0s in which a multiphase ISM is detected implies that the stellar population and abundances in this class of galaxies trace complex and varied star formation histories (SFHs). 
A recent burst of star formation in an E/S0 galaxy is expected to have a significant influence on the broad-band colours of the galaxy, even if the mass fraction of the young population is relatively small (see Fig.\ 4 in Schawinski et al.\ 2007). A recent analysis of optical spectra and of UV photometry in E/S0s shows that a significant fraction of these galaxies experienced low-level star formation within the past few Gyr  (e.g., Yi et al.\ 2005; Di Matteo et al.\ 2007; Kaviraj et al.\ 2007). 

The star formation activity in a certain galaxy can also be examined by studying emission-lines. 
The systematic search of IG in E/S0s with long-slit spectroscopic surveys allows the detection of various emission lines linked with excited and ionized gas (Caldwell 1984; Phillips et al.\ 1986; Rampazzo et al.\ 2005; Sarzi et al.\ 2006; Serra et al.\ 2008). Emission-line gas imaging enables the detecting of low levels of star formation even in faint galaxies with relatively modest integration times on small telescopes, and is therefore often used to reveal the full spatial extent of the IG.
In particular, observing a galaxy with a filter centered at the rest-frame H$\alpha$ line can, under reasonable assumptions, provide information on the amount of ionizing radiation emitted by stars.
Surveys focusing on H$\alpha$ emission imaging (Kim 1989; Shields 1991; Trinchieri \& di Serego Alighieri 1991; Buson et al.\ 1993; Goudfrooij et al.\ 1994a; Singh et al.\ 1995; Macchetto et al.\ 1996; Martel et al.\ 2004) detect IG in $\gtrsim50$\% of the E/S0s, and reveal various IG morphologies such as flattened disk-like or ring-like structures, or a filamentary structure. 

This paper studies a sample of E/S0s where dust lanes (DLs) have been reported in the literature and focuses on the dust and IG components and their interrelation, and on the stellar population within the gaseous disks.
The paper is organized as follows: \S~\ref{S:Obs_and_Red}
gives a description of all the observations and data reduction, \S~\ref{S:results} details the broad-band and narrow-band flux calibration, and the data analysis is presented in \S~\ref{S:analysis}. The results are discussed in \S~\ref{S:discuss} and a summary of our conclusions is given in \S~\ref{S:conclude}.
%
\section{Observations and data reduction}
\label{S:Obs_and_Red}
\subsection{The sample}
In most studies of the multiphase ISM in E/S0s the sample is selected by the presence of a particular ISM component, inferred by known tracers, such as X-ray emission, FIR emission or radio emission. Similarly, our sample consists of galaxies identified as dust-lane E/S0 galaxies. Most of these galaxies are chosen from the Bertola (1987) compilation where a close inspection of the presence of DLs was made.
Other objects are selected from more recent studies of E/S0 with DLs, including the search for cold gas (Gregorini, Messina \& Vettolani 1989; M\"{o}llenhoff, Hummel \& Bender 1992; Wang, Kenney \& Ishizuki 1992) and the determination of the extragalactic dust extinction law in the dark lanes (Goudfrooij et al.\ 1994b; Patil et al.\ 2007; Finkelman et al.\ 2008).
The sample galaxies are listed in Table \ref{t:Obs} with their coordinates, morphological classification, integrated blue luminosity, heliocentric velocity and optical size taken from the RC3 catalog (de Vaucouleurs et al.\ 1992) or from the HyperLEDA data base (Paturel et al.\ 2003).
\begin{table*}
 \centering
  \caption{Global parameters for galaxies in our sample.
  \label{t:Obs}}
\begin{tabular}{|lcclccc|}
\hline
Object     & RA       & DEC       & Morph.\       & B$^0_T$ & v$_{Helio}$  & Size \\
{}         & (J2000.0)& (J2000.0) &(RC3/HyperLEDA)&   {}    &   (km/s)     & (arcmin)\\
\hline 
IC1575     & 00:43:44 & -04:07:05 & S0            &  14.40  &    5797      & 0.9x0.8\\
NGC662     & 01:44:35 & +37:41:45 & S pec         &  14.63  &    5659      & 0.8x0.6\\
ESO477-7   & 01:49:24 & -26:44:48 & S0            &  15.19  &    5749      & 0.7x0.4\\
NGC708     & 01:52:46 & +36:09:06 & E Sy2         &  13.69  &    4855      & 2.6x1.2\\
ESO197-10  & 01:53:13 & -49:33:40 & E/S0          &  13.54  &    6173      & 1.7x1.5\\
ESO355-8   & 02:21:19 & -34:19:09 & S0            &  14.92  &    6287      & 1.2x0.8\\
NGC1199    & 03:03:39 & -15:36:51 & E3            &  12.39  &    2570      & 2.8x2.3\\
NGC1297    & 03:19:14 & -19:06:00 & S0            &  13.12  &    1586      & 2.4x2.0\\ 
ESO118-19  & 04:19:00 & -58:15:27 & S0            &  15.04  &    1239      & 0.9x0.8\\
NGC2534    & 08:12:54 & +55:40:19 & E1            &  13.70  &    3447      & 1.4x1.3\\
UGC4449    & 08:31:26 & +40:57:22 & pec           &  15.14  &    7310      & 1.1x0.9\\
NGC2968    & 09:43:12 & +31:55:41 & I0            &  12.78  &    1549      & 2.5x1.6\\
Mrk33      & 10:32:32 & +54:23:56 & Im pec        &  13.31  &    1437      & 1.1x0.9\\
UGC5814    & 10:42:38 & +77:29:41 & pec           &  14.87  &    10738     & 1.3x0.7\\
NGC3656    & 11:23:39 & +53:50:31 & pec LINER     &  13.52  &    2870      & 1.5x1.0\\
NGC3665    & 11:24:44 & +38:45:45 & SA            &  11.78  &    2062      & 4.1x3.0\\
NGC4370    & 12:24:55 & +07:26:38 & Sa            &  13.53  &     782      & 1.6x1.1\\
NGC4374    & 12:25:03 & +12:53:13 & E1 LINER      &  10.08  &     992      & 7.4x6.4\\
NGC4583    & 12:38:05 & +33:27:30 & S0/a Sy2      &  14.56  &    6920      & 1.5x1.5\\
NGC5249    & 13:37:38 & +15:58:18 & S0            &  14.32  &    7671      & 1.5x0.9\\
NGC5311    & 13:48:56 & +39:59:07 & S0            &  13.44  &    2649      & 2.0x1.7\\
NGC5363    & 13:56:07 & +05:15:17 & I0 LINER      &  11.10  &    1136      & 4.2x2.8\\
NGC5485    & 14:07:11 & +55:00:07 & SA0           &  12.40  &    1975      & 2.5x1.7\\
NGC6251    & 16:32:32 & +82:32:16 & E1 LINER      &  13.90  &    7402      & 1.7x1.4\\
NGC6314    & 17:12:39 & +23:16:15 & SAa           &  13.90  &    6632      & 1.4x0.7\\
NGC6702    & 18:46:58 & +45:42:20 & E LINER       &  13.23  &    4725      & 1.7x1.4\\
NGC7052    & 21:18:33 & +26:26:49 & E             &  13.90  &    4700      & 1.8x0.9\\
NGC7399    & 22:52:39 & -09:16:04 & S0/a          &  14.85  &    4465      & 1.1x0.7\\
NGC7432    & 22:58:06 & +13:08:00 & E             &  14.46  &    7627      & 1.6x0.9\\
NGC7625    & 23:20:30 & +17:13:35 & Sa            &  12.94  &    1631      & 1.5x1.5\\
\hline
\end{tabular}
\end{table*}
\subsection{Observations}
We present broad-band U, B, V, R, I and H$\alpha$ narrow-band observations of 30 galaxies acquired at the Mt.\ Graham International Observatory (MGIO), the South African Astronomical Observatory (SAAO) and the Wise Observatory (WO). Observations with the 1.8m Vatican Advanced Technology Telescope (VATT) at the MGIO were performed in two runs in June 2004 and May 2006 using the Loral CCD at the aplanatic f/9 Gregorian focus. 
A second set of observations was performed at the SAAO 1.9m telescope in Sutherland, South Africa, in November 2008. The images were obtained with the 3:1 ``Shara'' focal reducer attached between the f/18 Cassegrain focal plane of the telescope and the STE4 CCD.  
A third set of observations was performed with the PI VersArray camera mounted on the WO 1m telescope.

The basic characteristics of the narrow-band filters used for our sample galaxies, including the central wavelength, FWHM and peak transmission, are listed in Table \ref{t:Hafilters}. The dates of observations, the broad-band filters used in each run, the exposure times and typical seeing are given in Table \ref{t:ALLexp}.
\begin{table*}
 \centering
  \caption{Narrow-band filter specifications.
  \label{t:Hafilters}}
\begin{tabular}{|lccc|}
\hline
Filter name   & $\lambda_{cent}$ & FWHM          & Peak trans.\ \\
       {}     & ($\mbox{\AA}$)   & ($\mbox{\AA}$) & (\%)  \\
\hline
VATT/658           & 6585           & 66      & 87        \\
VATT/663           & 6633           & 72      & 87        \\
VATT/668           & 6683           & 66      & 92        \\
VATT/673           & 6736           & 65      & 91        \\
WO/H$\alpha$3	   & 6585           & 52      & 58        \\
WO/H$\alpha$4 	   & 6607           & 60      & 59        \\
WO/H$\alpha$5 	   & 6696           & 58      & 62        \\
WO/H$\alpha$6 	   & 6643           & 66      & 55        \\
WO/H$\alpha$7 	   & 6699           & 56      & 60        \\
WO/H$\alpha$8 	   & 6761           & 68      & 57        \\
WO/H$\alpha$9 	   & 6809           & 68      & 40        \\
WO/RGO67	   & 6674           & 86      & 46        \\
\hline
\end{tabular}
\end{table*}
\begin{table*}
\setlength{\tabcolsep}{5pt}
\caption{Observation log. The full table is only available in an electronic form. \label{t:ALLexp}}
\begin{tabular}{|l|lllllll|cccccc|}
\hline
{}         & \multicolumn{7}{|c|}{Exposure time (sec)} & \multicolumn{6}{|c|}{point source FWHM (")}\\
Galaxy     & Date[Observatory]       & U        & B        & V         & R        & I        & H$\alpha$[filter] & U & B & V & R   & I   & H$\alpha$\\ 
\hline
IC1575     & 14.11.2008[SAAO] & -        & -        & $600(3)$  & -        & - &1200(3)[WO/H$\alpha$5]   & -   & -   & 2.2 & -   & - & 2.3     \\ 
{}         & 15.11.2008[SAAO] & -        & $600(3)$ & $600(3)$  & $600(3)$ & - &-                     	& -   & 1.9 & 2.2 & 2.2 & - & -       \\
{}	   & 18.11.2008[SAAO] & $1200(3)$& -        & -         & -        & - &-       	& 2.0 & -   & -   & -   & - & -       \\
{}         & 19.11.2008[SAAO] & $120(2)$ & $60(2)$  & $60(2)$   & $60(2)$  & - &-       	& 2.1 & 2.2 & 1.9 & 2.1 & - & -       \\
NGC662     & 23.10.2008[WO] & $120(3)$ & $120(3)$ & $60(3)$   & $60(3)$  & $60(3)$  & -                  & 2.9 & 2.8 & 2.2 & 2.5 & 2.2 & -  \\
{}         & 23.10.2008[WO] & $1200(3)$& $1200(3)$& $600(3)$  & $600(3)$ & $600(3)$ & 1200(3)[WO/H$\alpha$5]   & 2.6 & 2.5 & 2.2 & 2.2 & 2.2 & 2.0\\ 
\hline
\end{tabular}
\begin{minipage}[]{6.5in}
\begin{footnotesize}
{\bf Note:} The number in parenthesis represents the number of individual frames used for individual galaxies in each band. The point source FWHM represents the typical FWHM of stars near each target combining seeing and image quality effects.
\end{footnotesize}
\end{minipage}
\end{table*}
\subsection{Data reduction}
Image reduction was performed with standard tasks within {\small IRAF}\footnote{{\small IRAF} is distributed by the National Optical Astronomy Observatories (NOAO), which is operated by the Association of Universities, Inc. (AURA) under co-operative agreement with the National Science Foundation}. These include bias subtraction, overscan subtraction and flatfield correction. 
Cosmic rays events were removed from single CCD exposures by using the L.A.Cosmic task in IRAF (van Dokkum 2001) while CCD hot pixels were removed with the FIXPIX task in {\small IRAF} using an appropriate mask. 
Interference fringes, which showed in the WO I-band images, were removed in a standard manner.

The reduced images are geometrically aligned by measuring centroids of several common stars in the galaxy frames and are then background-subtracted and combined with median scaling to improve the S/N ratio and remove cosmic ray events. This alignment procedure involves {\small IRAF} tasks for scaling, translation and rotation of the images, so that a small amount of blurring is introduced affecting the accuracy to be better than a few hundredths of a pixel. 
\section{Flux Calibration}
\label{S:results}
\subsection{Measuring the line emission}
\label{Measuring flux}
We observed each galaxy using two filters, a narrow-band filter (FWHM$\sim 70 \mbox{\AA}$) which covers the rest-frame H$\alpha$+[NII] emission, and a broad R-band filter with a band-width of $\sim 1000 \mbox{\AA}$ (see Table \ref{t:ALLexp}). 
The H$\alpha$ images include photons from the H$\alpha$ and the [NII] lines and from the continuum. Since we are interested in the H$\alpha$ line, the continuum signal should be removed. 
This is performed in several steps. The H$\alpha$ and R-band images are first geometrically aligned, convolved to the same resolution and background subtracted. The R image is later scaled to match the stellar continuum in the narrow-band image. 

The intensity scaling factor (ISF) is determined by comparing the light from the underlying galaxy (see Macchetto et al.\ 1996; B\"{o}ker et al.\ 1999) and the intensity of foreground stars in both the narrow-band and R images. Since different foreground stars may have different colours, we also plot the colour of the stars vs.\ their ISF to interpolate for the adequate ISF of a galaxy with a certain colour. The derived ISF is then used to obtain a continuum-subtracted image. 

The continuum-subtracted H$\alpha$+[NII] images are then used to measure the emission within an elliptical aperture enclosing the apparent IG disk-like region of each galaxy and within individual H$\alpha$ knots, if present. The aperture limits are set to include resolution elements with net H$\alpha$+[NII] emission $\sim 2\sigma$ above the noise as measured in the continuum-subtracted H$\alpha$+[NII] images. For galaxies with no apparent extended emission structure, a circular aperture with a radius twice the FWHM seeing size around the galactic center is examined in an attempt to detect inner line emission. Since the measured net H$\alpha$+[NII] emission in a galaxy may have high relative errors if the line emission is diffuse or weak, the detection criterion for the presence of IG in our sample was arbitrarily set to relative errors smaller than 75\%.    

Following Pogge \& Eskridge (1993), we specify in Table \ref{t:apertures} for the IG disk-like region in each galaxy the apparent major axis diameter $D_{disk}$ in arcmin, the major axis position angle $\theta_{disk}$, the axial ratio $\left( b/a\right) _{disk}$, and the projected radius of the disk $R(kpc)$. These are compared with the stellar isophotal major axis diameter at the $\mu_B$=25 mag arcsec$^{-2}$ isophote (taken from RC3 and HyperLEDA), $D_{25}$ in arcmin, the major axis position angle $\theta_{25}$, and the axial ratio of the limiting isophote $\left( b/a\right) _{25}$ of the host galaxy. We also list the ratio of the disk-to-galaxy isophotal diameter $D_{disk}/D_{25}$ to show the extension of the IG region with respect to the galaxy size.
Uncertainties in the galaxy characteristics are of order $\pm 10\%$, and the uncertainty in $\theta_{disk}$, which is estimated by eye due to the clumpy morphology, is typically $\pm 5 \deg$. 
\begin{table*}
 \centering
\caption{Ionized gas aperture properties. 
\label{t:apertures}
}
\begin{tabular}{|l|crc|crcc|c}
\hline
Galaxy & \multicolumn{3}{c}{Galaxy} & \multicolumn{4}{c}{HII disk} & {}\\
{}         & $D_{25}(')$ & $\theta_{25}$ & $\left( b/a\right)_{25}$ & $D_{disk}(')$ & $\theta_{disk}$ & $\left( b/a\right)_{disk}$ & $R(kpc)$ & $D_{disk}/D_{25}$\\      
\hline
IC1575     & 0.9        & 138           & 0.83     & 0.22          & 50              & 0.78       &  2.5       & 0.28          \\
NGC662     & 0.8        & 14            & 0.76     & 0.63          & 10              & 0.71       &  7.2       & 0.79          \\
ESO477-7   & 0.7        & 52            & 0.63     & 0.16          & 90              & 0.86       &  1.9       & 0.12          \\
NGC708     & 2.6        & 37            & 0.66     & 0.55          & 170             & 0.64       &  5.4       & 0.21          \\
ESO119-97  & 1.7        & 175           & 0.73     & 0.33          & 85              & 0.68       &  4.1       & 0.18          \\
ESO355-8   & 1.2        & 156           & 0.68     & 0.22          & 125             & 0.77       &  2.8       & 0.15          \\
NGC1199    & 2.8        & 56            & 0.83     & 0.45          & 60              & 0.54       &  2.5       & 0.19          \\
NGC1297    & 2.4        & 15            & 0.75     & 0.50          & 105             & 0.70       &  1.6       & 0.22          \\
ESO118-19  & 0.9        & 35            & 0.72     & 0.32          & 120             & 0.74       &  0.8       & 0.40          \\
NGC2534    & 1.4        & 80            & 0.85     & 0.38          & 5               & 0.47       &  2.8       & 0.27          \\
UGC4449*   & 1.1        & 62            & 0.78     & 0.15          & 0               & 1.00       &  2.2       & 0.25          \\
NGC2968*   & 2.5        & 69            & 0.70     & 0.15          & 0               & 1.00       &  0.5       & 0.03          \\
Mrk33      & 1.1        & 133           & 0.75     & 0.74          & 0               & 1.00       &  2.1       & 0.67          \\
UGC5814    & 1.3        & 137           & 0.66     & 0.25          & 120             & 0.59       &  5.4       & 0.16          \\
NGC3656    & 1.5        & 13            & 0.91     & 0.39          & 175             & 0.63       &  2.3       & 0.24          \\
NGC3665    & 4.1        & 33            & 0.76     & 0.79          & 35              & 0.35       &  3.3       & 0.19          \\ 
NGC4370    & 1.6        & 80            & 0.62     & 0.31          & 90              & 0.55       &  0.5       & 0.22          \\ 
NGC4374    & 7.4        & 130           & 0.94     & 0.46          & 105             & 0.63       &  1.0       & 0.07          \\
NGC4583    & 1.5        & 100           & 0.84     & 0.35          & 20              & 0.58       &  4.9       & 0.32          \\
NGC5249    & 1.5        & 0             & 0.43     & 0.40          & 0               & 1.00       &  6.2       & 0.27          \\
NGC5311    & 2.0        & 99            & 0.87     & 0.45          & 105             & 0.68       &  2.5       & 0.17          \\
NGC5363    & 4.2        & 130           & 0.67     & 1.56          & 140             & 0.65       &  3.5       & 0.38          \\
NGC5485    & 2.5        & 170           & 0.74     & 0.74          & 70              & 0.32       &  3.0       & 0.32          \\
NGC6251    & 1.7        & 9             & 0.80     & 0.35          & 10              & 0.86       &  4.9       & 0.19          \\
NGC6314*   & 1.4        & 174           & 0.56     & 0.15          & 0               & 1.00       &  2.0       & 0.11          \\
NGC6702    & 1.7        & 61            & 0.85     & 0.36          & 135             & 0.51       &  3.4       & 0.20          \\
NGC7052    & 1.8        & 64            & 0.45     & 0.30          & 65              & 0.49       &  3.0       & 0.12          \\
NGC7399    & 1.1        & 150           & 0.57     & 0.51          & 150             & 0.36       &  4.6       & 0.46          \\
NGC7432*   & 1.6        & 40            & 0.58     & 0.15          & 0               & 1.00       &  2.3       & 0.10          \\
NGC7625    & 1.5        & 25            & 0.95     & 0.80          & 60              & 0.65       &  2.7       & 0.50          \\
\hline
\end{tabular}
\begin{minipage}[]{13cm}
\begin{small}
*No apparent H$\alpha$+[NII] extended emission and therefore the aperture is selected to measure the possible central line emission.
\end{small}
\end{minipage}
\end{table*}
\subsection{Emission-line flux calibration}
\label{line_calib}
To convert the measured H$\alpha$+[NII] counts into physical units we calibrate the set of single R-band exposures by using standard stars (Landolt 1992) and determine the R-band flux density, $I_{\mbox{R}}$. 
The ratio between the H$\alpha$+[NII] flux and the R-band flux density is a measure of the effective equivalent width ($EW$) of these lines, $EW_{\mbox{H}\alpha+\mbox{[NII]}}$, and reflects the power of ionization with respect to the stellar light of the galaxy. 
Calculating this ratio should take into account not only the measured H$\alpha$+[NII] counts, but also the filters transmission. 

It can be shown that the calibrated H$\alpha$+[NII] flux corresponds to
\begin{equation}
f_{\mbox{H}\alpha+\mbox{[NII]}} = \dfrac{\mbox{cps}_{\mbox{N}} \cdot \mbox{ISF} -\mbox{cps}_{\mbox{W}}}{\mbox{T}_{\mbox{N}}\left( \lambda _ {\mbox{H}\alpha} \right) \cdot \mbox{ISF} - \mbox{T}_{\mbox{R}} \left( \lambda _ {\mbox{H}\alpha} \right)} \, \dfrac{\mbox{BW}_{\mbox{R}}}{\mbox{cps}_{\mbox{W}}} \, I_{\mbox{R}}
\end{equation}
where $\mbox{cps}_{\mbox{N}}$ and $\mbox{cps}_{\mbox{W}}$ are the measured count rates in instrumental ``analog-to-digital'' units of ADU s$^{-1}$ of the narrow-band and broad-band filters, respectively;
$\mbox{T}_{\mbox{R}}\left( \lambda _ {\mbox{H}\alpha} \right)$ and $\mbox{T}_{\mbox{N}}\left( \lambda _ {\mbox{H}\alpha} \right)$ are the rest-frame H$\alpha$ line transmission through the broad-band and narrow-band filters, respectively, considering also the CCD spectral responsivity; and $\mbox{BW}_{\mbox{R}}$ is the width of the R-band filter computed by 
integrating over $\mbox{T}_{\mbox{R}}$ for all wavelengths.
The derivation of this relation does not take into account possible variations due to subtle differences in the wavelength-dependent sensitivity of the two optical systems, nor any subtle variations due to the wavelength-dependent atmospheric extinction. 
We also assume that the emission is concentrated entirely in a single line, $\lambda_{\mbox{H}\alpha}$.
The measured H$\alpha$+[NII] flux, luminosity and $EW$ for each galaxy are listed in Table \ref{t:Halpha}.

Measuring the continuum-subtracted H$\alpha$+[NII] flux is subject to various errors (see Macchetto et al.\ 1996). The dominant error source in deriving the line fluxes is the continuum subtraction, i.e., the uncertainty in determining the ISF. 
Therefore, estimating the errors depends strongly on the $EW$ of the line emission. James et al.\ (2004) measured errors of $\sim 15$\% for $EW>20\mbox{\AA}$ and $\sim35$\% for galaxies with $EW<20\mbox{\AA}$ in their study of narrow-band H$\alpha$+[NII] imaging of a large sample of nearby galaxies with different morphologies. These are in agreement with the flux errors derived in our study (see Table \ref{t:Halpha}). A manifestation of the high uncertainty in determining the line emission flux in the literature is clearly shown for the case of NGC4374. Comparing the results of the H$\alpha$+[NII] measurements in five previous studies (Kim 1989; Trinchieri \& di Serego Alighieri 1991; Shields 1991; Goudfrooij et al.\ 1994a; Sarzi et al.\ 2006) we find a large scatter with values in the range $\left( 8-38\right) \times 10^{-14}$ erg s$^{-1}$ cm$^{-2}$, where our measurement also lies. This large scatter is further illustrated in Macchetto et al.\ (1996) where a comparison of the H$\alpha$ measurements of their sample galaxies with the literature data shows differences up to a factor of 3.
\begin{table*}
  \caption{H$\alpha$+[NII] emission data for the sample galaxies. 
  \label{t:Halpha}}
\begin{tabular}{|lcccccl|}
\hline
Object     & Dist.\ & $f_{\mbox{H}\alpha+\mbox{[NII]}}$               & $EW_{\mbox{H}\alpha+\mbox{[NII]}}$             & $L_{\mbox{H}\alpha+\mbox{[NII]}}$            & $\mbox{Log} \left(\frac{M_{HII}}{M_ \odot}\right)$       & Morph.\ \\
{}         & [Mpc]  & [$10^{-14}$ erg s$^{-1}$ cm$^{-2}$]  & [$\mbox{\AA}$] & [$10^{39}$ erg s$^{-1}$]  & {} & {} \\
\hline 
IC1575	   & 80.4 & $3.6 \pm 2.1$                        & $6.9\pm4.0$         & $28 \pm 16$     & $4.82\pm0.57$   & ED      \\
NGC662     & 79.4 & $93 \pm 12$                          & $97\pm13$           & $699 \pm 90$    & $6.20\pm0.13$   & CD+K    \\
ESO477-7   & 78.2 & $2.3 \pm 0.7$                        & $10.0\pm3.1$        & $16.7 \pm 5.1$  & $5.79\pm0.28$   & ED      \\
NGC708     & 68.4 & $49 \pm 14 $                         & $26.7\pm7.7$        & $273 \pm 78 $   & $4.62\pm0.31$   & ED      \\ 
ESO197-10  & 83.1 & $6.9 \pm 2.3$                        & $5.1\pm1.7$         & $57 \pm 19$     & $5.15\pm0.34$   & ED      \\
ESO355-8   & 85.2 & $6.8 \pm 1.9$                        & $15.6\pm4.1$        & $59 \pm 16$     & $5.16\pm0.27$   & ED      \\
NGC1199    & 34.5 & $4.9 \pm 2.4$                        & $4.4\pm2.0$         & $7.01 \pm 3.4$  & $4.28\pm0.49$   & ED      \\
NGC1297    & 20.7 & $8.1 \pm 2.2$                        & $8.9\pm2.4$         & $4.1 \pm 1.1$   & $4.03\pm0.28$   & ED      \\
ESO118-19  & 14.6 & $2.2 \pm 1.6$                        & $8.8\pm6.4$         & $0.56 \pm 0.41$ & $3.24\pm0.73$   & ED      \\
NGC2534    & 48.0 & $6.4 \pm 1.1$                        & $8.1\pm1.4$         & $17.6 \pm 3.0$  & $4.67\pm0.29$   & ED      \\
UGC4449    & 100.1& $0.9 \pm 2.7$                        & -                   & -               & -               & -       \\
NGC2968    & 20.9 & $1.0 \pm 5.6$                        & -                   & -               & -               & -       \\
Mrk33      & 20.4 & $255 \pm 19$                         & $144\pm13$          & $127 \pm 9$     & $5.47\pm0.07$   & ED      \\
UGC5814    & 149  & $11.4 \pm 4.5$                       & $42\pm17$           & $300 \pm 120$   & $5.85\pm0.39$   & ED      \\
NGC3656    & 40.5 & $33.7 \pm 4.3$                       & $19.8\pm2.7$        & $65.9 \pm 8.4$  & $5.18\pm0.13$   & CD+K+F  \\
NGC3665    & 28.2 & $36.3 \pm 9.2$                       & $7.5\pm1.4$         & $34.4 \pm 8.7$  & $4.92\pm0.25$   & ED      \\
NGC4370    &  9.7 & $2.0 \pm 1.5$                        & $3.2\pm2.4$         & $0.22 \pm 0.16$ & $2.75\pm0.75$   & CD      \\
NGC4374    & 13.8 & $35.8 \pm 8.5$                       & $2.6\pm0.6$         & $8.1 \pm 1.9 $  & $4.31\pm0.24$   & ED      \\ 
NGC4583    & 95.1 & $5.3 \pm 1.6$                        & $7.8\pm2.4$         & $57 \pm 17$     & $5.13\pm0.29$   & ED      \\ 
NGC5249    & 105.0& $17 \pm 11$                          & $17\pm11$           & $220 \pm 150$   & $5.75\pm0.65$   & ED      \\
NGC5311    & 38.0 & $10.8 \pm 9.1$                       & -                   & -               & -               & -       \\
NGC5363    & 15.3 & $144.1 \pm 37.9$                     & $10.7\pm2.9$        & $40 \pm 11$     & $4.99\pm0.26$   & ED+SB   \\
NGC5485    & 29.1 & $12.3 \pm 3.8$                       & $4.4\pm1.4$         & $12.4 \pm 3.8 $ & $4.42\pm0.31$   & ED      \\   
NGC6251    & 103.9& $10.1 \pm 3.4$                       & $6.7\pm2.9$         & $130 \pm 44$    & $5.41\pm0.33$   & ED      \\
NGC6314    & 92.9 & $0.5 \pm 4.8$                        & -                   & -               & -               & -       \\
NGC6702    & 67.7 & $3.3 \pm 1.5$                        & $2.6\pm1.2$         & $18.0 \pm 8.2$  & $4.60\pm0.45$   & ED      \\ 
NGC7052    & 67.0 & $6.0 \pm 1.6$                        & $4.8\pm1.3$         & $32.1 \pm 8.6$  & $4.90\pm0.27$   & ED      \\
NGC7399    & 62.7 & $11.0 \pm 8.7$                       & $17\pm13$           & $52 \pm 41$     & $5.07\pm0.78$   & ED      \\
NGC7432    & 106.6& $2.6 \pm 10.5$                       & -                   & -               & -               & -       \\
NGC7625    & 24.7 & $157.9 \pm 8.3$                      & $45.5\pm3.2$        & $114.8 \pm 6.0$ & $5.39\pm0.05$   & CD+K+F  \\
\hline
\end{tabular}
\begin{minipage}[]{6.5in}
\begin{footnotesize}
{\bf Morph.:}\\
ED = extended disk; CD = clumpy disk; K = multiple knots; F = filamentary structure; SB = bar with weak spiral arms; 
\end{footnotesize}
\end{minipage}
\end{table*}
\subsection{Broad-band flux calibration}
The broad-band images were flux-calibrated using Landolt (1992) standard stars observed during the different runs.
For some targets, only short $\sim60-120$ sec exposures were obtained during photometric weather conditions. These were later used to calibrate the flux measured in longer exposure frames. 
This was done by measuring the ratio between the intensities within a $\sim20''$ wide elliptical aperture of both long and short exposures, or by measuring the intensity ratio for many foreground stars which were used later as secondary standard stars to calibrate the galaxy flux.

\begin{figure}
  \caption{B$_{25}$-magnitude values for the sample galaxies.}
  \centering
   \includegraphics[]{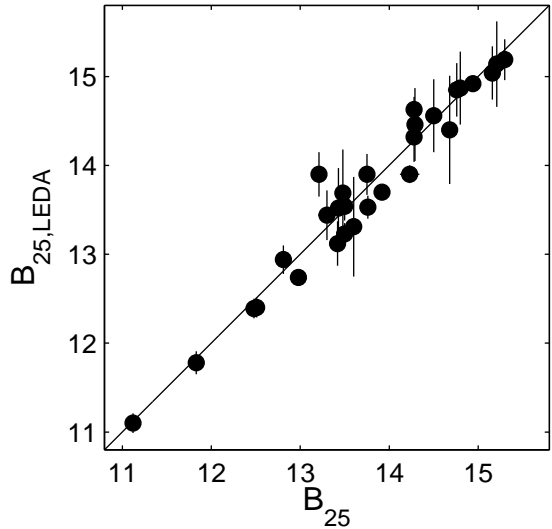}
    \label{f:LEDA}
\end{figure}

We measure with the ELLIPSE task the B-magnitude and colours for an elliptical aperture with the semi-major axis as given in the literature (see HyperLEDA, Paturel et al.\ 2003). The measured B-band magnitudes are plotted vs.\ the values from the HyperLEDA database in Fig.\ \ref{f:LEDA}. We also measure the B-magnitude and colours within the aperture enclosing the apparent IG disk-like region of each galaxy.
To improve the wavelength coverage we retrieved data from the SDSS Data Release 7 (DR7) and 2MASS archives for galaxies where they are available.  We then calculate the flux by matching the apertures to those used for our observations and calibrating the data with the formal pipelines. The results are listed in Tables \ref{t:WOphot} and \ref{t:VATTphot}.

In addition we present in APPENDIX A a R-band contour map image, a continuum-subtracted  H$\alpha$+[NII] narrow-band image and a B-R colour image for each galaxy. 
\begin{table*}
 \centering
  \caption{Photometry results (WO and SAAO observations). The full table is only available in an electronic form. \label{t:WOphot}}
\begin{tabular}{|lcccccccc|}
\hline
 Galaxy   & B & U-B & B-V & V-R & R-I & R-J & J-H & H-K \\
\hline 
IC1575    & $14.47 \pm 0.02$ & - & $1.11 \pm 0.04$ & $0.66 \pm 0.04$ & $0.94 \pm 0.06$ & -               & -               & -               \\
          & $16.22 \pm 0.02$ & - & $1.37 \pm 0.04$ & $0.82 \pm 0.04$ & $0.83 \pm 0.06$ & $2.20 \pm 0.03$ & $0.79 \pm 0.02$ & $0.33 \pm 0.03$ \\
NGC662    & $14.30 \pm 0.04$ & $-0.12\pm0.12$ & $0.59 \pm 0.06$ & $0.46 \pm 0.06$ & $0.61 \pm 0.05$ &  - & -               & -               \\
          & $14.45 \pm 0.04$ & $-0.12\pm0.12$ & $0.61 \pm 0.06$ & $0.47\pm 0.03$ & $0.61 \pm 0.05$ & $1.45\pm 0.05$& $0.69 \pm 0.04$ & $0.31 \pm 0.04$ \\
ESO477-7  & $15.30 \pm 0.02$ & - & $1.00 \pm 0.03$ & $0.70 \pm 0.03$ & $0.79 \pm 0.03$ &  -              & $0.76 \pm 0.04$ & $0.30 \pm 0.05$ \\
          & $16.98 \pm 0.02$ & - & $1.20 \pm 0.03$ & $0.85 \pm 0.03$ & $0.68 \pm 0.03$ & $1.82 \pm 0.03$ & $0.79 \pm 0.04$ & $0.26 \pm 0.05$ \\
\hline
\end{tabular}
\begin{minipage}[]{6.5in}
\begin{small}
{\bf Note:} For each galaxy the values in the first line are measured from the total flux enclosed in an elliptical region as defined in the RC3/HyperLEDA database. The values in the second line are measured from the total flux enclosed in the apertures specified in Table \ref{t:apertures}.\\
\end{small}
\end{minipage}
\end{table*}
\begin{table*}
 \centering
  \caption{Photometry results (VATT observations). The full table is only available in an electronic form. \label{t:VATTphot}}
\begin{tabular}{|lcccccccc|}
\hline
 Galaxy & B & U-B & B-R & r-i (SDSS)& i-z (SDSS) & R-J & J-H & H-K \\
\hline 
NGC2534  & $13.92 \pm 0.03$ & $0.18 \pm 0.05$ & $1.31 \pm 0.04$ & -               & -               & -               & -               & -                             \\
         & $15.13 \pm 0.03$ & $0.30 \pm 0.05$ & $1.54 \pm 0.04$ & $0.38 \pm 0.01$ & $0.30 \pm 0.04$ & $1.80 \pm 0.04$ & $0.70 \pm 0.03$ & $0.28 \pm 0.04$ \\
UGC5814  & $14.80 \pm 0.03$ & $0.38 \pm 0.03$ & $1.59 \pm 0.04$ & -               & -               & -               & -               & -                             \\
         & $16.58 \pm 0.03$ & $0.37 \pm 0.05$ & $1.84 \pm 0.04$ & $0.43 \pm 0.01$ & $0.26 \pm 0.04$ & $2.07 \pm 0.04$ & $0.85 \pm 0.04$ & $0.41 \pm 0.05$ \\
NGC3656  & $13.43 \pm 0.03$ & $0.24 \pm 0.05$ & $1.37 \pm 0.04$ & -               & -               & -               & -               & -                             \\
         & $14.25 \pm 0.03$ & $0.25 \pm 0.05$ & $1.50 \pm 0.04$ & $0.41 \pm 0.01$ & $0.33 \pm 0.01$ & $1.87 \pm 0.03$ & $0.90 \pm 0.03$ & $0.35 \pm 0.03$ \\
\hline
\end{tabular}
\begin{minipage}[]{6.5in}
\begin{small}
{\bf Note:} For each galaxy the values in the first line are measured from the total flux enclosed in an elliptical region as defined in the RC3/HyperLEDA database. The values in the second line are measured from the total flux enclosed in the apertures specified in Table \ref{t:apertures}.\\
\end{small}
\end{minipage}
\end{table*}
\section{Analysis}
\label{S:analysis}
In this section we discuss the analysis of the dust and IG components based on the imaging data and photometric results. 
We first study the ISM by deriving the extinction law in the dark lanes, and by estimating the mass of the dust and IG components. We then combine the extinction-corrected colours with the measured H$\alpha$+[NII] flux to determine an approximate SFH for each object. 

\subsection{Dust extinction}
\label{S:dustext}
Dark lanes are produced by the absorption and scattering of light by dust grains in the dusty structure. These processes depend strongly on the size distribution, structure and chemical composition of the grains. The attenuation of starlight by the dust is wavelength-dependent and is often referred to as the ``extinction law'' or ``reddening''. 

In order to estimate the dust extinction in galaxies where the dust covering factor is relatively small, a dust-free model is constructed by fitting the unextinguished parts of the galaxy with an underlying light distribution of a spheroid following a S\'{e}rsic profile. This is done by fitting the galaxies with elliptical isophotes using the ISOPHOTE package in {\small IRAF} and subtracting the observed galaxy image from the dust-free model to measure the extinction. 

\begin{figure*}
\begin{minipage}{165mm}
\caption{Extinction curves for the sample galaxies (solid lines) together with the canonical Galactic curve (dashed lines) for comparison. Extinction values derived from previous papers (Goudfrooij et al.\ 1994b; Patil et al.\ 2007; Finkelman et al.\ 2008) are plotted as open circles. The error bars are $1\sigma$ errors.     \label{f:ALLext}}
\includegraphics[trim=15mm 5mm 30mm 25mm, clip]{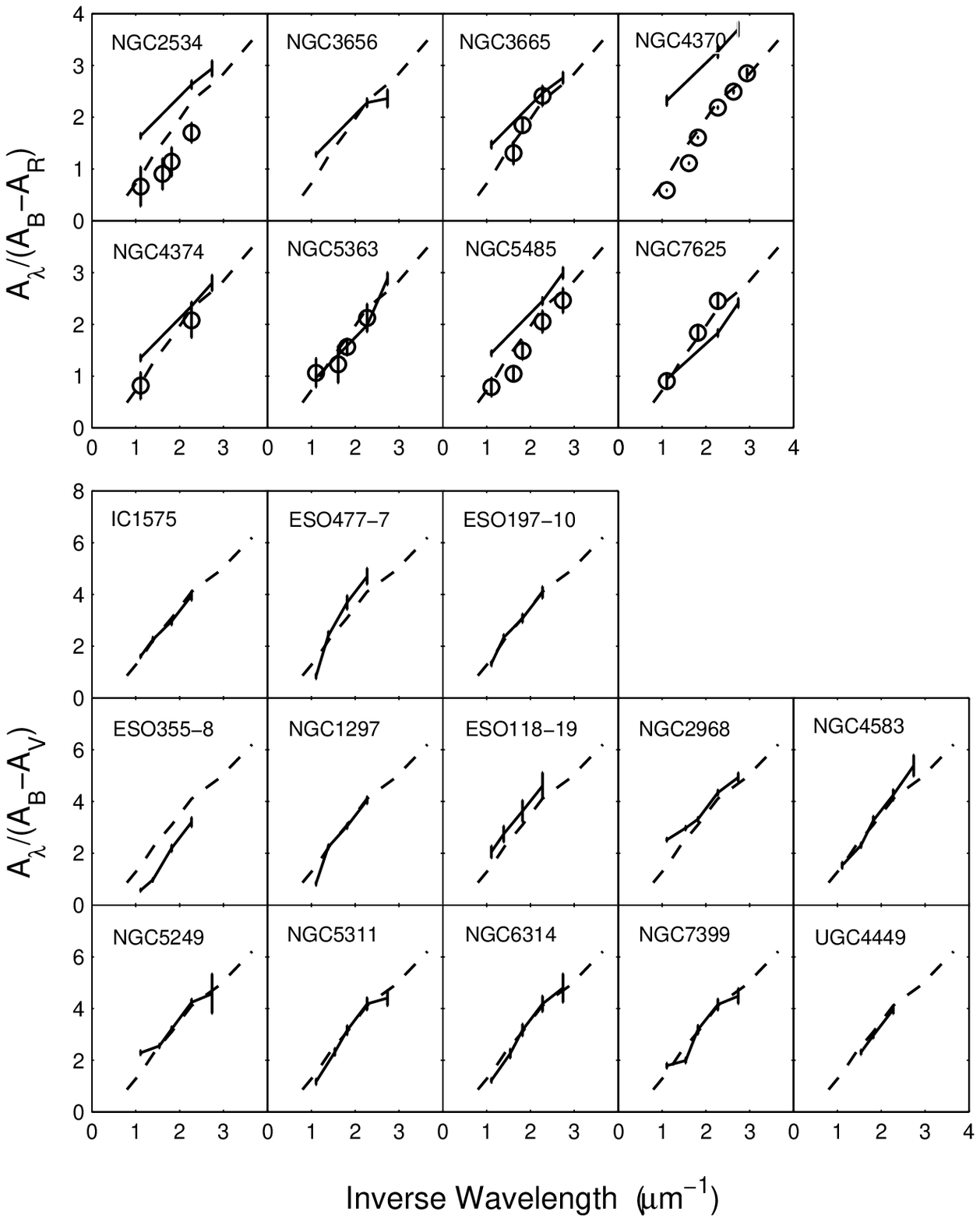}
\end{minipage}
\end{figure*}

Since this process requires the dusty regions to be well-sampled, we could not derive an extinction law for cases where the apparent dust coverage is restricted to the inner parts of the galaxies or where the S\'{e}rsic profile fits badly the galaxy. As a result, we derive extinction curves for 21 of our 30 sample galaxies where the dusty regions could be properly modelled and sampled (see Table \ref{t:RvaluesALL} and Fig.\ \ref{f:ALLext}). We found the extinction curves to lie parallel to the Galactic extinction curve and to match on average the standard Galatic extinction law. 

The total extinction was inferred from the reddening by comparing the observed colour with the colour of the unextinguished parts of the galaxy and assuming no intrinsic colour variations. The extinction was then measured by invoking a Galactic-like extinction law. 
We considered only elements with colour $\gtrsim 2 \sigma$ redder than the average colour to calculate the total extinction. At locations where the extinction could also be measured using our ellipse fitting model, it was compared with the values inferred from reddening and was found to match. 
\begin{table*}
 \centering
\caption{Extinction values for the sample galaxies. \label{t:RvaluesALL}}
\begin{tabular}{|l|ccccc|}
\hline
Object     & $\dfrac{A_U}{E(B-V)}$  & $\dfrac{A_B}{E(B-V)}$  & $\dfrac{A_V}{E(B-V)}$ & $\dfrac{A_R}{E(B-V)}$ & $\dfrac{A_I}{E(B-V)}$\\
\hline
MW galaxy  &  4.7            &    4.1          &   3.1            & 2.32             & 1.5               \\
IC1575     & -               & $3.96 \pm 0.18$ & $ 2.96 \pm 0.14$ & $ 2.26 \pm 0.10$ &  $ 1.61 \pm 0.06$ \\
ESO477-7   & -               & $4.70 \pm 0.31$ & $ 3.69 \pm 0.27$ & $ 2.41 \pm 0.16$ &  $ 0.84 \pm 0.10$ \\
ESO197-10  & -               & $4.08 \pm 0.23$ & $ 3.08 \pm 0.17$ & $ 2.33 \pm 0.13$ &  $ 1.33 \pm 0.09$ \\
ESO355-8   & -               & $3.20 \pm 0.18$ & $ 2.19 \pm 0.13$ & $ 0.97 \pm 0.09$ &  $ 0.57 \pm 0.07$ \\
NGC1297    & -               & $4.05 \pm 0.14$ & $ 3.05 \pm 0.10$ & $ 2.24 \pm 0.09$ &  $ 0.83 \pm 0.07$ \\
ESO118-19  & -               & $4.60 \pm 0.51$ & $ 3.63 \pm 0.41$ & $ 2.74 \pm 0.31$ &  $ 2.05 \pm 0.23$ \\
UGC4449    & -               & $3.96 \pm 0.17$ & $ 2.96 \pm 0.12$ & $ 2.31 \pm 0.10$ &  -                \\
NGC2968    & $4.93 \pm 0.17$ & $4.34 \pm 0.10$ & $ 3.31 \pm 0.08$ & $ 2.97 \pm 0.07$ &  $2.52 \pm 0.06$  \\
NGC4583    & $5.38 \pm 0.40$ & $4.25 \pm 0.21$ & $ 3.26 \pm 0.17$ & $ 2.31 \pm 0.09$ &  $1.53 \pm 0.12$  \\
NGC5249    & $4.57 \pm 0.76$ & $4.24 \pm 0.12$ & $ 3.20 \pm 0.09$ & $ 2.55 \pm 0.07$ &  $2.29 \pm 0.08$  \\
NGC5311    & $4.40 \pm 0.29$ & $4.18 \pm 0.23$ & $ 3.16 \pm 0.18$ & $ 2.32 \pm 0.13$ &  $1.17 \pm 0.60$  \\
NGC6314    & $4.79 \pm 0.55$ & $4.18 \pm 0.30$ & $ 3.17 \pm 0.23$ & $ 2.26 \pm 0.17$ &  $1.22 \pm 0.08$  \\
NGC7399    & $4.48 \pm 0.29$ & $4.15 \pm 0.21$ & $ 3.18 \pm 0.17$ & $ 1.99 \pm 0.09$ &  $1.79 \pm 0.09$  \\
\hline 
\hline
Object    & $\dfrac{A_U}{E(B-R)}$  & $\dfrac{A_B}{E(B-R)}$ & $\dfrac{A_V}{E(B-R)}$ & $\dfrac{A_R}{E(B-R)}$ & $\dfrac{A_I}{E(B-R)}$ \\
\hline
MW galaxy  &    2.65         &   2.30           & 1.72 & 1.30          & 0.83             \\
NGC2534    & $2.94 \pm 0.15$ & $ 2.63 \pm 0.08$ & - & $ 1.64 \pm 0.05$ & - \\
NGC3656    & $2.36 \pm 0.17$ & $ 2.28 \pm 0.08$ & - & $ 1.28 \pm 0.04$ & - \\
NGC3665    & $2.76 \pm 0.11$ & $ 2.47 \pm 0.10$ & - & $ 1.47 \pm 0.07$ & - \\
NGC4370    & $3.70 \pm 0.14$ & $ 2.27 \pm 0.12$ & - & $ 2.32 \pm 0.09$ & - \\
NGC4374    & $2.80 \pm 0.15$ & $ 2.35 \pm 0.08$ & - & $ 1.35 \pm 0.06$ & - \\
NGC5363    & $2.88 \pm 0.12$ & $ 1.98 \pm 0.08$ & - & $ 0.97 \pm 0.04$ & - \\
NGC5485    & $2.99 \pm 0.11$ & $ 2.45 \pm 0.09$ & - & $ 1.44 \pm 0.05$ & - \\
NGC7625    & $2.41 \pm 0.09$ & $ 1.83 \pm 0.07$ & - & $ 0.94 \pm 0.05$ & - \\ 
\hline
\end{tabular}
\end{table*}
\subsection{Dust mass estimation}
\label{S:dustmass}
The total extinction values are used to obtain one estimate for the dust mass in the DLs, as described in detail in Finkelman et al.\ (2008). We assume that the chemical composition of the extragalactic dust grains is uniform throughout each galaxy and is similar to that of the dust in the MW. For the chemical composition and the size of the dust grain we use the Mathis, Rumpl \& Nordsieck (1977) two-component model consisting of individual spherical silicate and graphite grains with an adequate mixture of sizes.
The dust mass is then calculated by integrating the dust column density over the image areas S occupied by DLs, yielding (see Finkelman et al.\ 2008):
\begin{equation}
M_d=\mbox{S}{\times}\Sigma_d=\mbox{S}{\times}l_d{\times}\int\limits_{a_-}^{a_+}\frac{4}{3}{\pi}a^3{\rho}_dn\left(a\right)da .
\end{equation}
where $a$ represents a certain grain size; $n\left( a\right)$ represents the size distribution of dust grains; $a_-$ and $a_+$ represent the lower and upper cutoffs of the size distribution, respectively; ${\rho}_d$ is the specific grain densities; $l_d$ represents the dust column length along the line of sight; and $n_H$ is the hydrogen number density.

Although $l_d$ and $n_H$ cannot be directly measured from our data, the value of the product $l_d \times n_H$ can be inferred from the measured total extinction and by calculating the extinction efficiency (Finkelman et al.\ 2008). We also assume that the dust forms a foreground screen for the galaxy and neglect any intermix of the dust with stars within the galaxy or the possible twisting of dark lanes around or through the host galaxies. Therefore, estimating the dust mass from total extinction values provides only a lower limit to the true dust content of the host galaxies. We also note that uncertainties in the average grain size or in the lower and upper cutoffs result in an uncertainty in dust mass, which can lead to an error of tens of percent, depending on the deviation from the Galactic extinction law. 

An alternative method for estimating the dust mass is based on the expected far-IR emission of the dust grains. Assuming thermal equilibrium, the relation between the dust mass and the observed far-IR flux density, $I_{\nu}$, follows (Hildebrand 1983)
\begin{equation}
M_d=\frac{4}{3} a \rho _d D^2 \frac{I_{\nu}}{Q_{\nu}B_{\nu}(T_d)}
\end{equation}
where ${\rho}_d$ and $D$ are the specific grain mass density and distance of the galaxy in Mpc (assuming $H_0=72$ km s$^{-1}$ Mpc$^{-1}$), respectively; $Q_{\nu}$ and $B_{\nu}(T_d)$ are the grain emissivity and the Planck function for the temperature $T_d$ at frequency $\nu$, respectively.
The dust grain temperature can be calculated from the IRAS flux densities at 60 $\mu$m and 100 $\mu$m using $T_d=\left( \frac{S_{60}}{S_{100}}\right)^{0.4}$  (Young et al.\ 1989). Hot circumstellar dust may also contribute to the 60 $\mu$m and 100 $\mu$m passbands, although it seems this effect on the 60 $\mu$m and 100 $\mu$m flux densities is only a few percent (Goudfrooij \& de Jong 1995).
The quantity $a{\rho}_d/Q_{\nu}$ can be calculated for the silicate and graphite grains assuming it is independent of $a$ for  $\lambda \gg a$ (see Hilderbrand 1983).

These dust mass estimates represent lower limits, since IRAS was not sensitive to dust cooler than about 20K, which emits mostly at $\lambda>100$ $\mu$m.
Table \ref{t:dmass} lists the estimated dust mass from the total optical extinction and the estimated dust mass based on IRAS flux densities taken from the catalog of Knapp et al.\ (1989) for bright early-type galaxies.

\begin{table*}
 \centering
\caption{Dust mass and morphology. \label{t:dmass}}
\begin{tabular}{|l|cccc|c|}
\hline
Object     & \multicolumn{2}{c}{IRAS flux (mJy)} &  $ \mbox{Log} \left(\frac{M_d}{M_ \odot}\right)_{\mbox{IRAS}}$ & $ \mbox{Log}\left( \frac{M_d}{M_ \odot}\right) _{\mbox{optical}} $ & Dust Morph.\ \\
   {}      & 60$\mu$m     & 100$\mu$m      &     {}           &    {}          & {}\\
\hline
IC1575	   & -            & -              & -                & $6.73 \pm 0.02$ & minor DL\\ 
NGC662     & $2125 \pm234$& $4589\pm505$   & $6.65\pm0.15$    & $6.16 \pm 0.01$ & complex\\ 
ESO477-7   & $0 \pm 38$   & $0 \pm 119$    & $<6.5$*          & $5.51 \pm 0.02$ & intermediate DL\\
NGC708     & $210 \pm 42$ & $590\pm183$    & $5.82\pm0.41$    & $5.64 \pm 0.02$ & nuclear\\ 
ESO197-10  & $90  \pm 28$ & $170 \pm 99$   & $5.16\pm0.68$    & $6.19 \pm 0.03$ & minor DL\\
ESO355-8   & $0 \pm 42$   & $0 \pm 210$    & $<6.8$*          & $6.31 \pm 0.03$ & intermediate DL\\
NGC1199    & $0 \pm 42$   & $0 \pm 80$     & $<5.7$*          & -               & nuclear \\
NGC1297    & $<38$        & $<128$         & $<5.4$*          & $4.66 \pm 0.07$ & minor DL\\
ESO118-19  & $733  \pm 37$& $1391 \pm 125$ & $4.56\pm0.11$    & $4.60 \pm 0.02$ & minor DL \\
NGC2534    & $380  \pm 50$& $440 \pm 184$  & $4.79\pm0.86$    & $4.98 \pm 0.24$ & minor DL\\ 
UGC4449    & -            & -              & -                & $6.67 \pm 0.01$ & minor DL \\ 
NGC2968    & -            & -              & -                & $5.93 \pm 0.01$ & major DL \\ 
Mrk33      & $4786 \pm479$& $5488 \pm 549$ & $5.12\pm0.11$    & -               & -       \\ 
UGC5814    & $1095 \pm 66$& $2588 \pm 155$ & $7.00\pm0.08$    & $6.88 \pm 0.01$ & major DL\\
NGC3656    & $2333\pm117$ & $5442\pm272 $  & $6.19\pm0.07$    & $5.60 \pm 0.01$ & minor DL \\ 
NGC3665    & $1917\pm115$ & $6344\pm 317$  & $6.22\pm0.08$    & $5.04 \pm 0.03$ & major DL\\
NGC4370    & $990 \pm 40$ & $2900 \pm 77$  & $4.91\pm0.05$    & $4.57 \pm 0.02$ & major DL  \\ 
NGC4374    & $502 \pm 5$  & $980 \pm 215$  & $4.38\pm0.26$    & $3.77 \pm 0.16$ & nuclear \\ 
NGC4583    & -            & -              & -                & $5.34 \pm 0.07$ & minor DL\\ 
NGC5249    & -            & -              & -                & $6.41 \pm 0.01$ & minor DL  \\ 
NGC5311    & $522 \pm 52$ & $1868 \pm 187$ & $6.00\pm0.16$    & $5.98 \pm 0.01$ & major DL  \\
NGC5363    & $1693 \pm118$& $5150\pm 464$  & $5.52\pm0.13$    & $5.41 \pm 0.01$ & spiral disk  \\
NGC5485    & $150 \pm 34$ & $850 \pm 88$   & $5.86\pm0.28$    & $4.80 \pm 0.01$ & minor DL  \\ 
NGC6251    & $0 \pm 23$   & $ 0 \pm 96$    & $<6.6$*          & $4.62 \pm 0.05$ & nuclear \\
NGC6314    & $500 \pm 45$ & $ 1659 \pm 20$ & $6.66\pm0.17$    & $6.06 \pm 0.09$ & major  \\ 
NGC6702    & $0 \pm 48$   & $2380 \pm 209$ & $<7.6$*          & $4.40 \pm 0.02$ & nuclear  \\ 
NGC7052    & $448 \pm 45$ & $1477 \pm 118$ & $6.33\pm0.14$    & $5.70 \pm 0.03$ & nuclear \\ 
NGC7399    & -            &  -             & -                & $6.14 \pm 0.01$ & major DL \\ 
NGC7432    & -            &  -             & -                & -               & nuclear  \\ 
NGC7625    & $9326 \pm560$& $17770\pm1066$ & $6.12\pm0.08$    & $5.73 \pm 0.01$ & complex  \\ 
\hline
\end{tabular}
\label{t:DMass}
\begin{minipage}[]{13cm}
\begin{small}
*Upper limits assuming $\mbox{T}=20\mbox{K}$
\end{small}
\end{minipage}
\end{table*}
\subsection{Hydrogen gas mass}
Assuming case B recombination (Osterbrock 1989), we can roughly estimate the total IG mass from the H$\alpha$ luminosities. For a given electron temperature and density the mass can be written as:
\begin{equation}
M_{{\mbox{HII}}}  = \left( L_{\mbox{H}\alpha} m_H / n_e\right) / \left( 4 \pi j_{\mbox{H}\alpha} / n_e n_p \right) 
\end{equation}
where $m_H$ is the mass of the hydrogen atom; $n_e$ and $n_p$ are the number of electrons and protons per $\mbox{cm}^{3}$ and $j_{\mbox{H}\alpha}$ is the emission coefficient of the H$\alpha$ line. Assuming an electron temperature of $\sim10^4$ K the expression above can be simplified as:
\begin{equation}
M_{{\mbox{HII}}} = 2.33\times 10^3 \, \left(\dfrac{L_{\mbox{H}\alpha}}{10^{39}\mbox{ erg s}^{-1}}\right) \left( \dfrac{10^3 \mbox{ cm}^{-3}}{n_e} \right) M_\odot.
\end{equation}
We estimate the HII masses assuming a typical electron density of  $\sim 10^3$ cm$^{-3}$ (see Goudfrooij et al.\ 1994a) and list the results in the sixth column of Table \ref{t:Halpha}. 

To properly derive the H$\alpha$ luminosity requires measuring the ratios of the [NII] and H$\alpha$ lines, which are not available here. The HII masses are therefore upper limits, while the true values are likely lower by a factor of $\sim2-3$ (see Goudfrooij et al.\ 1994a; Macchetto et al.\ 1996);
\subsection{Star formation histories}
\label{S:sfh}
In order to determine the stellar population of each galaxy we fit two-component star formation histories to the galaxy derived extinction-corrected colours. The old component, the bulk of the stellar population, is modelled as an instantaneous starburst. We add to this a young component represented by either an instantaneous burst or an exponentially declining star formation rate with an e-folding time of 100 Myr (see Schawinsky et al.\ 2007). By varying these parameters we take into account various star formation histories, including single bursts. 
The library model colours are calculated based on the simple and complex stellar population generated by the galaxy evolution models (GEMs) of Bruzual \& Charlot (2003, hereafter BC03).
We use the default models pre-calculated in BC03 using the Padova 1994 evolutionary tracks and the Salpeter (1955) initial mass function (IMF) with lower and upper mass cutoffs of 0.1$M_\odot$ and 100$M_\odot$, respectively. The models calculate spectra and galactic colours at 220 time steps, which range from 0.1 Myr and 20 Gyr, for six different metallicities in the range $0.02Z_\odot<z<1.5Z_\odot$. 
Our library of models for each GEM is created from linear combinations of two bursts, where the variables are the age and metallicity of each burst and the mass fraction of the young component. In addition, we allow for a variation of the stellar formation scenario of the younger population, thus our model to fit consists of six independent parameters. 

While our observations focus on the optical regime, the spectral coverage of most of the objects does not always include the entire  U, B, V, R \& I broad-band filter set. To improve the wavelength coverage, we retrieve 2MASS J, H \& K$_s$ band imaging (see Skrutskie et al.\ 2006) and data from SDSS DR7 for the galaxies for which they are available. This is done by matching the aperture to that used for our observations and calibrating the data with the formal pipelines. This allows in most instances to determine fully the two-component stellar population of the galaxy. The average errors for the photometric calibration are typically $\leq 0.02$ mag for the SDSS and 2MASS observations, $\leq 0.05$ mag for the VATT and SAAO observations, and $\leq 0.10$ mag for the WO observations, but may increase after correcting for extinction, most notably for the bluer bands. 

\begin{figure*}
\caption{The observed optical and near-IR colours and the $EW_{\mbox{H}\alpha}$ plotted vs.\ the corresponding values obtained from the adopted models. The upper panels correspond to the fitting results to a model with a second instantaneous burst, while the lower panels correspond to the fitting results to a model with a second exponentially decaying burst with an e-folding time of 0.1 Gyr. The plots illustrate the difficulty in distinguishing between the two formation scenarios. \label{f:fit}}
\begin{center}
\begin{tabular}{rlrlrl}
{} & \includegraphics[width=5cm]{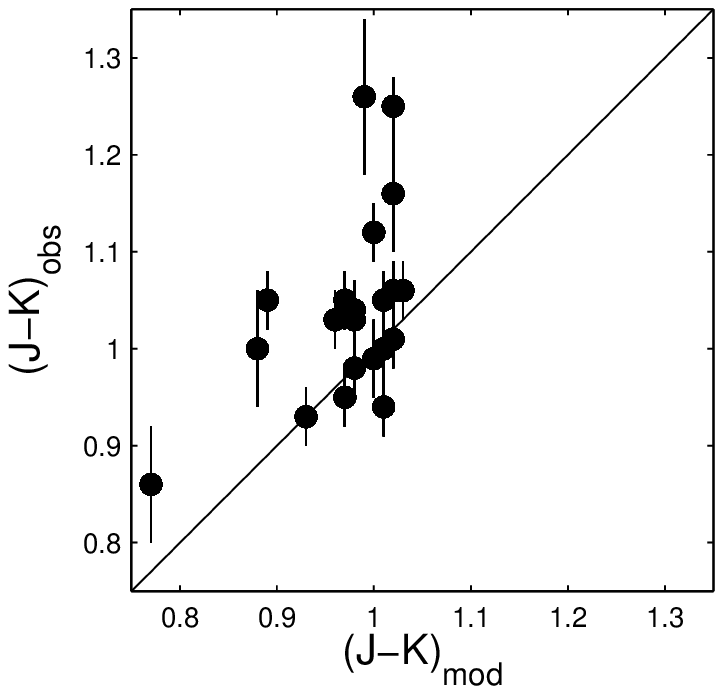} & {} & \includegraphics[width=5cm]{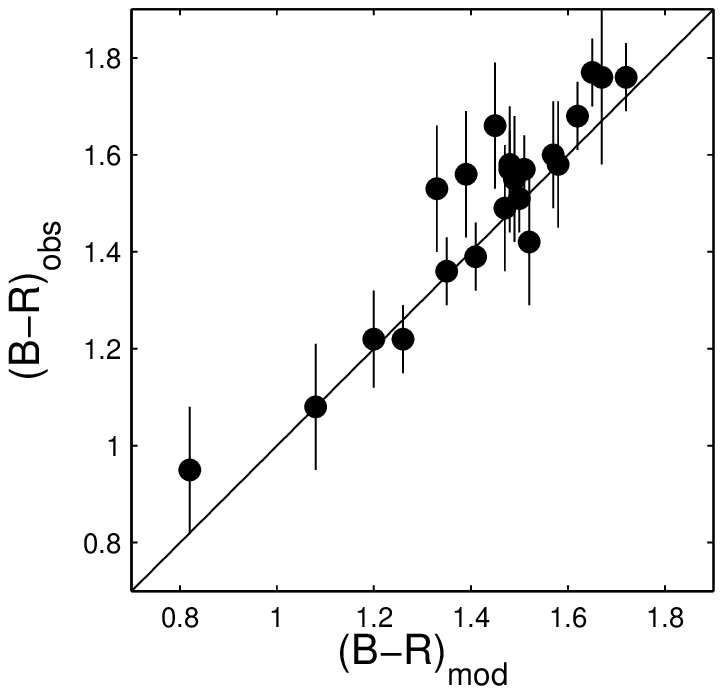} & {} & \includegraphics[width=5cm]{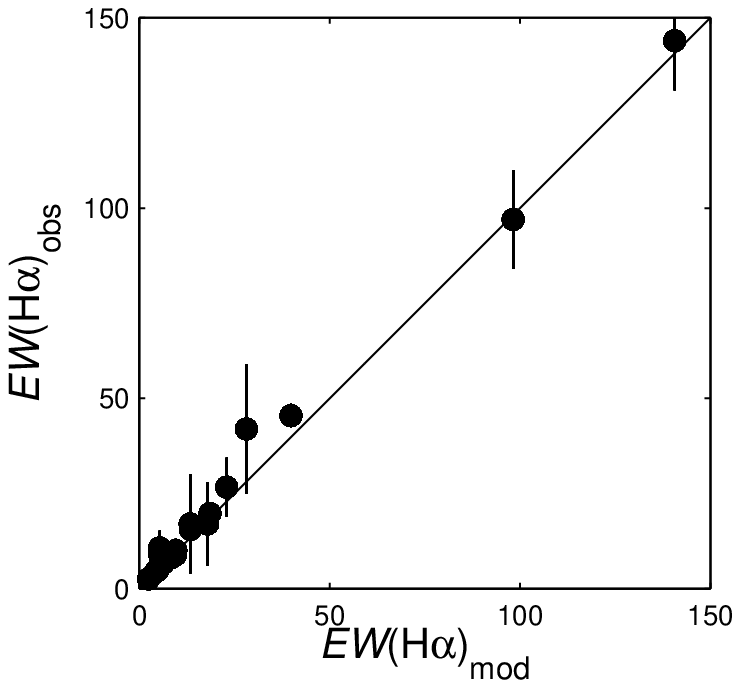} \\
{} & \includegraphics[width=5cm]{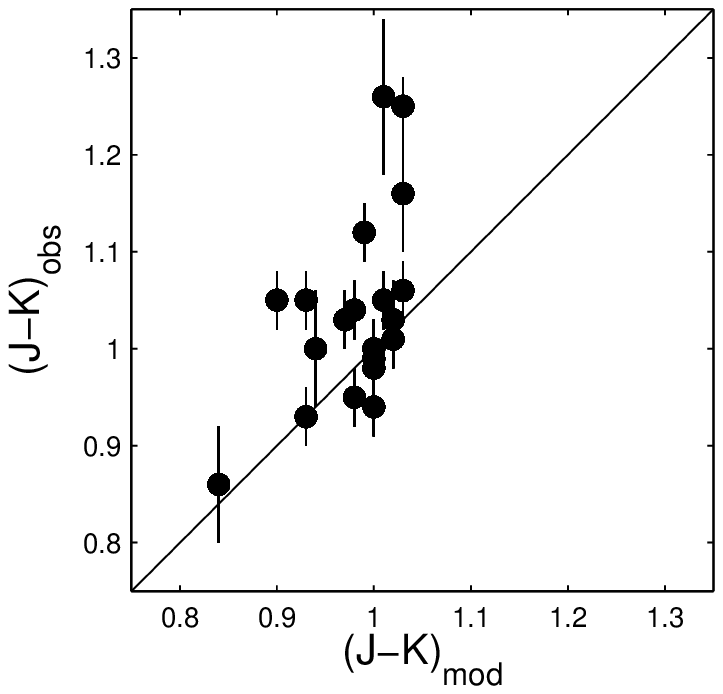} & {} & \includegraphics[width=5cm]{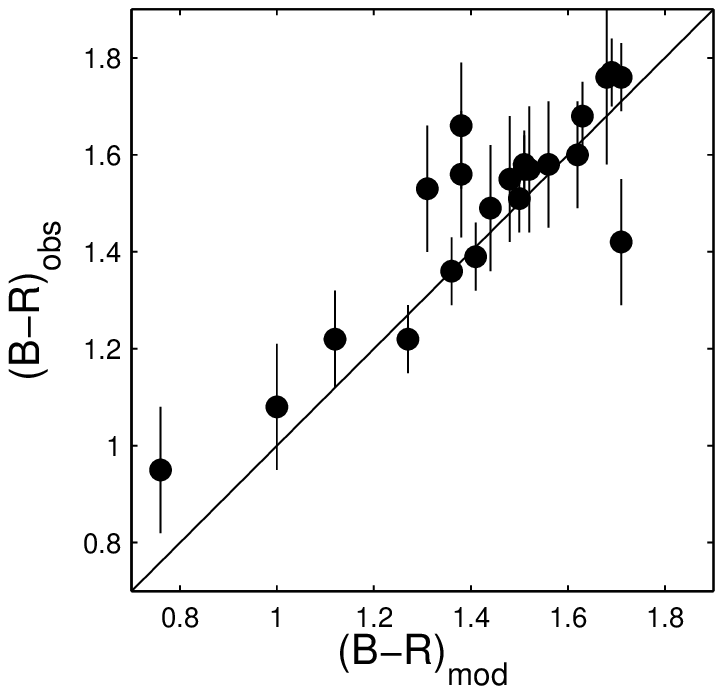} & {} & \includegraphics[width=5cm]{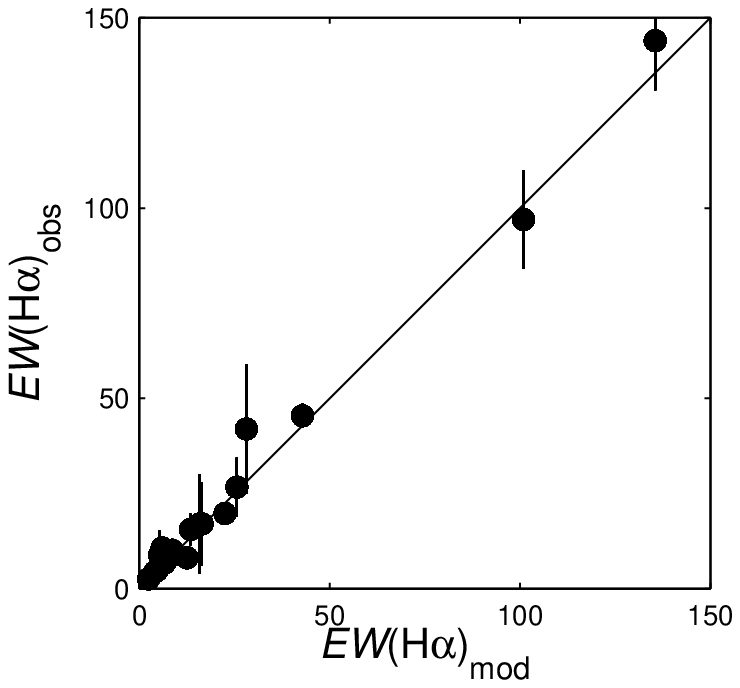} 
\end{tabular}
\end{center}
\end{figure*}

The best fit to the observed colours is obtained by performing a $\chi ^2$ test by minimizing the following expression: 
\begin{equation}
\chi ^2 = \Sigma _i \dfrac{\left( \mbox{colour}_{i,obs} - \mbox{colour}_{i,mod} \right)^2}{\sigma_i ^2} .
\end{equation}
where $\sigma_i$ is the estimated error for the observed $\mbox{colour}_i$.
To enhance the fit reliability, the $EW_{\mbox{H}\alpha}$ is also derived from the models. Since BC03 essentially predicts the stellar spectra and not the nebular emission lines, we calculate the $EW_{\mbox{H}\alpha}$ from the number of ionizing Lyman-continuum ($N_{\mbox{LyC}}$) photons predicted by the GEMs. Assuming a simple case B hydrogen recombination theory, the measured H$\alpha$ luminosity is related to the LyC photon flux (Osterbrock 1989) as follows:
\begin{equation}
 N_{\mbox{LyC}}=7.43 \times 10^{11} \, L\left(\mbox{H} \alpha \right) 
\end{equation}
where $L\left( \mbox{H} \alpha \right)$ is in erg s$^{-1}$. This leads to
\begin{equation}
EW_{\mbox{H}\alpha} = 1.13 \times 10^{-52} \, \dfrac{N_{\mbox{LyC}}}{I_{\mbox{R}}}
\end{equation}
where $I_{\mbox{R}}$ is calculated from the R-band absolute magnitude computed in the models so as to be consistent with the method described in \S~\ref{line_calib}. We note that no extinction correction is applied to the observed $EW_{\mbox{H}\alpha+\mbox{[NII]}}$ values. The model fitting results are listed in Table \ref{t:fitresults}. In addition, to illustrate the agreement between the observations and models we plot in Fig.\ \ref{f:fit} the observed optical and near-IR colours and the $EW_{\mbox{H}\alpha}$ vs.\ the corresponding values obtained from the adopted models.
\begin{table*}
 \centering
  \caption{Stellar population synthesis models fitting results. Upper and lower case numbers represent upper and lower values at 1$\sigma$. 
  \label{t:fitresults}}
\begin{tabular}{|l|ccccc|}
\hline
{}         & \multicolumn{2}{|c}{Instantaneous Burst} & \multicolumn{3}{c|}{Instantaneous Burst/Exponential decay ($\tau=100$Myr)}\\
Galaxy     & First Burst (Gyr)    & Metal.\ [$Z_\odot$]   & Second Burst (Myr) & Metal.\ [$Z_\odot$] & fraction [\%]\\
\hline
IC1575	   & $1.7_{1.1}^{3.7}$ & 1       & $2.4_{0.6}^{6.6}$/$21.9_{6.0}^{102.0}$   & 1.5/1            & $1$/$1$ \\[4pt]
NGC662     & $4.3_{1.7}^{12.5}$   & 1.5     & $10.0_{8.7}^{12.5}$/$53_{17}^{102}$ & $0.2$/$1_{0.4}^{1.5}$ & $25_{20}^{30}$/$28_{23}^{33}$ \\[4pt]
ESO477-7   & $20_{16.3}$          & 1.5     & $2.7_{2.4}^{3.3}$/$57.1_{20.1}^{180.1}$ & $0.02$/$0.2_{0.02}^1$ & $1^5$/$3_1^5$ \\ [4pt]
NGC708     & $3.5_{3.0}^{5.5}$    & 1.5    & $3.3_{2.2}^{4.0}$/$4.4_{1.6}^{14}$ & $1.5_{1.0}$/$1.5_{1.0}$ & $2_1$/$2_1$ \\[4pt]
ESO197-10  & $6.3_{2.6}^{20.0}$ & $1_{0.4}^{1.5}$  & $21.8_{13.2}^{180.2}$/$255_{72}^{321}$ & $0.05_{0.02}^{0.4}$/$0.4_{0.02}^{1.5} $ & 1/$2_1^3$ \\[4pt]
ESO355-8   & $8.8_{6.3}^{12.0}$   & 1       & $4.2_{3.2}^{5.0}$/$48_{21}^{72}$ & $0.02^{1}$/$0.02^{0.4}$ & $2_1^4$/$1$ \\[4pt]
NGC1199    & $20_{15}$            & 1.5     & $37.0_{20.9}^{101.5}$/$404_{360}^{719}$ & $1.5_{0.2}$/$1.5$ & $7_4^{10}$/$9_8^{10}$ \\[4pt]
NGC1297    & $20.0_{5.0}$ & 1.5 & $6.6_{5.8}^{12.6}$/$160_{43}^{321}$ & $1.5$/$1.5$ & $16_{10}^{20}$/$6_{2}^{10}$ \\[4pt]
ESO118-19  & $6.3_{3.0}^{8.5}$    & 1.5     & $20.1_{10.0}^{100.0}$/$453_{360}^{640}$ & 0.02/$0.02^{0.05}$ & $4_3^5$/$10_{9}^{11}$ \\[4pt]
NGC2534    & $4.3_{2.7}^{6.3}$    & 1.5     & $6.6_{6.0}^{7.6}$/$227_{114}^{360}$ & 1.5/1.5 & $16_{11}^{9}$/$15_{11}^{19}$ \\[4pt]
Mrk33      & $6.5_{1.7}^{20.0}$    & 1.5     & $10.0_{8.3}^{12.0}$/$55_{17}^{102}$ & 0.4/$0.4_{0.2}^{1.0}$ & $38_{32}^{42}$/$37_{22}^{42}$ \\[4pt]
UGC5814    & $1.8_{1.1}^{3.7}$    & 1.5   & $3.0_{2.6}^{7.9}$/$4.3_{3.3}^{57}$ & $0.2_{0.02}^{1.5}$/$0.2_{0.02}^{1.5}$&$1^2$/$1^2$ \\[4pt]
NGC3656    & $7.5_{4.3}^{11.3}$   & 1.5     & $5.8_{5.2}^{6.3}$/$114_{37}^{227}$ & 1.5/1.5 & $13_7^{15}$/$11_6^{14}$ \\[4pt]
NGC3665    & $3.5_{2.8}^{5.0}$    & 1.5     & $6.0_{4.6}^{6.9}$/$57_{7}^{181}$ & 1.5/$1.5_{0.2}$ & $4_3^{6}$/$2_1^{5}$ \\[4pt]
NGC4370    & $1.8_{1.7}^{2.6}$    & 1.5     & $6.0_{4.8}^{7.6}$/$114_{72}^{161}$ & 1.5/$1.5$ & $1$/$1$ \\[4pt]
NGC4374    & $17.3_{11.3}^{20.0}$ & 1.5     & $13.6_{8.3}^{20.9}$/$286_{45}^{-}$ & $1.5$/$1.5_{0.02}$ & $10_7^{13}$/$1_0^5$ \\[4pt]
NGC4583    & $8.5_{2.6}^{20.0}$   & 1.5     & $4.8_{3.8}^{6.6}$/$203_{57}^{360}$ & 1.5/$1.5_{0.02}$ & $3_1^6$/$6_2^{10}$ \\[4pt]
NGC5249    & $8.0_{2.6}^{20.0}$   & 1.5     & $4.0_{3.0}^{5.5}$/$90_{3}^{168}$ & $0.4_{0.02}^{1.5}$/$0.4_{0.02}^{1.5}$& $3_2^{6}$/$3_2^{5}$ \\[4pt]
NGC5363    & $8.5_{7.0}^{12.0}$   & 1       & $12.0_{9.6}^{14.5}$/$81_{13}^{227}$ & $1$/$1_{0.2}^{1.5}$ & $2_1^3$/$2_{1}^{3}$ \\[4pt]
NGC5485    & $4.5_{3.3}^{5.3}$    & 1.5     & $9.6_{7.6}^{12.6}$/$360_{286}^{509}$ & $0.4_{0.2}^{1.5}$/$1_{0.2}^{1.5}$ & $7_6^8$/$12_{10}^{14}$ \\[4pt]
NGC6251    & $7.0_{5.5}^{9.0}$    & 1.5     & $6.3_{5.2}^{6.6}$/$38_{11}^{102}$ & 1.5/$1_{0.2}^{1.5}$ & $3_1^5$/$1$ \\[4pt]
NGC6702    & $4.5_{3.3}^{5.5}$    & 1.5     & $6.6_{5.8}^{11.5}$/$181_{72}^{360}$ & $0.4^{1.5}$/$1_{0.05}^{1.5}$ & $1$/$1$ \\[4pt]
NGC7052    & $14.3_{11.3}^{18.8}$ & 1.5    & $6.6_{6.3}^{6.9}$/$50_{17}^{181}$ & $1.5$/$1.5$ & $5_4^6$/$1$ \\[4pt]
NGC7399    & $12.5_{3.7}^{20.0}$  & 1.5    & $6.3_{5.0}^{26.3}$/$25_{7}^{57}$ & $0.2_{0.02}$/$0.2_{0.02}$ & $5_2^8$/$8_5^{10}$ \\[4pt]
NGC7625    & $1.8_{1.6}^{2.1}$    & 1.5     & $10.0_{9.6}^{10.5}$/$29_{7}^{101}$ & $0.4$/$1.5$ & $9_8^{10}$/$10_8^{12}$ \\
\hline 
\end{tabular}
\end{table*}
\subsection{An example: stellar populations of NGC5363}
\label{S:ngc5363}
NGC5363 has by far the most extended spectral coverage among our sample galaxies including sufficiently deep GALEX and SST/IRAC observations to measure the flux in the IG region. We therefore use this galaxy as an example for our SFH fitting scheme for a case where an optimal data set is available.

Previous studies of the stellar population of NGC5363 assumed that it contains an almost coeval stellar population and therefore can be described by a simple stellar population (Denicol\'{o} et al.\ 2005; Sansom et al.\ 2006; Annibali et al.\ 2007). 
A more detailed inspection of the optical characteristics of NGC5363 suggests that at least some stars formed after $z\sim1$ (see \S~\ref{S:dngc5363}). However, the observed optical colours are degenerate with respect to the competing stellar population models in the sense that they cannot discriminate between them. 

Our fitting scheme for NGC5363 included the colours [(U-B), (B-R), (r-i), (i-z), (FUV-NUV), (NUV-U), (R-J), (J-H), (H-K), (R-3.6$\mu$m), (3.6$\mu$m-4.5$\mu$m)] and the $EW_{\mbox{H}\alpha}$ produced as described above. 
The average errors for the photometric calibration of the SST/IRAC magnitudes are approximately $\sim 0.01$ mag, and for the GALEX FUV and NUV magnitudes they are $\sim0.25$ mag and $\sim 0.15$ mag, respectively. 

The old component that best fits this galaxy is an $\sim 8.5$ Gyr population with Solar metallicity. This result is reproduced by both the double instantaneous burst scenarios and by the early burst followed by an exponential decay scenario, with similar likelihood. However, while both formation scenarios clearly favour the existence of a younger population in the IG disk, the double instantenous burst scenario estimates its age at $\sim 12$ Myr while the early burst followed by an exponential decay scenario estimates a $\sim 81$ Myr population. Both scenarios agree that the metallicity of the younger population is $\sim$Solar and that this population constitutes $\sim 2$\% of the overall stellar mass in the galaxy. 

To verify that our analysis is not biased by the GEM we adopt, we computed also a similar model library based on the PEGASE GEM (Fioc \& Rocca-Volmerange 1997). The fitted results matches those of BC03, with an old population of $\sim9$ Gyr age given by both scenarios, and a young population of $\sim5$ Myr given by the double instantaneous bursts scenario and of $\sim80$ Myr given by the instantaneous burst followed by an exponential decay scenario.

\begin{figure*}
\caption{$\chi^2$ contour maps. The maps represent the results of the $\chi^2$ minimum fitting with PEGASE (right) and BC03 (left). The solid lines represent the fitting results to a model with a second instantaneous burst, while the dashed lines represent the fitting results to a model with a second exponentially decaying burst with an e-folding time of 0.1 Gyr. The $\chi^2$ minimum is marked with 'x' and we indicate the 68\% ($1 \sigma$), 95\% ($2 \sigma$) and 99\% ($3\sigma$) confidence contours. \label{f:chi2m}}
\begin{center}
\begin{tabular}{rlrl}
{} & \includegraphics[]{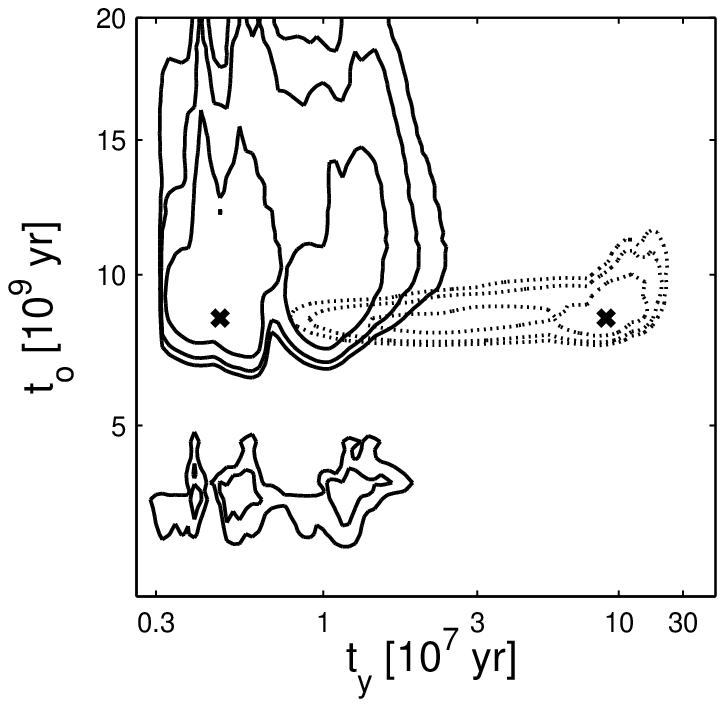} & {} & \includegraphics[]{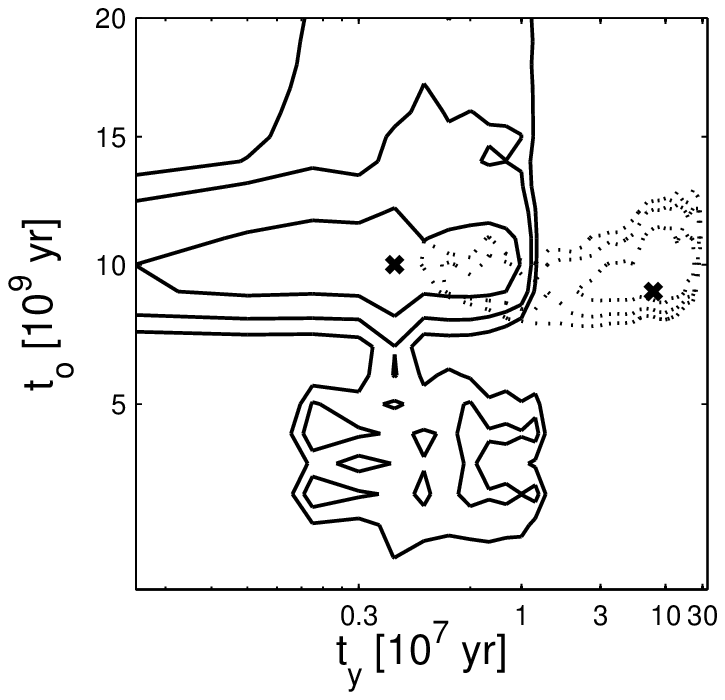} 
\end{tabular}
\end{center}
\end{figure*}

Fig.\ \ref{f:chi2m} shows $\chi ^ 2$ contour map representing the results of the $\chi^2$ minimum fitting with PEGASE (right) and BC03 (left). The solid lines represent the fitting results to a model with a second instantaneous burst, while the dashed lines represent the fitting results to a model with a second exponentially decaying burst with an e-folding time of 0.1 Gyr. The $\chi^2$ minimum is marked with 'x' and we indicate the 68\% ($1 \sigma$), 95\% ($2 \sigma$) and 99\% ($3\sigma$) confidence contours. The figure implies that one cannot distingush with confidence between the two formation scenarios tested. 

\begin{figure}
  \caption{The NGC5363 extinction law (filled circles) with comparison to the Galactic (dash-dot line), SMC (dashed line) and LMC (dotted line) extinction laws. Extinction values for the MW, SMC and LMC are taken from Gordon et al.\ (2003).}
  \centering
    \includegraphics[]{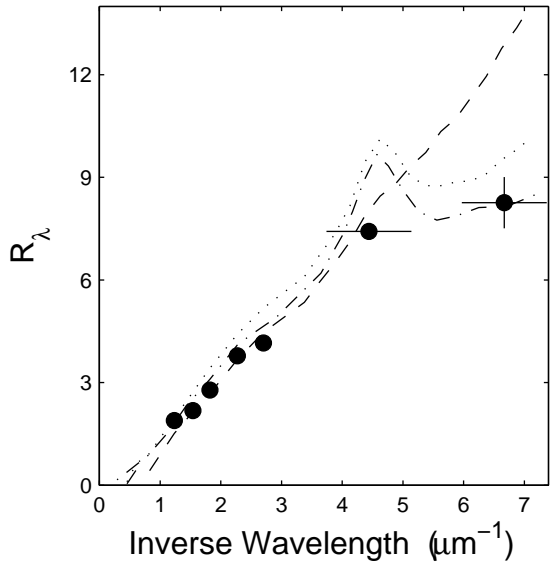}
    \label{f:NGC5363extlaw}
\end{figure}

The deep GALEX observations also allow the extension of the measurements of the extinction law to the space-UV region. 
Fig.\ \ref{f:NGC5363extlaw} shows the NGC5363 extinction law compared with the Galactic, Small Magellanic Clouds (SMC) and Large Magellanic Clouds (LMC) extinction laws (see Gordon et al.\ 2003). 
\section{Discussion}
\label{S:discuss}
\subsection{Dust and ionized gas correlations}
\subsubsection{Morphological relation}
Although our survey is targeted to observe E/S0s with $\sim$kpc sized DLs, and the objects were selected from lists where they are classified as such, we find that in eight of the 30 sample galaxies the dust is restricted to the central region where its study requires better angular resolution than our study offers. 

A close inspection of the dust morphology reveals an inner disk in most of our objects, while a large fraction also shows an extended component with a variety of spatial distributions. 
The dust structures or DLs often lie along axes different from those of the stellar body. 
We identify extended DLs along the minor-axis in 10 of the sample galaxies and major-axis DLs in six galaxies, while in five objects the DL appears to lie along intermediate-axes. Three other galaxies show a more complex dust distribution, two with clumps and one with a barred spiral-like structure. 
Since the observed objects were chosen according to visibility conditions and availability of observing nights and not as part of a complete sample, we do not attempt to account for the relative distribution of galaxies within this classification.

Several imaging surveys (e.g., Buson et al.\ 1993; Goudfrooij et al.\ 1994a; Macchetto et al.\ 1996) showed that IG is detected in nearly every E/S0 where dust is found, and that it is virtually always morphologically associated with dust absorption.
H$\alpha$ emission was detected here in 25 of the 30 studied galaxies. Our sample galaxies typically show an IG distribution characterized by strong central emission surrounded by a weak, extended diffuse emission component in the shape of an inclined disk. However, a close inspection of the line-emitting regions reveals differences in size and morphology. While the gas morphology appears to be rather smooth for most  galaxies, some objects show individual knots, clumps and filamentary structures (NGC662, NGC3656 and NGC7625) where we detect strong H$\alpha$ emission. These features coincide with the DLs and are known sites of current star formation believed to be triggered by a recent merger event (see APPENDIX \ref{S:IndivGal}). 

We could not detect with confidence H$\alpha$ emission in five objects (UGC4449, NGC2968, NGC5311, NGC6314 and NGC7432). To check this, we compare our results with data from the SDSS/DR7 archive. The SDSS spectra, obtained through 3$\arcsec$ fibers placed at the optical centers of the galaxies, confirms that the H$\alpha$ emission in these objects is below our detection limit. 

While IG features are found in all the minor- and intermediate-axis DL galaxies in our sample, they are less common among the major axis DL galaxies. Examining the dust and IG distributions shows that in objects where both dust and IG are detected, the two components lie in disks with similar inclinations with respect to the galactic plane (see figures in APPENDIX \ref{S:IndivGal}). It is therefore reasonable to assume that the extended DLs are, in most cases, projected dust disks or rings.

We find that the HII disks have mean radii of 3 kpc, and that the ratio of their size with respect to the galaxy size has a large scatter (average 0.28 and median 0.22, see Table \ref{t:apertures}). This implies that ionization sources are typically restricted to the inner regions of the galaxy, in agreement with Macchetto et al.\ (1996) who found that 54\% of the Es showed IG disk structures with mean diameters $<4$ kpc, 
while only 25\% of the sample Es had a more extended gas distribution.
\subsubsection{Mass relation}
The relation between the dust and IG components can also be tested quantitatively by searching for a correlation between the mass of dust in the DLs and the estimated IG mass. 

Our method for measuring the extragalactic dust extinction restricted its derivation to only 21 of the sample galaxies. The $R_B$ values derived in \S~\ref{S:dustext} are close to the standard Galactic value, $R_B=4.1$, at least for most of the derived extinction curves, with a mean value of $4.07\pm0.35$. The characteristic grain size responsible for the optical extinction in these galaxies is therefore similar to that in the MW, as found for a large number of dust-lane E/S0s (Goudfrooij et al.\ 1994b; Patil et al.\ 2007; Finkelman et al.\ 2008). Using the optical extinction values and considering MW-like grains we evaluated the dust mass for each galaxy, and found it in the range $\sim 10^4$ to $10^7M_{\odot}$.
This is in good agreement with the earlier estimates for E/S0s with dark lanes (see Finkelman et al.\ 2008 and references therein). 

In addition, we compared dust masses derived using reddening and optical extinction with those using IRAS flux densities and found that the latter are up to one order of magnitude higher than the first, also in agreement with previous studies (see Finkelman et al.\ 2008). As suggested by Goudfrooij \& de Jong (1995), this discrepancy may indicate the presence of a large component of diffuse dust distributed over the galaxy, or of DLs or dust rings that are optically thin in the visible and are located far from the galactic center.
\begin{figure}
  \caption{Dust mass vs.\ HII mass. Our sample galaxies are represented by filled circles, open circles are from Goudfrooij et al.\ (1994a, 1994b) and filled triangles are from Ferrari et al.\ (1999).
  The dashed line is the fitted linear trend with a slope of $\sim0.8$. }
  \centering
    \includegraphics[]{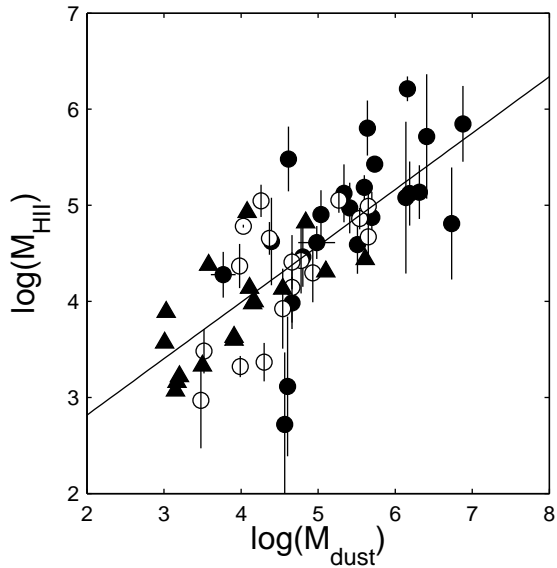}
    \label{f:DustHa}
\end{figure}

In Fig.\ \ref{f:DustHa} we plot the logarithm of the dust mass measured within the DLs (not the diffuse component) vs.\ the logarithm of the IG mass. Although the points are somewhat scattered, the correlation between the two components is clear, with a slope of $\sim0.8$.
While the H$\alpha$ luminosity is not corrected for extinction, the scattering is also likely the result of large uncertainties in estimating the electron densities and the [NII]/H$\alpha$ line ratio values, not measured in this study. 

Since the gas-to-dust ratio in early-type galaxies is typically between a few hundred to a few thousand, the low $M_{\mbox{HII}}$/$M_d$ values detected in the IG disks of the sample galaxies imply that most of the gas in the disks is not ionized but in atomic or molecular form, as argued for other such galaxies by Oosterloo et al.\ (2002). 
\subsection{Ionization mechanisms}
The gas ionization and the distribution of the IG and dust were investigated in many spectroscopic and imaging surveys. However, it seems that the question of what powers the observed nebular emission in early-type galaxies is far from being solved.
The current status of the study of IG in E/S0 galaxies suggests a large variety of mechanisms possibly responsible for the ionization in different objects and within single galaxies (see Macchetto et al.\ 1996; Sarzi et al.\ 2006; Sarzi et al.\ 2009). Below we discuss which of the mechanisms are likely to account for the apparent ionization in our sample galaxies. However, we note that disentangling with confidence the relative importance of these mechanisms may require more detailed observations than reported here, and a broader wavelength coverage. 

\subsubsection{The effect of a central AGN}
\label{S:AGN}
At least some of the sample galaxies host an active galactic nucleus (AGN; see Table \ref{t:Obs}). 
We therefore consider the possible effect of non-stellar continuum radiation from an AGN on the observed flux of a galaxy. 
A useful diagnostic tool to distinguish between different ionization sources and to probe AGN activity is the line-ratio diagram (Baldwin, Phillips, \& Terlevich 1981; hereinafter BPT). 

\begin{figure}
  \caption{BPT diagram for the inner 3$\arcsec$ of our galaxies. The dominant ionization source can be distinguished with sufficient information of the emission-lines, which enables to put the sample galaxies on the BPT diagnostic diagram. The data points were taken from SDSS/DR7 and Ho et al.\ (1997). The demarcation between ionization by an AGN or starburst is taken from Kauffmann et al.\ (2003).}
  \centering
    \includegraphics[]{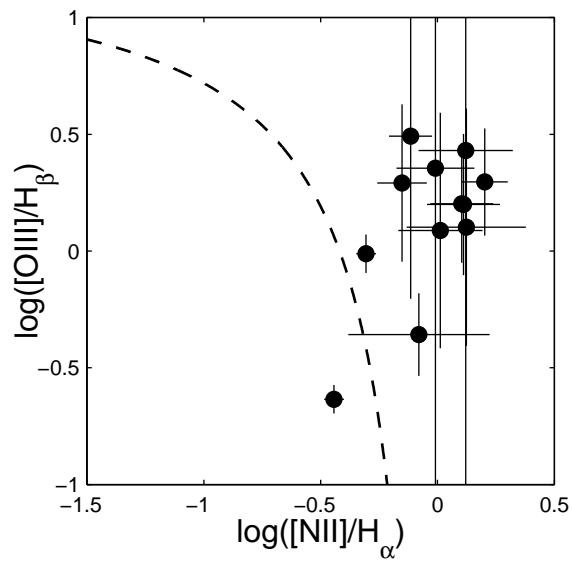}
    \label{f:BPT}
\end{figure}

Extracting SDSS/DR7 spectroscopic data for the central 3$\arcsec$ of the galaxies from the publicly available MPA/JHU catalogs we find in most cases line-ratios typical of LINERs (Ho 2008; see Fig.\ \ref{f:BPT}). 
In addition, we find that the central 3$\arcsec$ H$\alpha$ flux measurements do not exceed $\sim5$\% percent of the total emission line flux measured from the total IG aperture. 
We therefore conclude that, at least for most of the objects, the extended nature of the H$\alpha$ line emission implies that an AGN, if present, cannot produce the bulk of the ionizing flux.

Nuclear emission from an AGN is also expected to affect the observed colours of a galaxy. However, since we cannot properly test this effect in our galaxies, there is no attempt here to correct any of the colours.
\subsubsection{Non-stellar ionization mechanisms}
Hot gas can, in principle, ionize colder gas through thermal conduction (Sparks, Macchetto \& Golombek 1989; de Jong et al.\ 1990), emission of extreme-UV photons (see O'Connell 1999 for a review), or by induced shocks (Ford \& Butcher 1979; Heckman et al.\ 1989). 
However, the conditions required for these to occur are not consistent with the observations for E/S0s (Sparks et al.\ 1989; Sarzi et al.\ (2009). In addition, galaxies with strong X-ray emission do not necessarily show H$\alpha$ emission and vice versa (Macchetto et al.\ 1996). 

Furthermore, a detailed morphological comparison between X-ray and H$\alpha$ emission has only been established for a small number of E/S0s (Trinchieri \& Goudfrooij 2002; Sparks et al.\ 2004). The latter revealed highly asymmetric, clumpy and filamentary X-ray emission, remarkably similar to the IG and dust distribution, which does not resemble the smooth distribution observed in our sample galaxies.

An alternative ionization source was suggested by Serra et al.\ (2008) in their study of E/S0s with large HI masses. The authors claimed that a low-level residual star formation process cannot produce the observed LINER-like line ratios, and suggested that, at least in some cases, the ionization can be accounted for by the shocks experienced by cold gas flowing towards the inner stellar body.
\subsubsection{Stellar ionization mechanisms}
Several studies suggested that hot evolved stars provide an adequate number of Lyman continuum photons to account for the $EW$ of the H$\alpha$ emission line and for the excitation of IG in Es (di Serego Alighieri, Trinchieri \& Brocato 1990; Binette et al.\ 1994; Macchetto et al.\ 1996). 
The tight correlation in E/S0s between the detected emission-line flux and the host galaxy blue luminosity implies that the ionization is dominated by a diffuse and old stellar component (Macchetto et al.\ 1996; Sarzi et al.\ 2009). Sarzi et al.\ further noted that in most of their sample galaxies the correlation is a result of the IG emission following closely the stellar surface brightness pattern. 

However, a young stellar component, even of minor mass, is expected to dominate the optical spectrum, thus the observed blue luminosity of the IG region cannot be associated with the host galaxy alone. In addition, in many of our objects the extended structures of IG and dust do not follow the stellar surface brightness pattern, arguing against a correlation between the host galaxy and the source of ionization.

Several studies of photoionization by hot young stars in E/S0s found it to be inconsistent with the optical colours (Goudfrooij et al.\ 1994a) and the integrated optical spectra (Sadler 1987; Heckman et al.\ 1989). However, a growing line of recent evidence indicates that the presence of a young population in a significant fraction of E/S0 galaxies cannot be ruled out as previously thought.
For instance, V\'{i}lchez \& Iglesias-P\'{a}ramo (1998) studied a sample of E/S0s in compact groups and detected enhanced H$\alpha$ emission several times higher than expected from an evolved stellar population, supporting a recent star formation episode.
Kaviraj et al.\ (2007) and Schawinski et al.\ (2007) studied large samples of E/S0s with multiwavelength photometry from near-UV to near-IR and found in some galaxies evidence for current star formation contributing $\sim1-10$\% of the current stellar mass in these galaxies. 

Furthermore, a group of galaxies within the SAURON sample (de Zeeuw et al.\ 2002) shows extremely low [OIII]/H$\beta$ values and settled gaseous disks where on-going star formation is occurring (Sarzi et al.\ 2006, 2009). These objects, which constitute 4\% of the sample, are characterized by regular disk-like gas distributions and by very regular and circularly symmetric DLs. 
Studies of cold gas content (Combes, Young \& Bureau 2007) and PAH emission (Shapiro et al.\ 2009) in the SAURON sample suggested that the low-level star formation in E/S0s may actually be even more frequent than suggested by Sarzi et al.\ (2009). 
\subsection{The stellar population}
The observational evidence discussed above indicates that the ionization of gas is probably dominated by stellar sources, although its nature is still debated. 
The [NII]/H$\alpha$ values of $\sim1$, together with the low $EW_{\mbox{H}\alpha}\lesssim 3 \mbox{\AA}$ measured for some of our sample galaxies in SDSS/DR7, indicate that an old stellar population, rather than a current star formation episode, is more likely to dominate the ionization in the inner part of the galaxies (see Cid Fernandez et al.\ 2009). However, assuming [NII]/H$\alpha \sim1$ we find $EW_{\mbox{H}\alpha}>3 \mbox{\AA}$ for the total extended line emission. This result implies that the power of ionization with respect to the underlying galaxy is stronger on average than in the central galactic region. 

Fitting our data with stellar population models clearly indicates the presence of a small fraction of a young population linked with the IG region, while the bulk of the stellar component consists of a significantly older population.
Such a late low-level star formation episode in E/S0s was also found in previous studies (e.g., Kaviraj et al.\ 2007; Schawinski et al.\ 2007). 

However, we cannot determine with confidence the age of the young population, since our fitting scheme for a secondary population produces similar minimal $\chi ^2$ values for the two adopted scenarios (see Fig.\ \ref{f:fit} and Fig.\ \ref{f:chi2m}). An instantaneous burst would have produced a population of $\sim 1-10$ Myr, while an exponentially-decaying star formation process, with an e-folding time of $100$ Myr, would have produced a population ten times older ($\sim 10-100$ Myr). While for most galaxies this uncertainty may be explained by the restricted spectral coverage, which may be insufficient to fully constrain the models, we show that similar results are obtained for NGC5363 where deep UV and near-IR observations were combined with the optical information. This emphasizes the need to identify an observational parameter that could distinguish between the alternatives.

In addition, we use the Maraston (2005)\footnote{http://www-astro.physics.ox.ac.uk/~maraston} models to calculate the number of Lyman continuum photons and the expected $EW_{\mbox{H}\alpha}$ values for a stellar population evolving from an instantenous burst.
The Maraston et al.\ models account for the late evolutionary stages of stars which has significant influence on the flux of $\sim1$Gyr old stellar population (Maraston et al.\ 2005). Comparing our measurements with the model predictions indicates, in agreement with the BC03 fitting, the presence of a young $\sim10$Myr stellar population. 

While the detection of a younger population seems real, the relatively young age of $\sim1-5$ Gyr found for the old component in some of the galaxies is somewhat suspicious. This can be explained if the intrinsic stellar population of the galaxies does not follow the assumed model. Due to the observational limitations we cannot rule out the presence of a third population of $\sim10$ Myr stars within the gaseous disks or, alternatively, a more complex continuous star formation scenario.  

\subsection{The case of NGC5363}
\label{S:dngc5363}
Perhaps more puzzling is the discovery of a new and relatively rare class of Es with tightly wound star-forming spiral arms. Known examples in the literature include NGC5128, IC1459 (Goudfrooij et al.\ 1990), NGC5173 (see Vader \& Vigroux 1990) and NGC3108 (Hau et al.\ 2008). IG in a loosely-wound spiral feature, similar to an integral sign, was also found in four of the SAURON sample galaxies (Sarzi et al.\ 2009).

NGC5363 is a known LINER and a strong radio galaxy which belongs to a group of seven galaxies (Garcia 1993).
Carefully examining the continuum-subtracted H$\alpha$+[NII] image of NGC5363, we note that the spatial distribution of the HII regions also resembles a barred spiral. The colour map in Fig.\ \ref{fig:BImaps6} clearly shows the heavily obscured inner bar to be part of a more complex dust structure that follows the spiral structure and extends further along the major axis. 
Morphologically, the NGC5363 dust structure bears some resemblance to NGC5128, a prototype minor-axis dust-lane E, which appears as a barred spiral residing inside an elliptical (Mirabel et al.\ 1999). 

The NGC5363 extinction law does not show a strong 2174$\mbox{\AA}$ bump, but is otherwise in agreement with the MW extinction law (see Fig.\ \ref{f:NGC5363extlaw}). The 2174$\mbox{\AA}$ bump is generally believed to be linked with the presence of PAH particles (Li \& Greenberg 2003), and although its absence may imply a low graphite abundance compared to that of the MW dust grains, the detection of near-IR dust emission in NGC5363 supports the presence of such particles (Pahre et al.\ 2004). Therefore, if the dust particles within the dust disk of NGC5363 are truly similar to those in the MW, then this deviation in the near-UV might be the consequence of the spectral coverage of the GALEX NUV band. The NUV band is much broader than the 2174$\mbox{\AA}$ feature width and therefore transmits photons at longer wavelength, where the extinction is significantly lower. 

Analysis of SST/IRAC images shows that the stellar body of NGC5363 matches a model consisting of a de Vaucouleurs bulge plus an exponential disk (Pahre et al.\ 2004). 
The spiral-like spatial distribution shown by the 8.0$\mu$m image does not follow the stellar 3.6$\mu$m emission indicating that most of the 8.0$\mu$m emission is probably non-stellar. 
This also indicates that the dust is not produced by stellar mass loss from an AGB population (see also Shapiro et al.\ 2009), 
implying that internal mechanisms could not account for the observed dust mass (Patil et al.\ 2007).
These findings, combined with the existence of shells and CO emission (Sanders \& Mirabel 1985) and the orthogonal angular momentum of the gas and stars (Sharples et al.\ 1983), suggest that this galaxy is the result of an interaction with another galaxy by either accretion or merger, and provide strong evidence for the external origin of the ISM in NGC5363

The extended 8.0$\mu$m emission, which perfectly coincides with the HII emission, rules out a significant contribution of non-stellar continuum emission from an AGN. The close relation between the 8.0$\mu$m emission and the distribution of the IG and dust implies that the near-IR non-stellar emission may be dominated by PAH features. The UV radiation field required for the PAH excitation could be produced by a recent star formation event (Peeters, Spoon \& Tielens 2004; Wu et al.\ 2005) or by post-AGB stars and planetary nabulae (Peeters et al.\ 2002). 

The $\mbox{[NUV-r]}$ colour of NGC5363, $5.40\pm0.15$, is only marginally consistent with a recent star formation episode (Kaviraj et al.\ 2007; Hern\'{a}ndez \& Bruzual 2009) suggesting that UV colours alone may not indicate with confidence the presence of a young population. 
In order to examine the contribution of evolved stars, we measure the near-IR flux, where these stars are expected to dominate the spectrum, and compare the $[3.6-4.5]=-0.03\pm0.02$ colour, measured within the IG region, with the expected values from the Maraston et al.\ (2005) models. We conclude that this value can be accounted for by the presence of a $\sim$Myr stellar population with a Solar-like metallicity, and not by an old population alone. For lower metallicities, the age determination is less decisive due to the minor variation of the [3.6-4.5] colour with age.

Furthermore, fitting the UV to near-IR colours and the detected H$\alpha$ emission with the GEMs indicates the presence of a young stellar population in NGC5363, which constitutes a few percent of the mass of the galaxy. We therefore conclude that it is highly likely that this galaxy experienced a recent star formation episode triggered by accretion of external material.
\section{Conclusions}
\label{S:conclude}
We reported the properties of dust and ionized gas in a sample of 30 dust-lane E/S0 galaxies based on multicolour and H$\alpha$ narrow-band imaging observations. For most galaxies, extended physically-associated dust and ionized gas structures are observed.
The correlation between the dust and ionized gas is further illustrated by the correlation between the masses of these two components.

The extragalactic extinction law by the dust in the sample galaxies indicates similar properties to those canonical Galactic grains. This property is used to correct the flux measured within the ionized gas aperture for galactic extinction. The derived colours and H$\alpha$ fluxes are fitted with stellar population synthesis models, with results indicating the presence of a small fraction of $\sim10-100$ Myr old stars within the old stellar bulge. 

We also discussed in detail the case of NGC5363 where we detected a unique spiral-like gas and dust distribution. The ISM characteristics combined with the wide spectral coverage available for this galaxy further support the late star-forming episode often observed in E/S0s.
\subsection*{Acknowledgments}
We thank the anonymous referee for providing constructive comments and help improving the content of this paper.
IF wishes to thank Hagai Netzer, Oded Spector and Benny Trakhtenbrot for many useful discussions. JGS would like to thank Chris Krall and Ray Axline for their participation in the observations.
AYK and PV acknowledge support from the National Research Foundation of South Africa. 

Based on observations with the VATT: the Alice P. Lennon Telescope and the Thomas J. Bannan Astrophysics Facility.

This publication makes use of data products from the Two Micron All Sky Survey, which is a joint project of the University of Massachusetts and the Infrared Processing and Analysis Center/California Institute of Technology, funded by the National Aeronautics and Space Administration and the National Science Foundation.

Funding for the SDSS and SDSS-II has been provided by the Alfred P. Sloan Foundation, the Participating Institutions, the National Science Foundation, the U.S. Department of Energy, the National Aeronautics and Space Administration, the Japanese Monbukagakusho, the Max Planck Society, and the Higher Education Funding Council for England. The SDSS Web Site is http://www.sdss.org/.

The SDSS is managed by the Astrophysical Research Consortium for the Participating Institutions. The Participating Institutions are the American Museum of Natural History, Astrophysical Institute Potsdam, University of Basel, University of Cambridge, Case Western Reserve University, University of Chicago, Drexel University, Fermilab, the Institute for Advanced Study, the Japan Participation Group, Johns Hopkins University, the Joint Institute for Nuclear Astrophysics, the Kavli Institute for Particle Astrophysics and Cosmology, the Korean Scientist Group, the Chinese Academy of Sciences (LAMOST), Los Alamos National Laboratory, the Max-Planck-Institute for Astronomy (MPIA), the Max-Planck-Institute for Astrophysics (MPA), New Mexico State University, Ohio State University, University of Pittsburgh, University of Portsmouth, Princeton University, the United States Naval Observatory, and the University of Washington.

\appendix
\section{Description of individual galaxies}
\label{S:IndivGal}
\begin{figure*}
\begin{center}
\begin{tabular}{ccc}
\includegraphics[width=6cm]{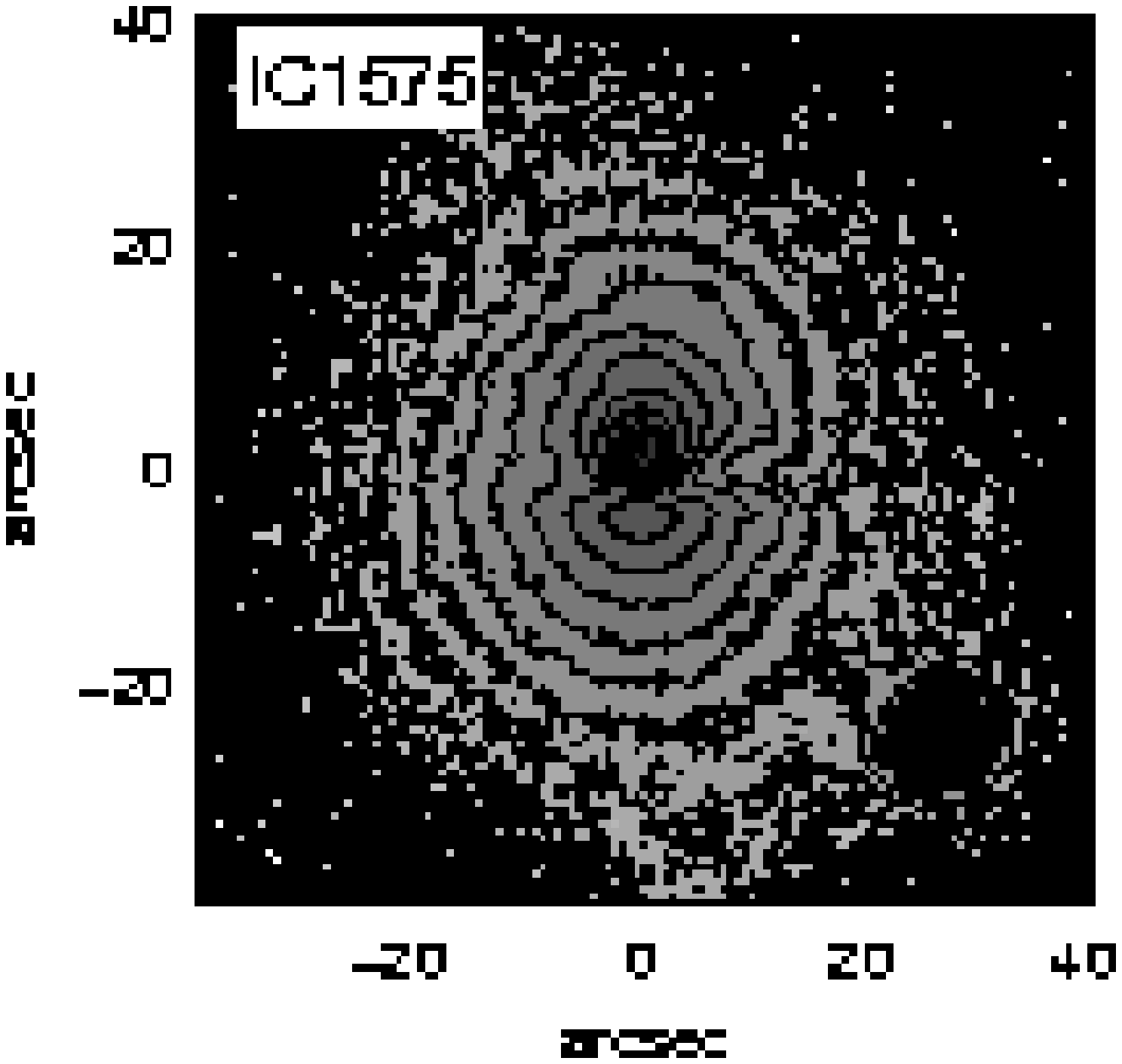} & \includegraphics[]{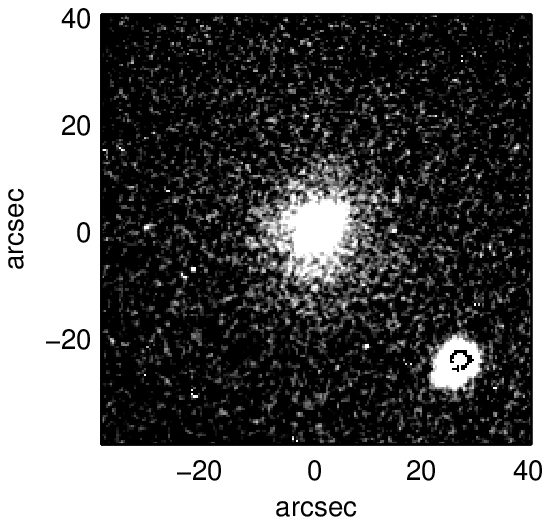} & \includegraphics[]{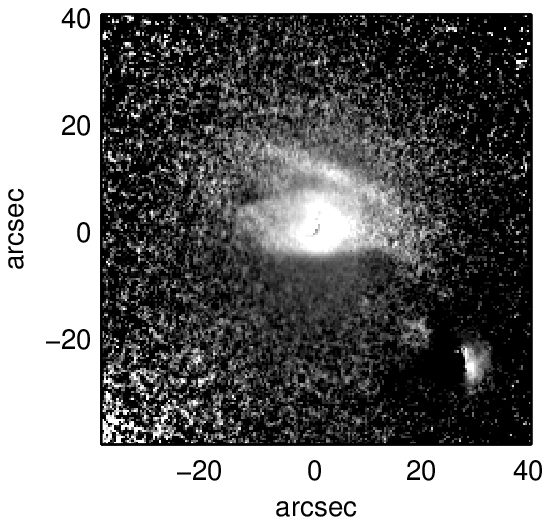}\\
  \vspace{-3mm}
\includegraphics[width=6cm]{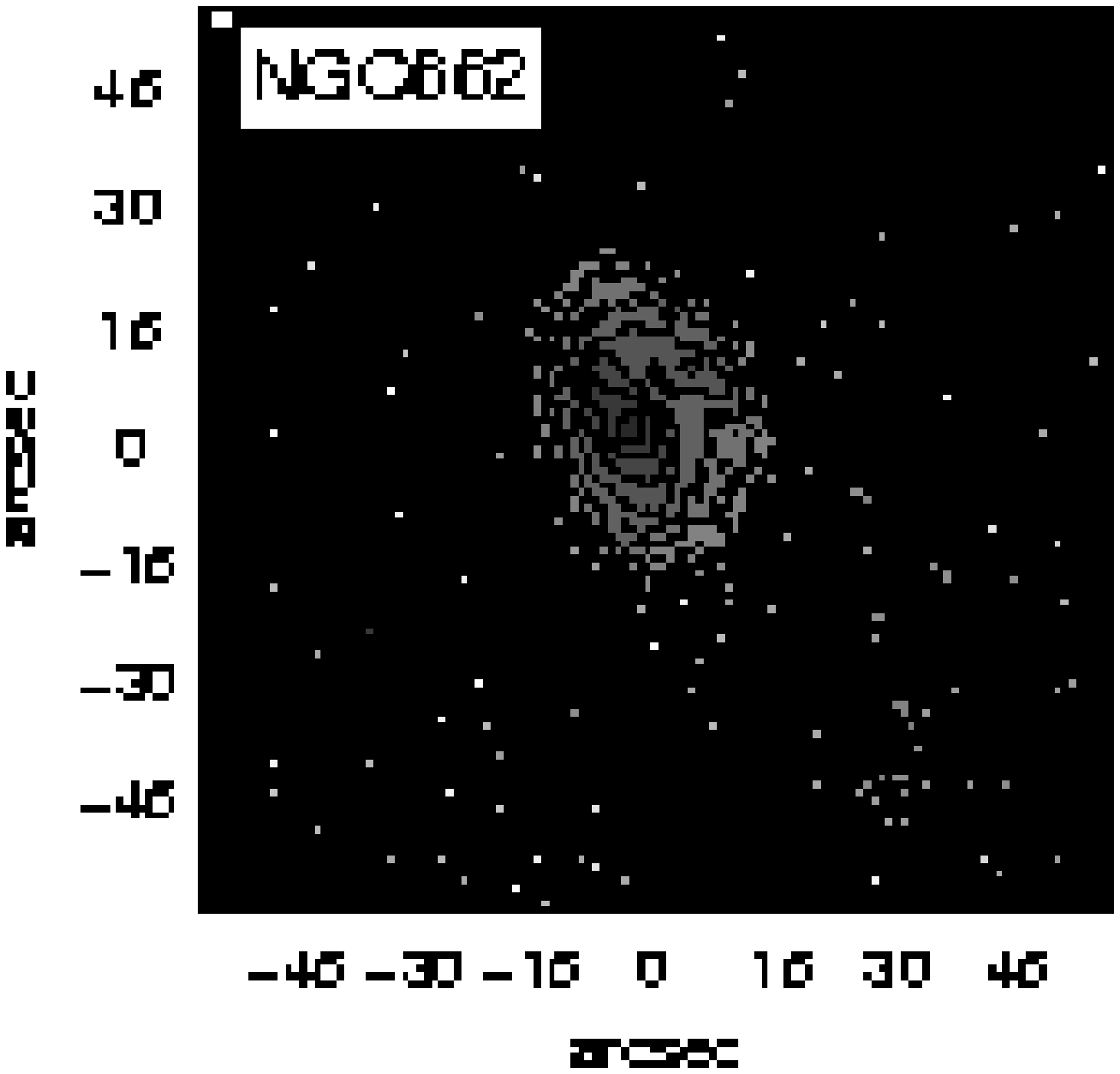} & \includegraphics[]{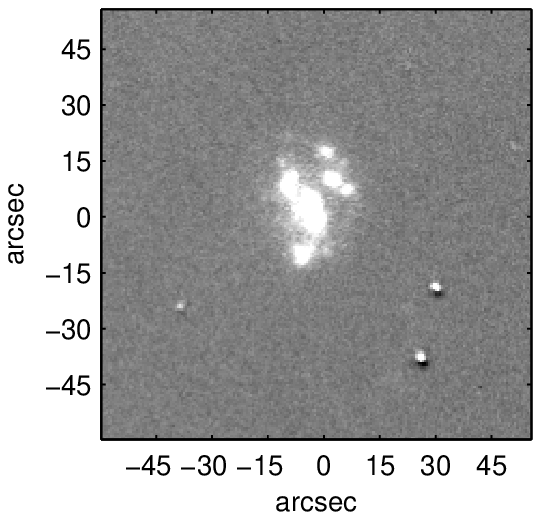} & \includegraphics[]{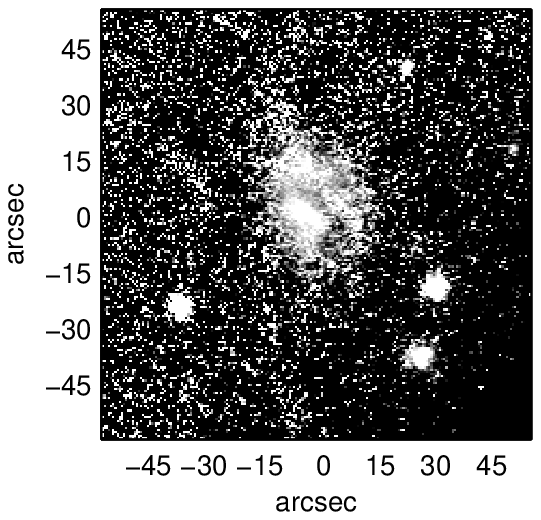}\\
  \vspace{-3mm}
\includegraphics[width=6cm]{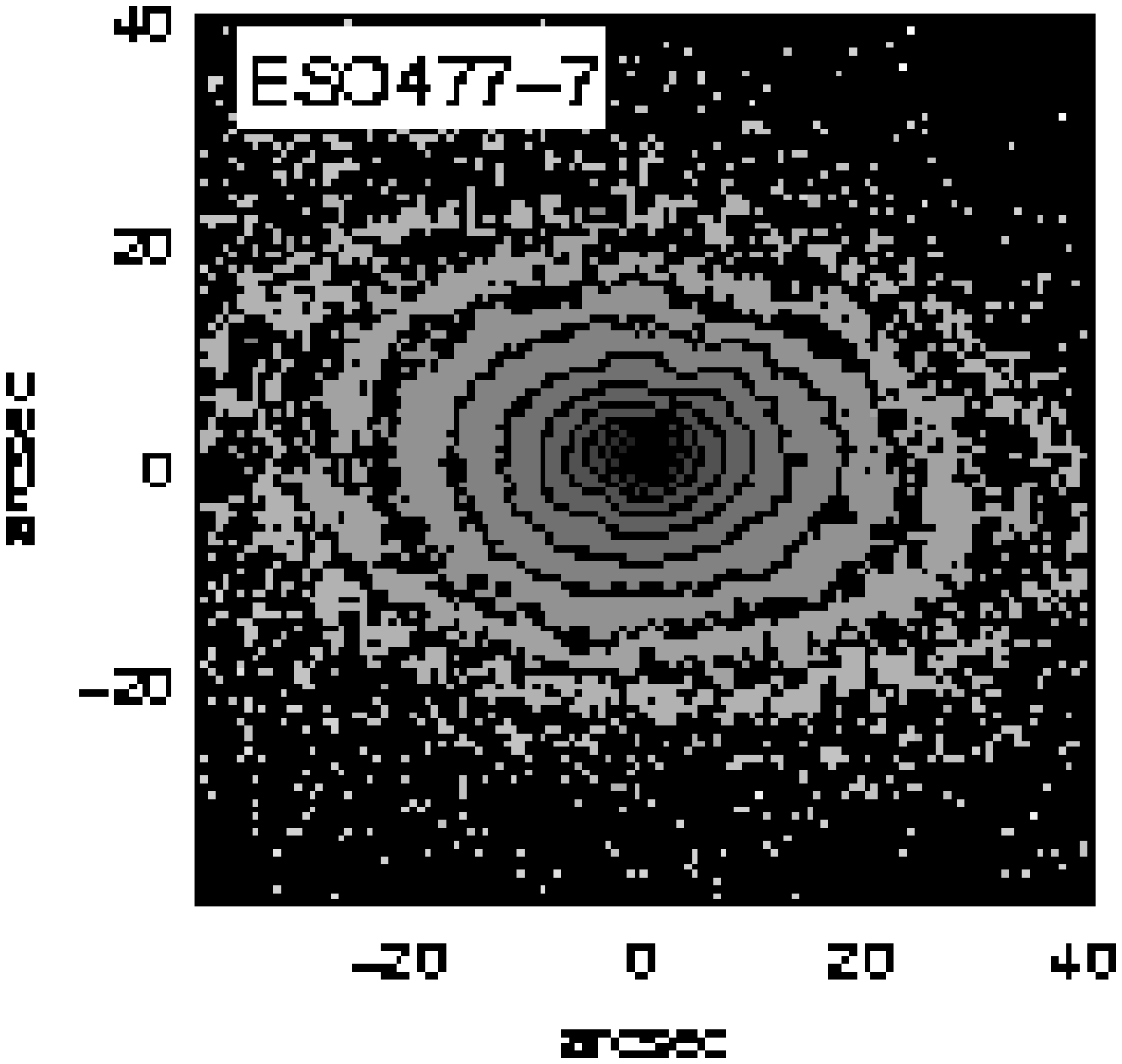} & \includegraphics[]{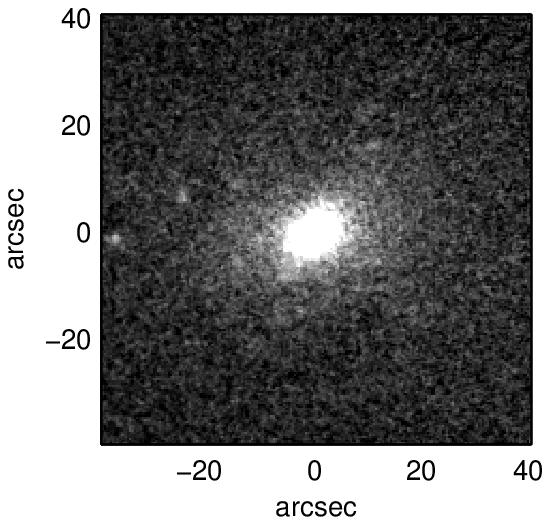} & \includegraphics[]{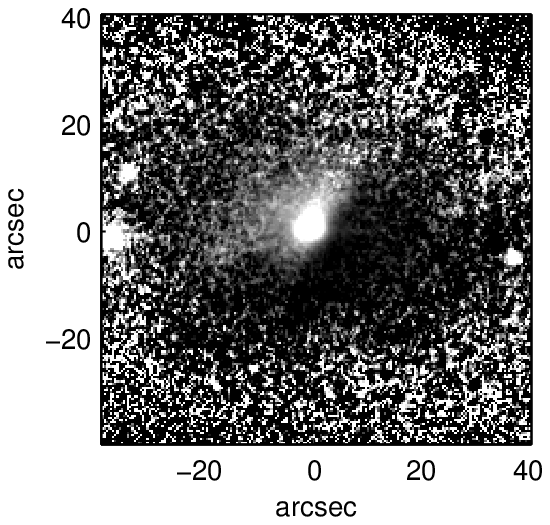}\\
  \vspace{-3mm}
\includegraphics[width=6cm]{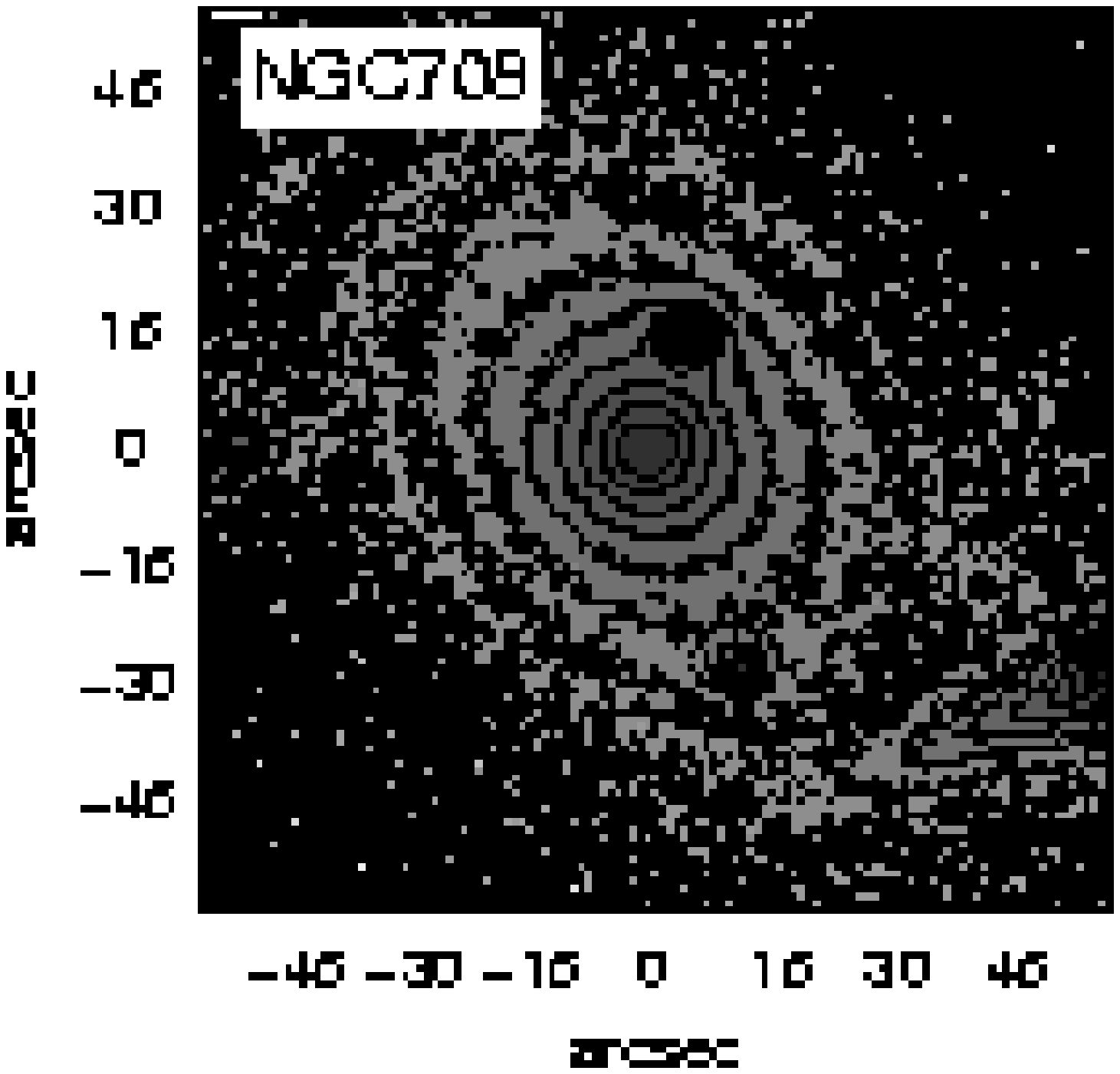} & \includegraphics[]{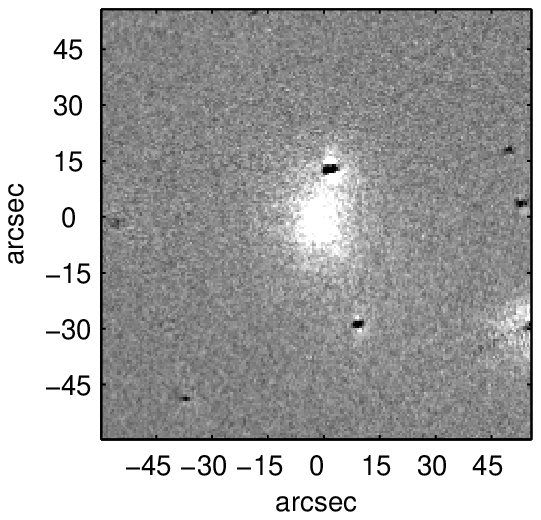} & \includegraphics[]{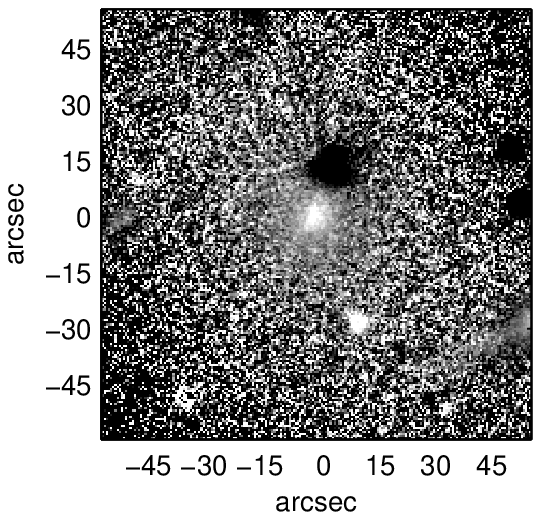}
\end{tabular}
\end{center}
\caption{R-band contour maps, continuum-subtracted H$\alpha$+[NII] images and B-R colour-index maps.}
\label{fig:BImaps1}
\end{figure*}

\begin{figure*}
\begin{center}
\begin{tabular}{ccc}
 \includegraphics[width=6cm]{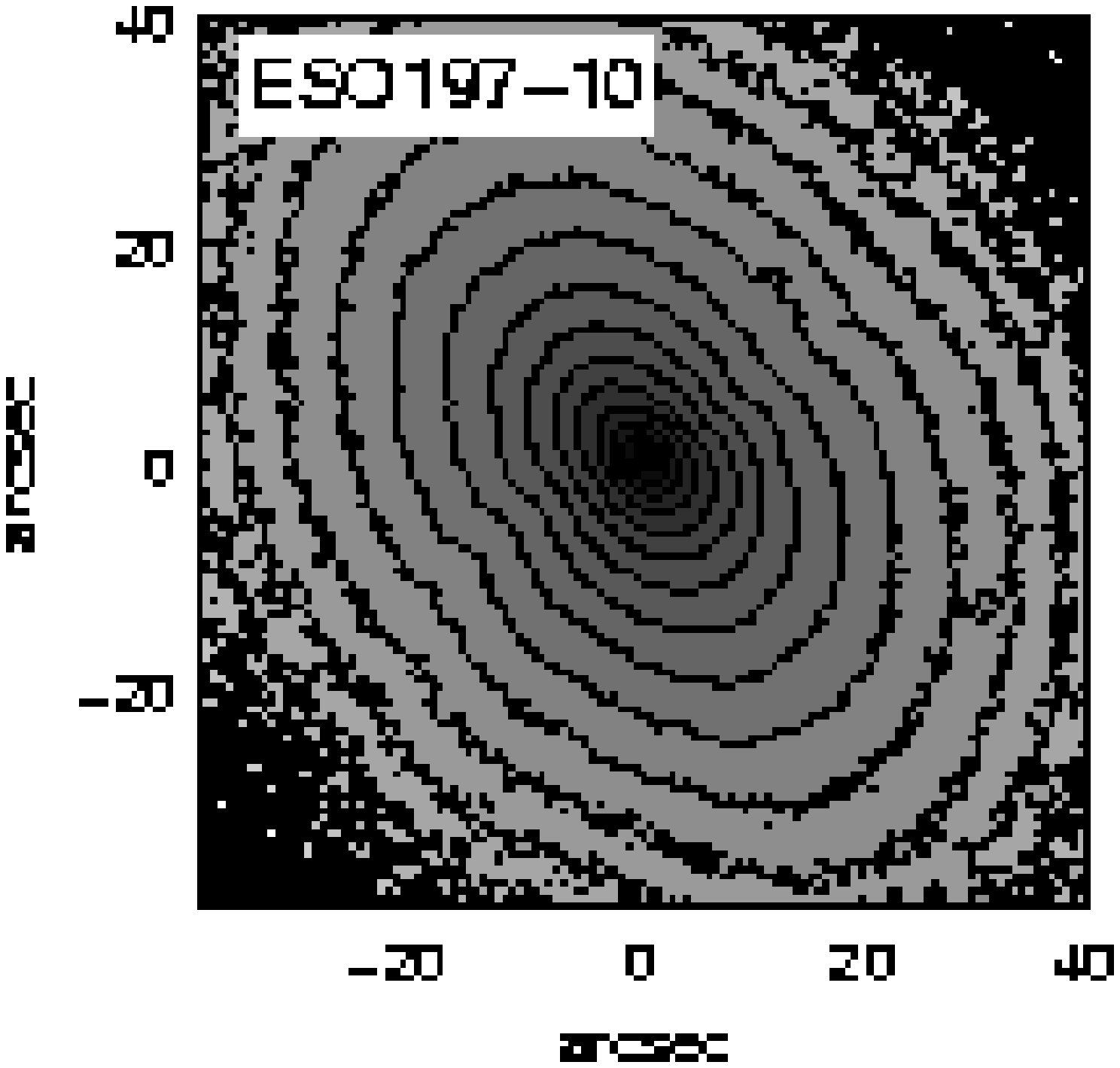} &  \includegraphics[]{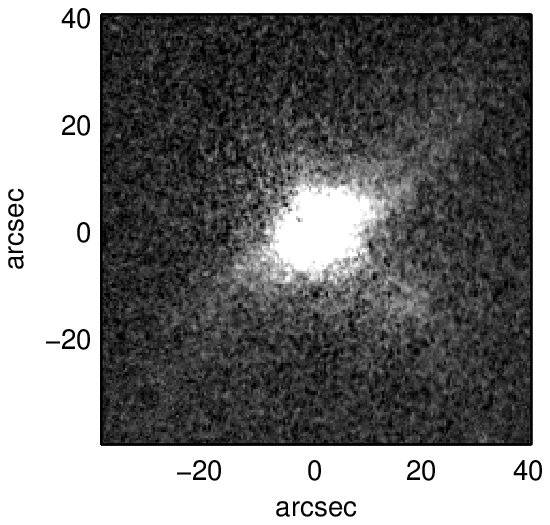} &\includegraphics[]{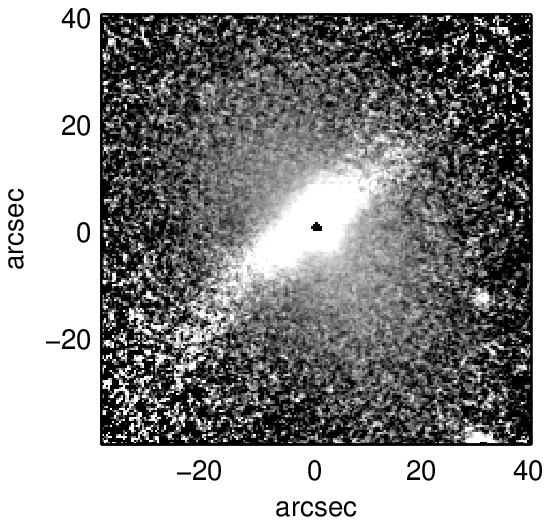}\\
   \vspace{-3mm}
 \includegraphics[width=6cm]{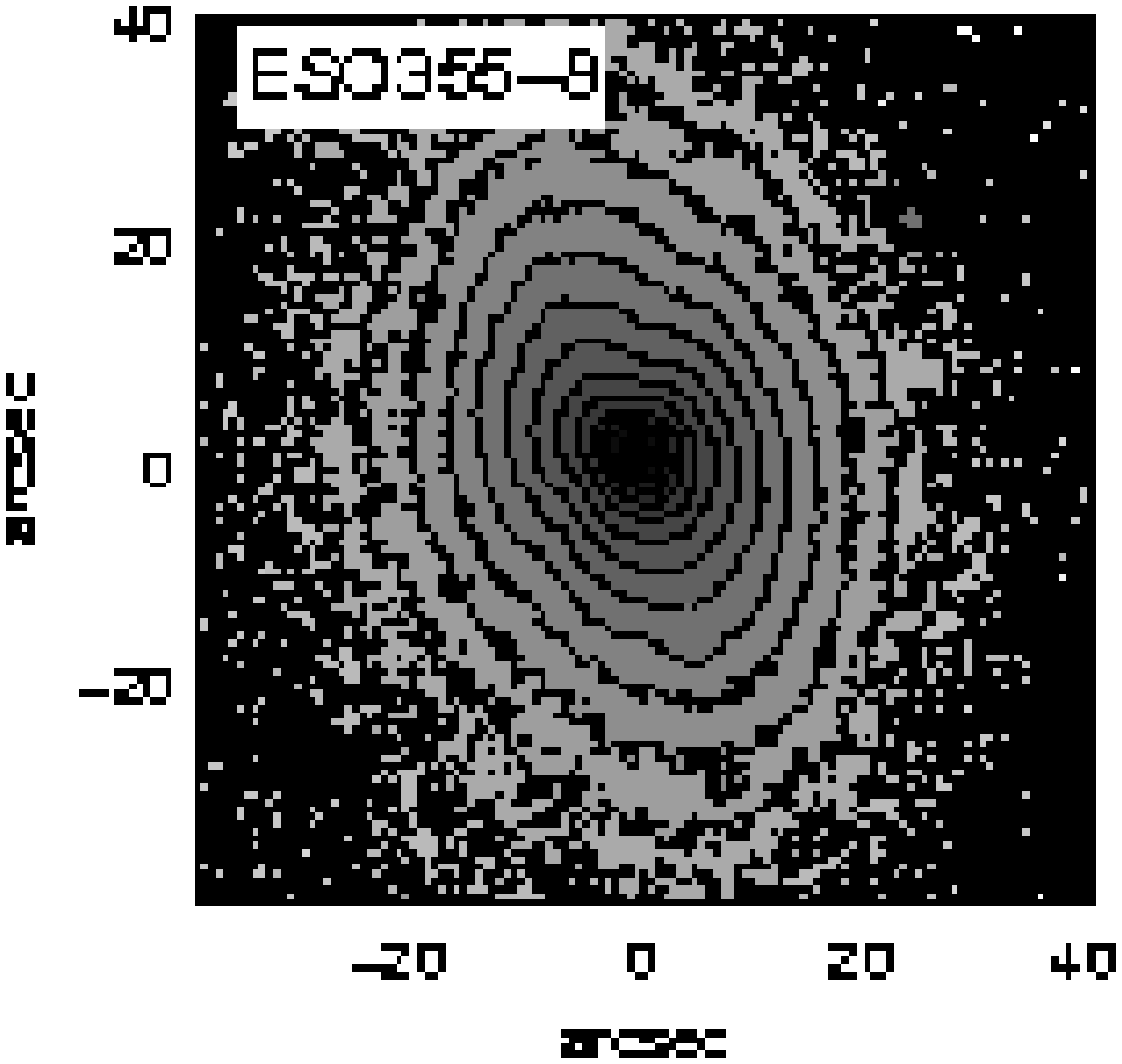} & \includegraphics[]{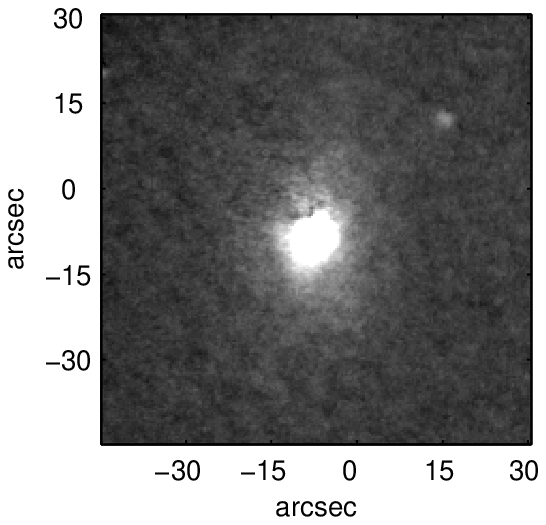}  & \includegraphics[]{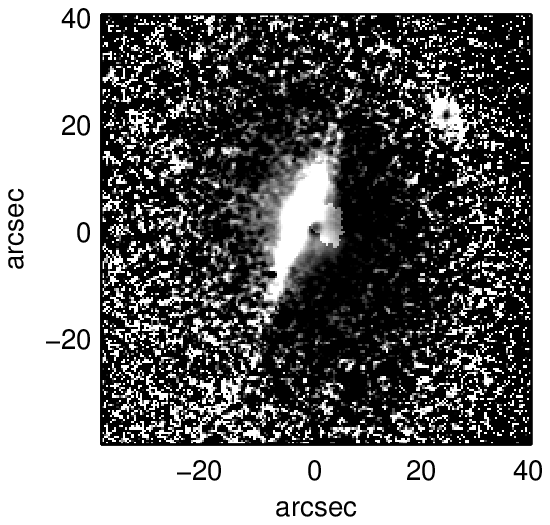}\\
   \vspace{-3mm}
\includegraphics[width=6cm]{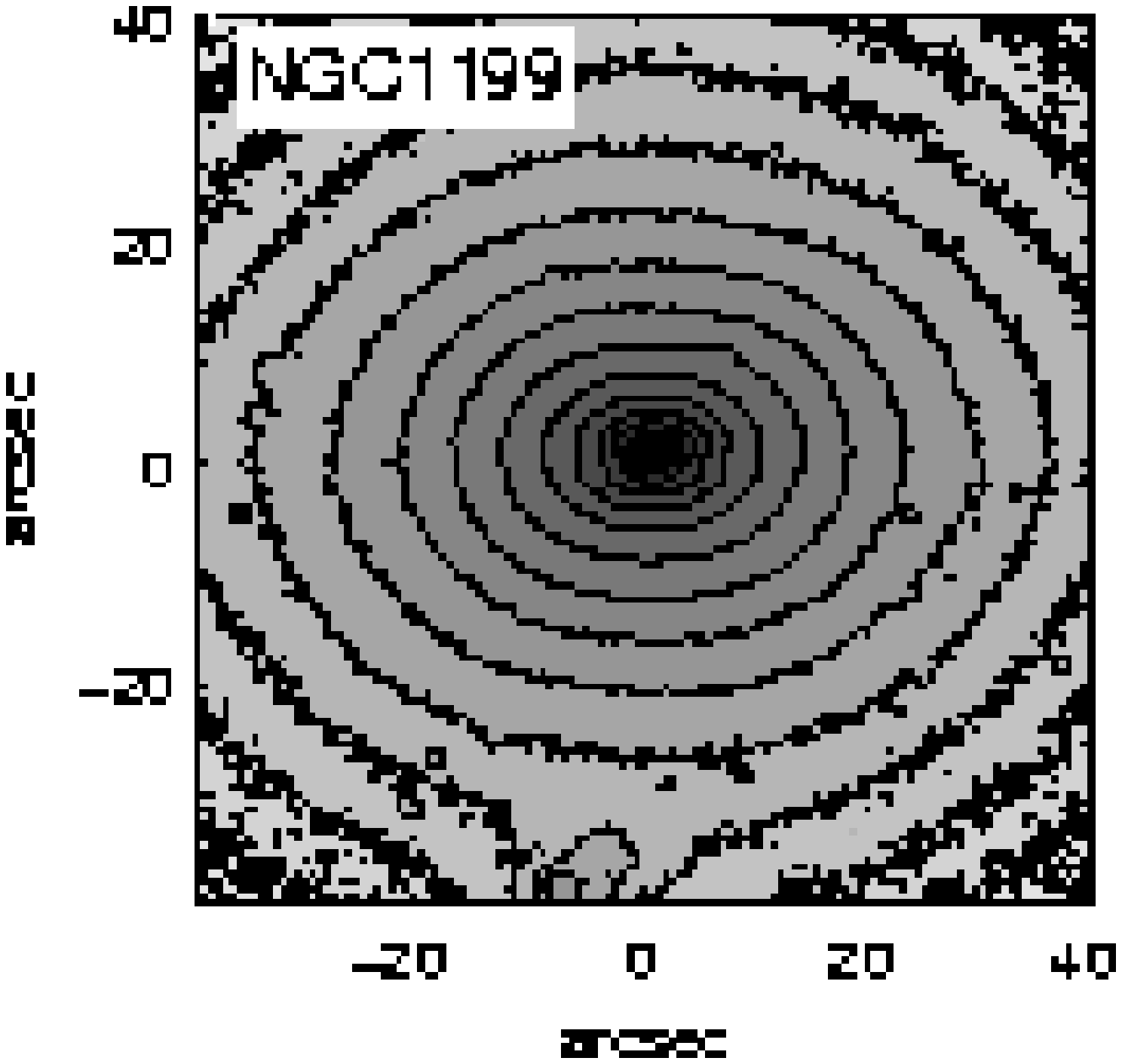} & \includegraphics[]{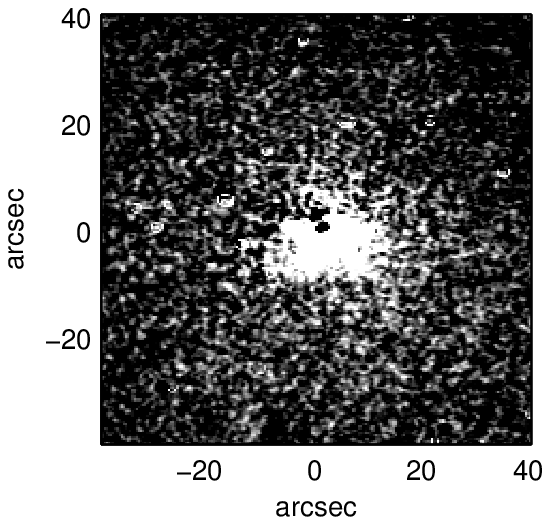} & \includegraphics[]{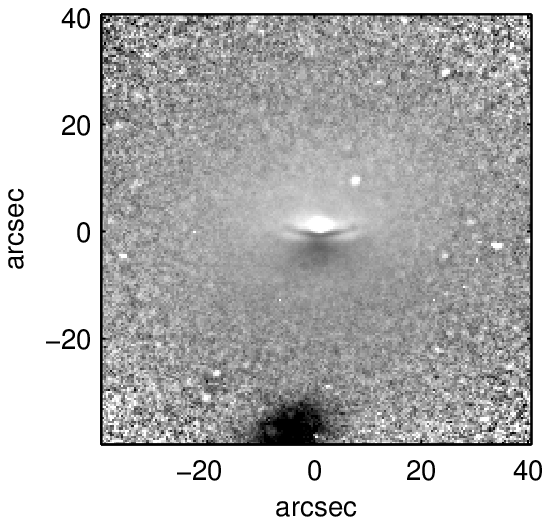}\\
  \vspace{-3mm}
\includegraphics[width=6cm]{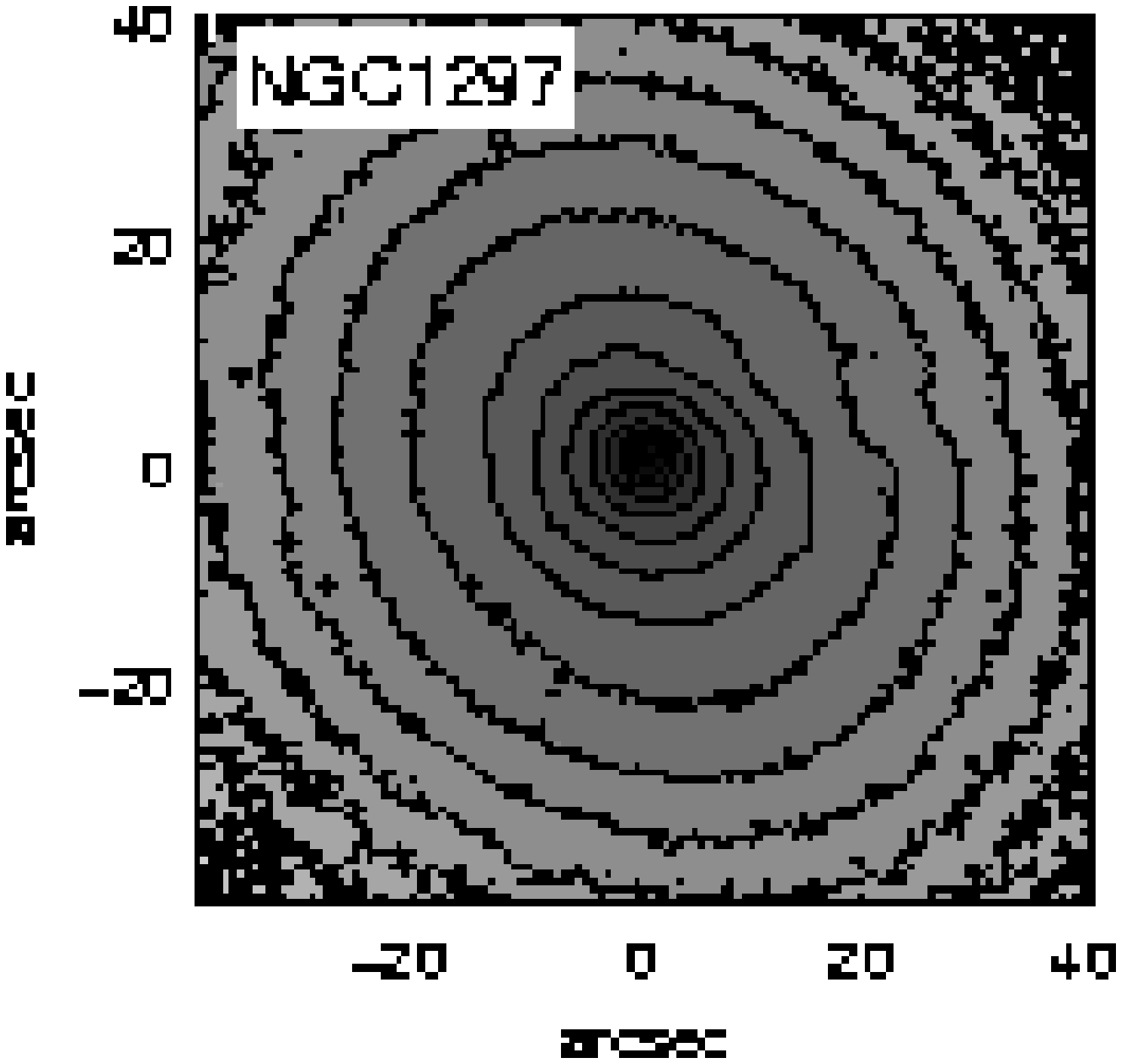} & \includegraphics[]{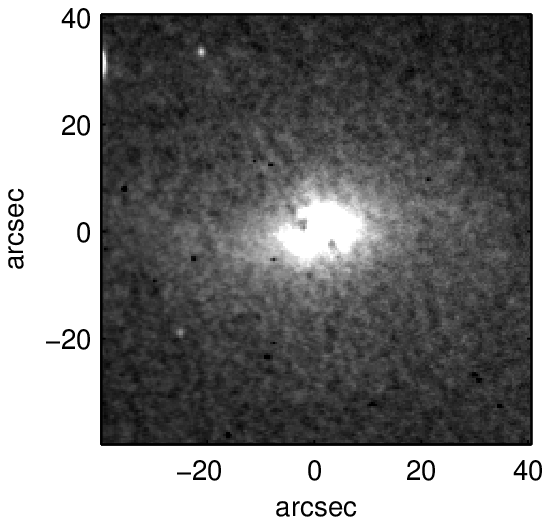} &\includegraphics[]{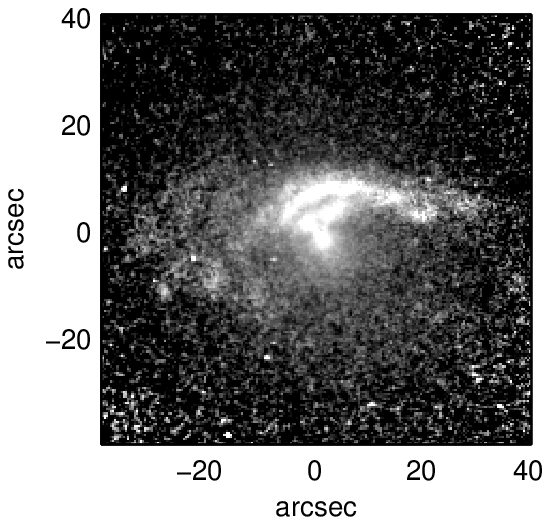}
\end{tabular}
\end{center}
\caption{R-band contour maps, continuum-subtracted H$\alpha$+[NII] images and B-R colour-index maps.}
 \label{fig:BImaps2}
\end{figure*}

\begin{figure*}
\begin{center}
\begin{tabular}{ccc}
 \includegraphics[width=6cm]{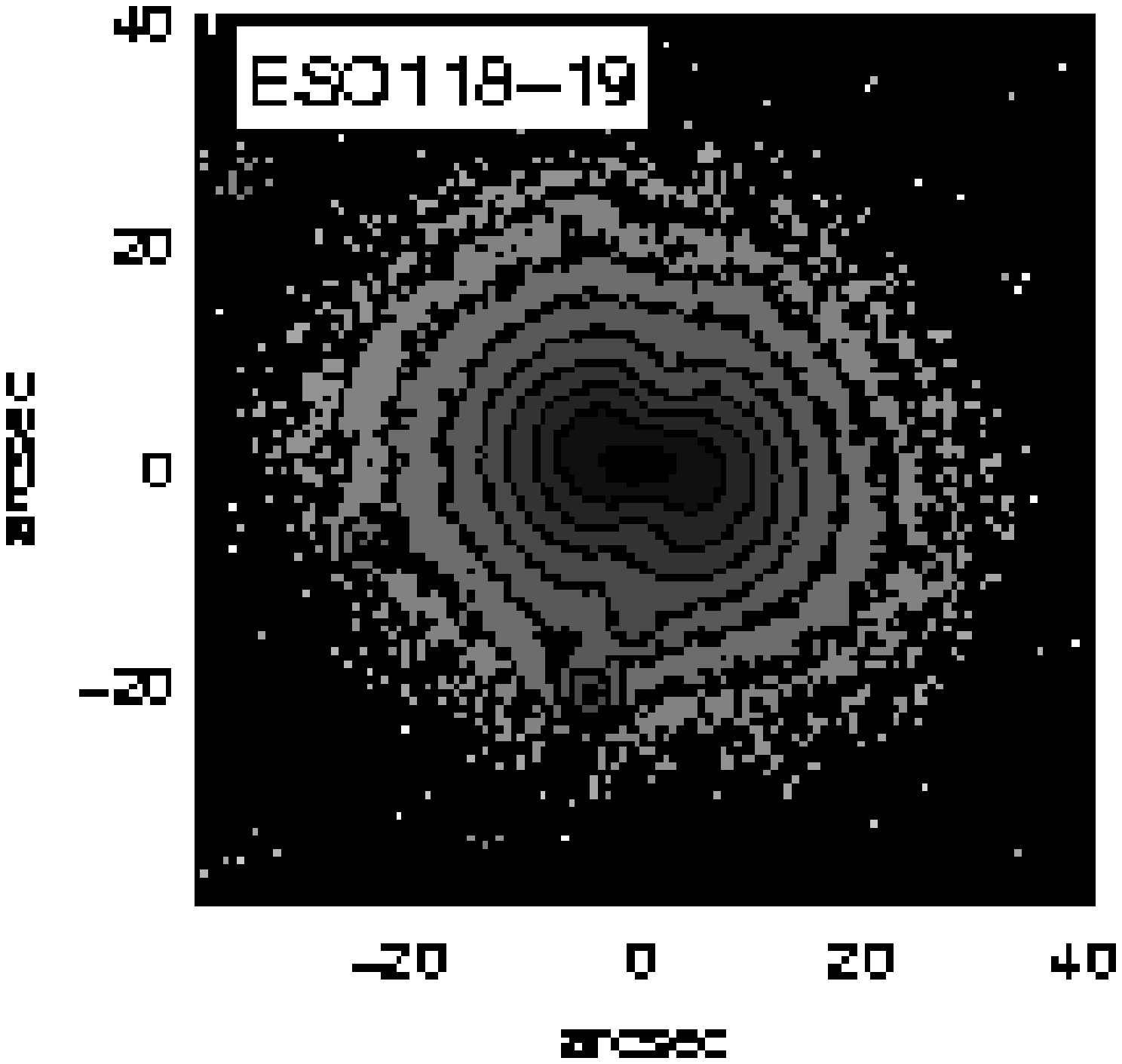} &  \includegraphics[]{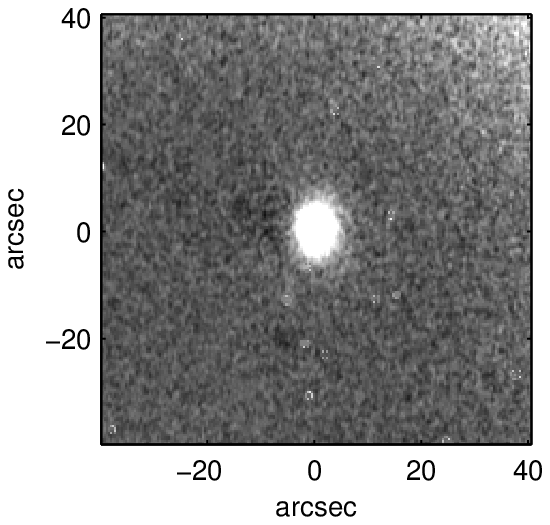} &\includegraphics[]{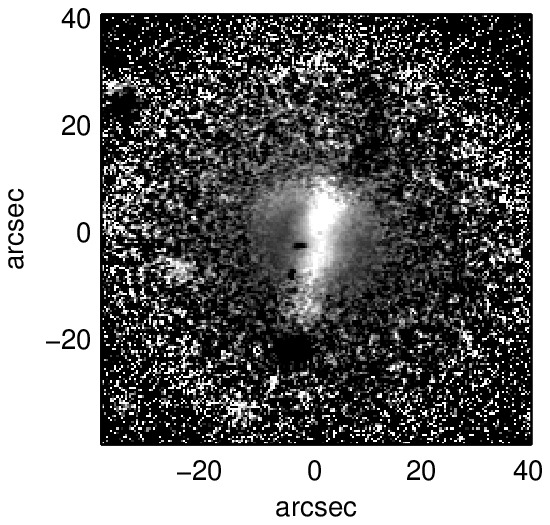}\\
   \vspace{-3mm}
 \includegraphics[width=6cm]{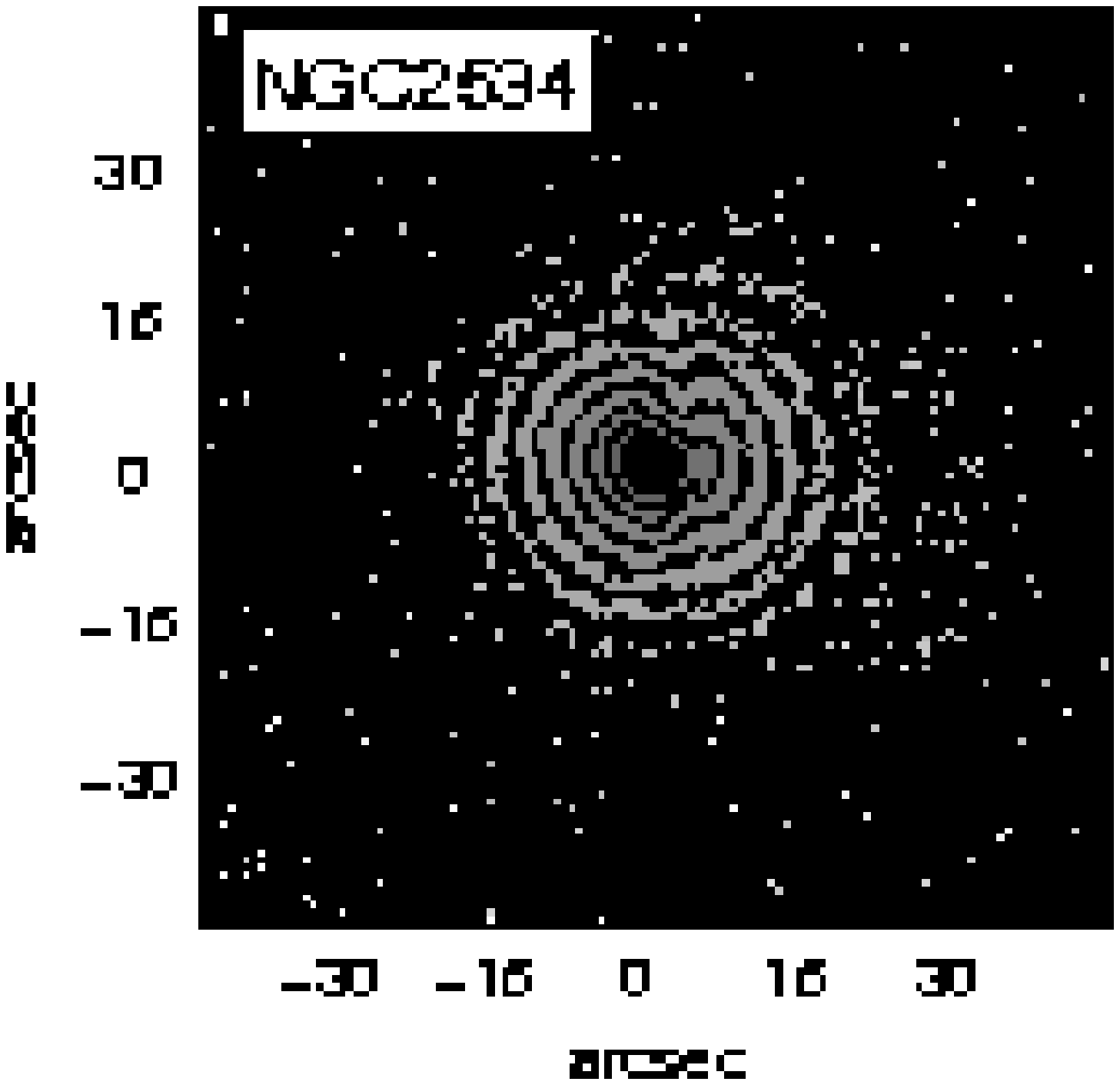} & \includegraphics[]{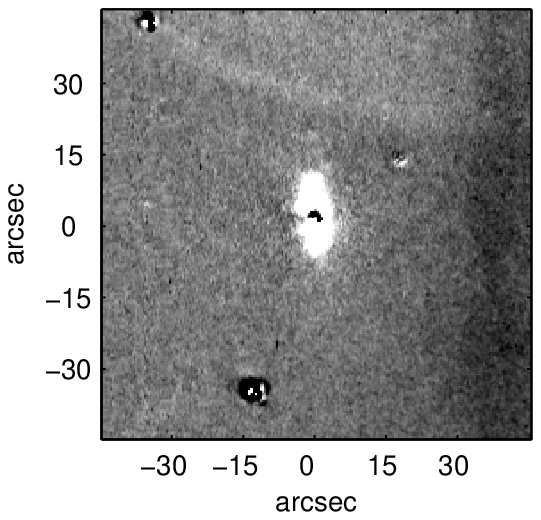}  & \includegraphics[]{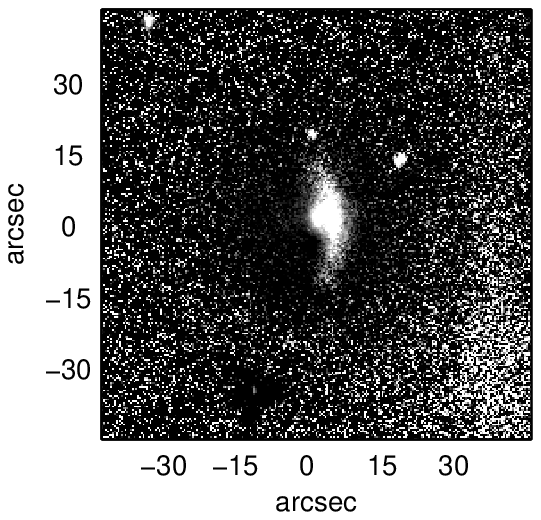}\\
   \vspace{-3mm}
\includegraphics[width=6cm]{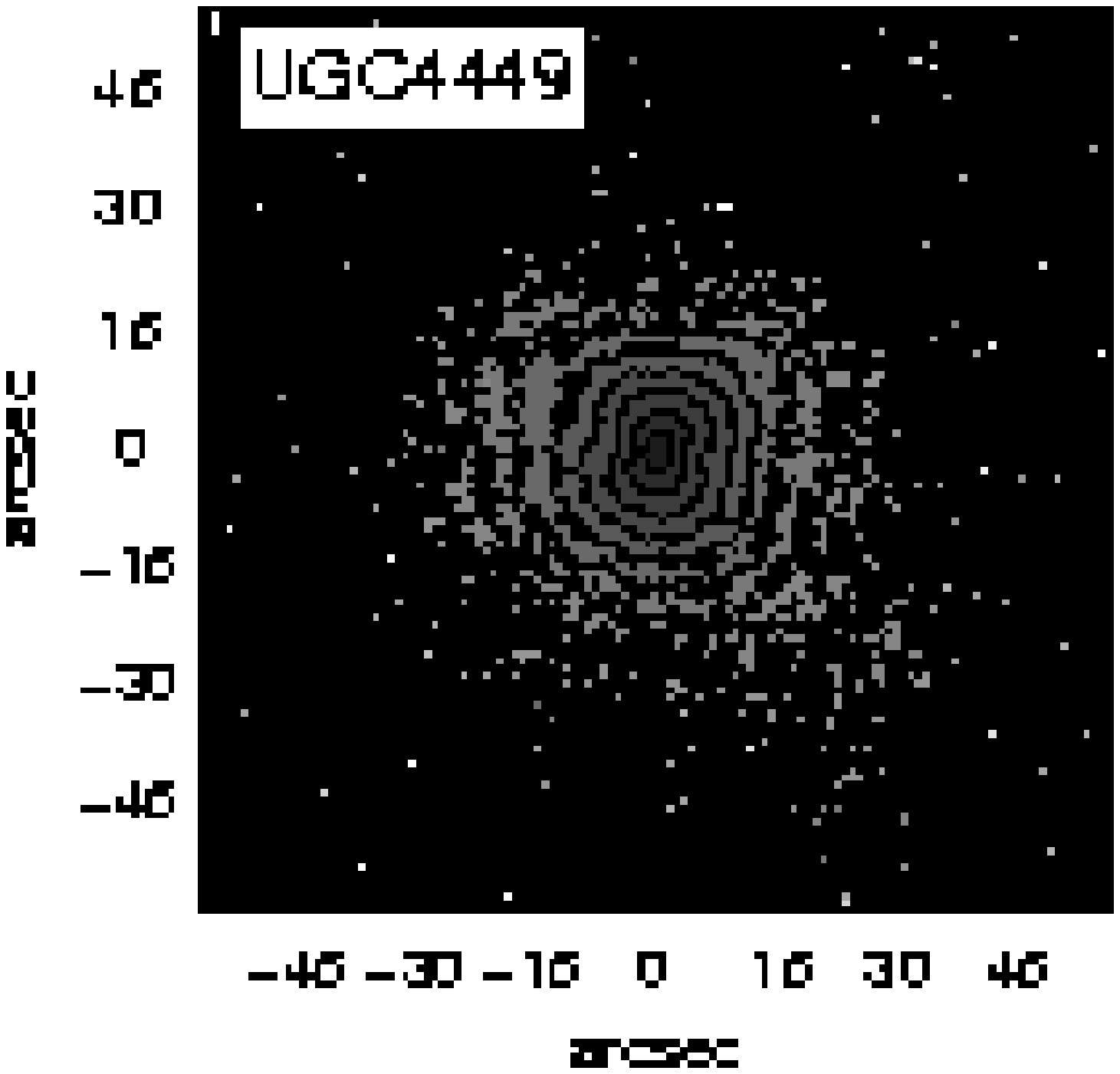} & \includegraphics[]{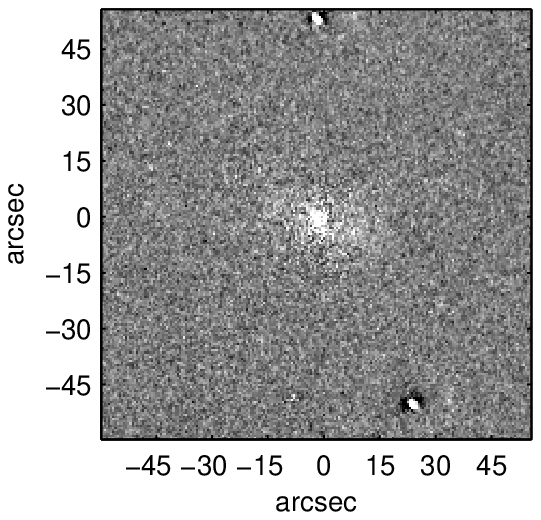} & \includegraphics[]{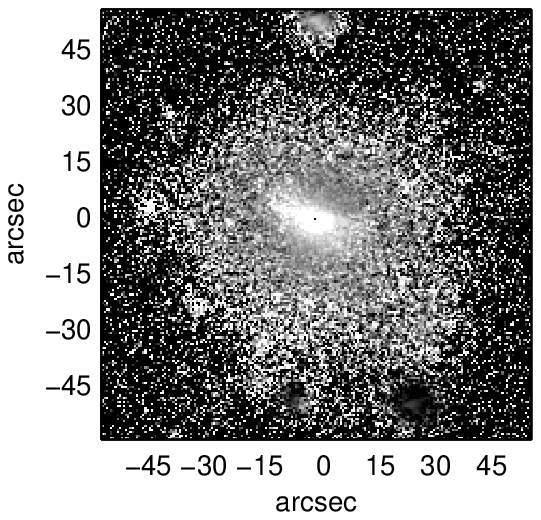}\\
  \vspace{-3mm}
\includegraphics[width=6cm]{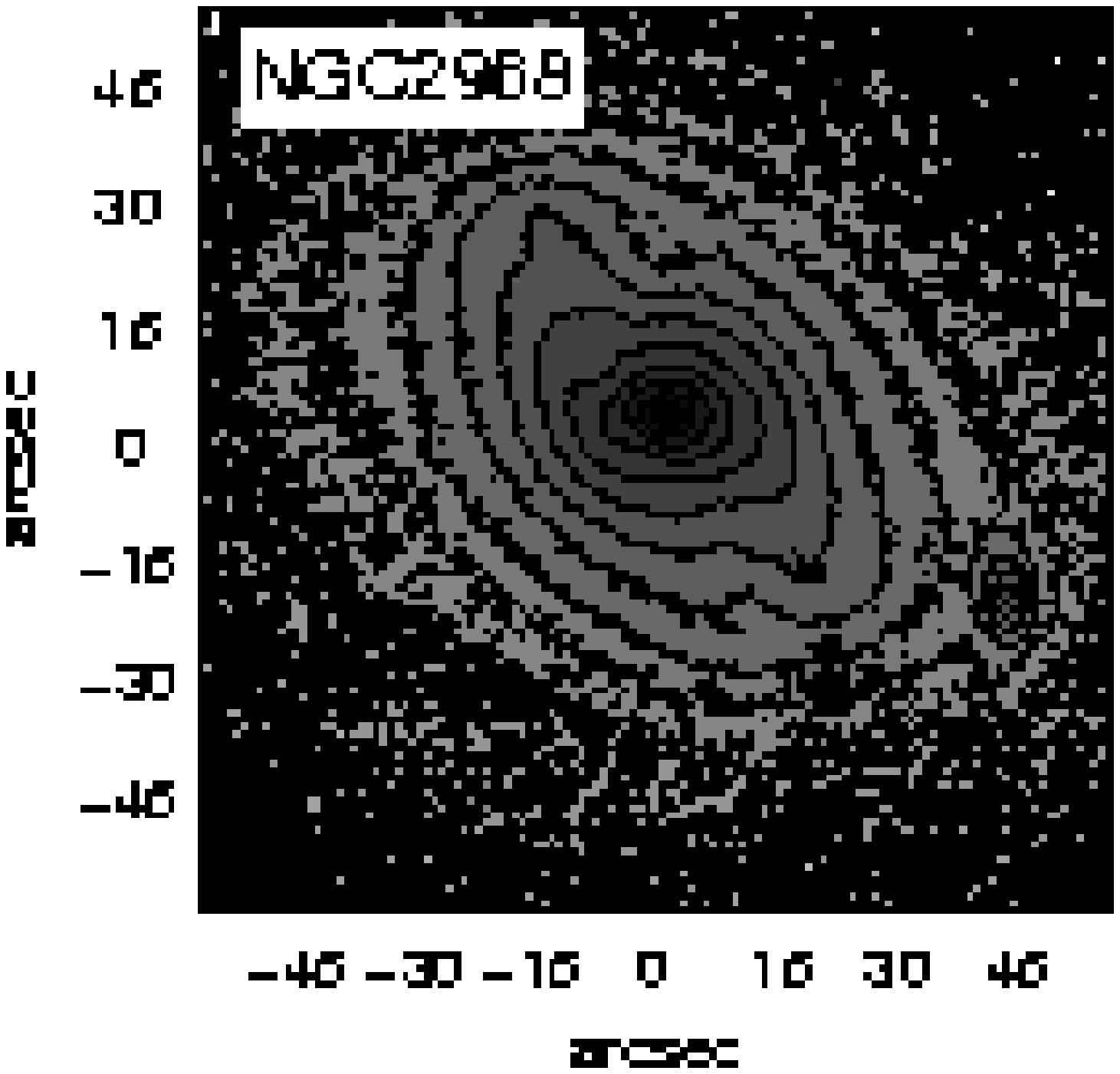} & \includegraphics[]{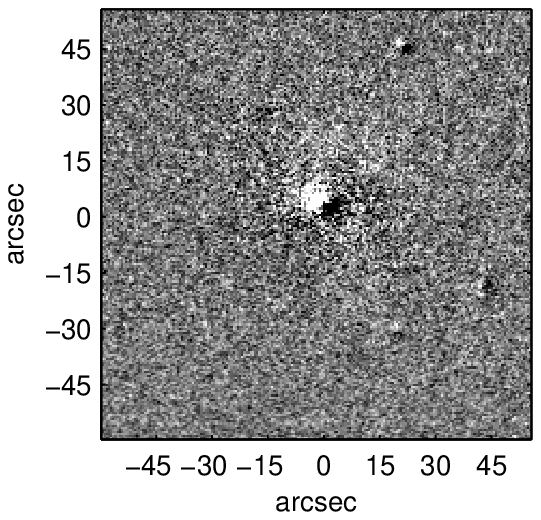} &\includegraphics[]{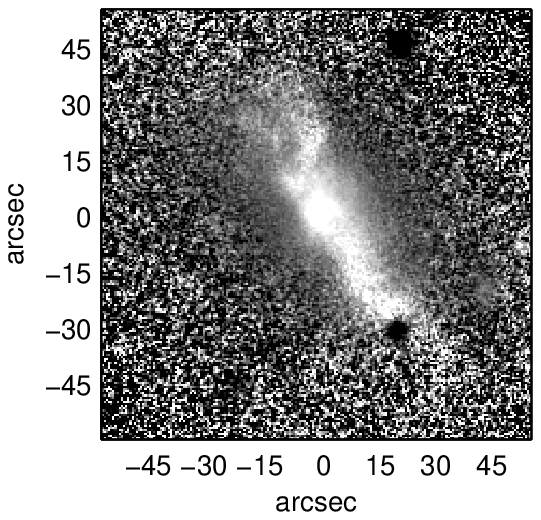}
\end{tabular}
\end{center}
\caption{R-band contour maps, continuum-subtracted H$\alpha$+[NII] images and B-R colour-index maps.}
 \label{fig:BImaps3}
\end{figure*}

\begin{figure*}
\begin{center}
\begin{tabular}{ccc}
 \includegraphics[width=6cm]{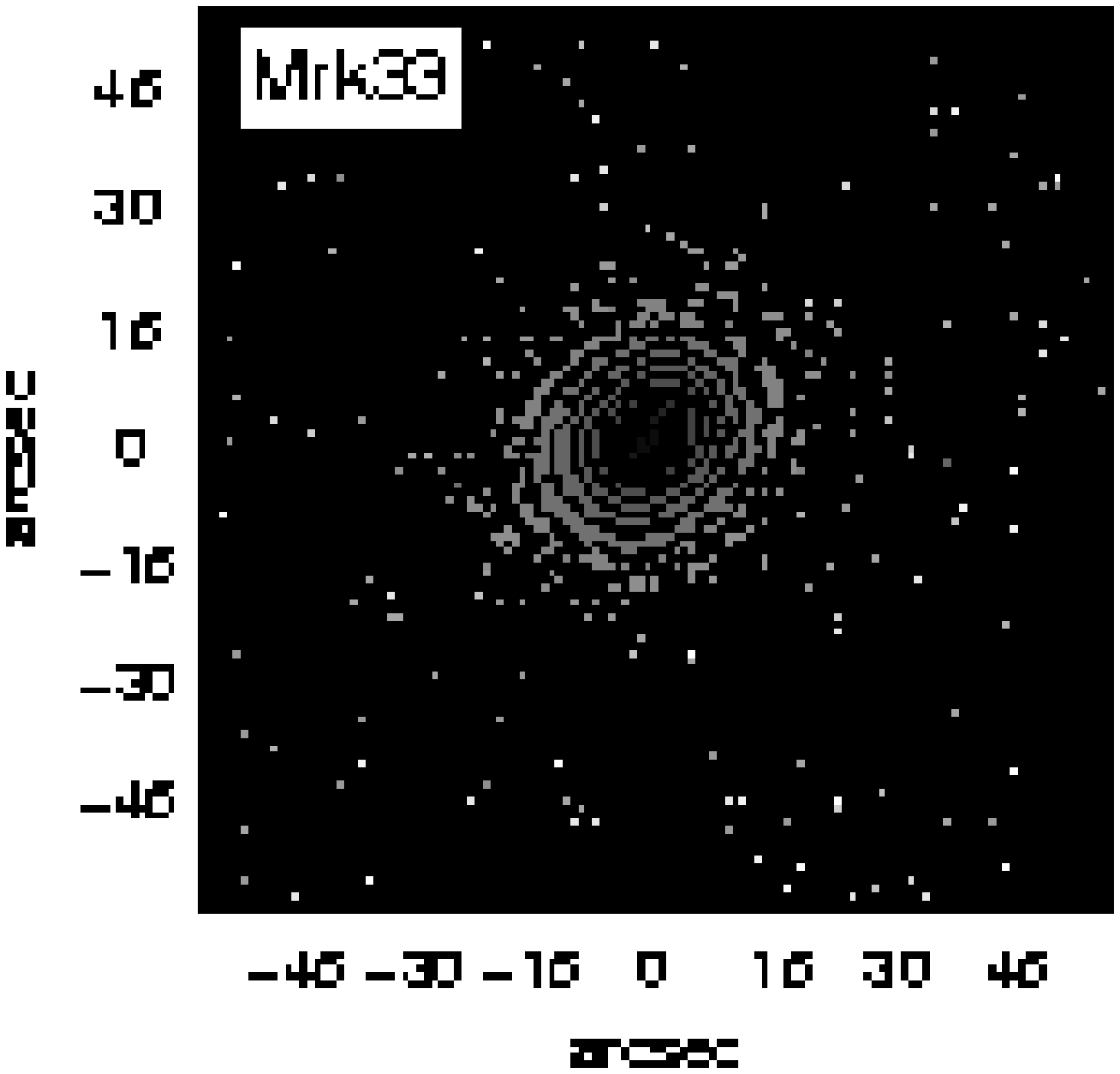} &  \includegraphics[]{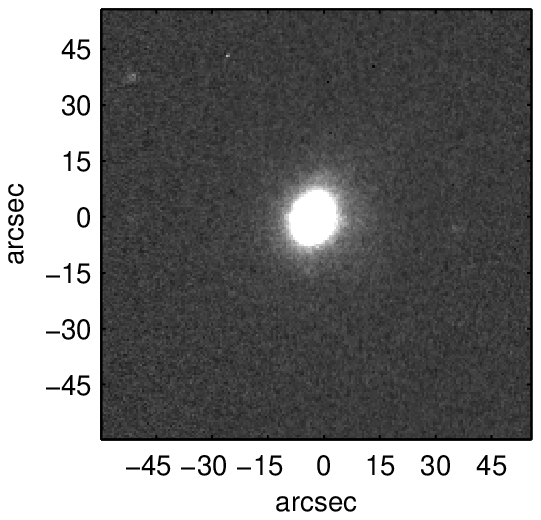} &\includegraphics[]{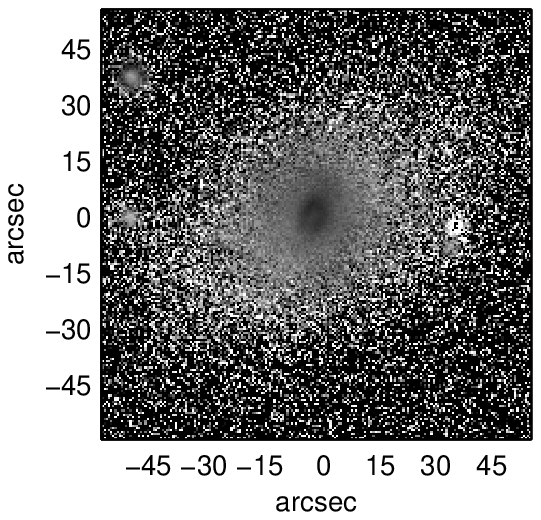}\\
   \vspace{-3mm}
 \includegraphics[width=6cm]{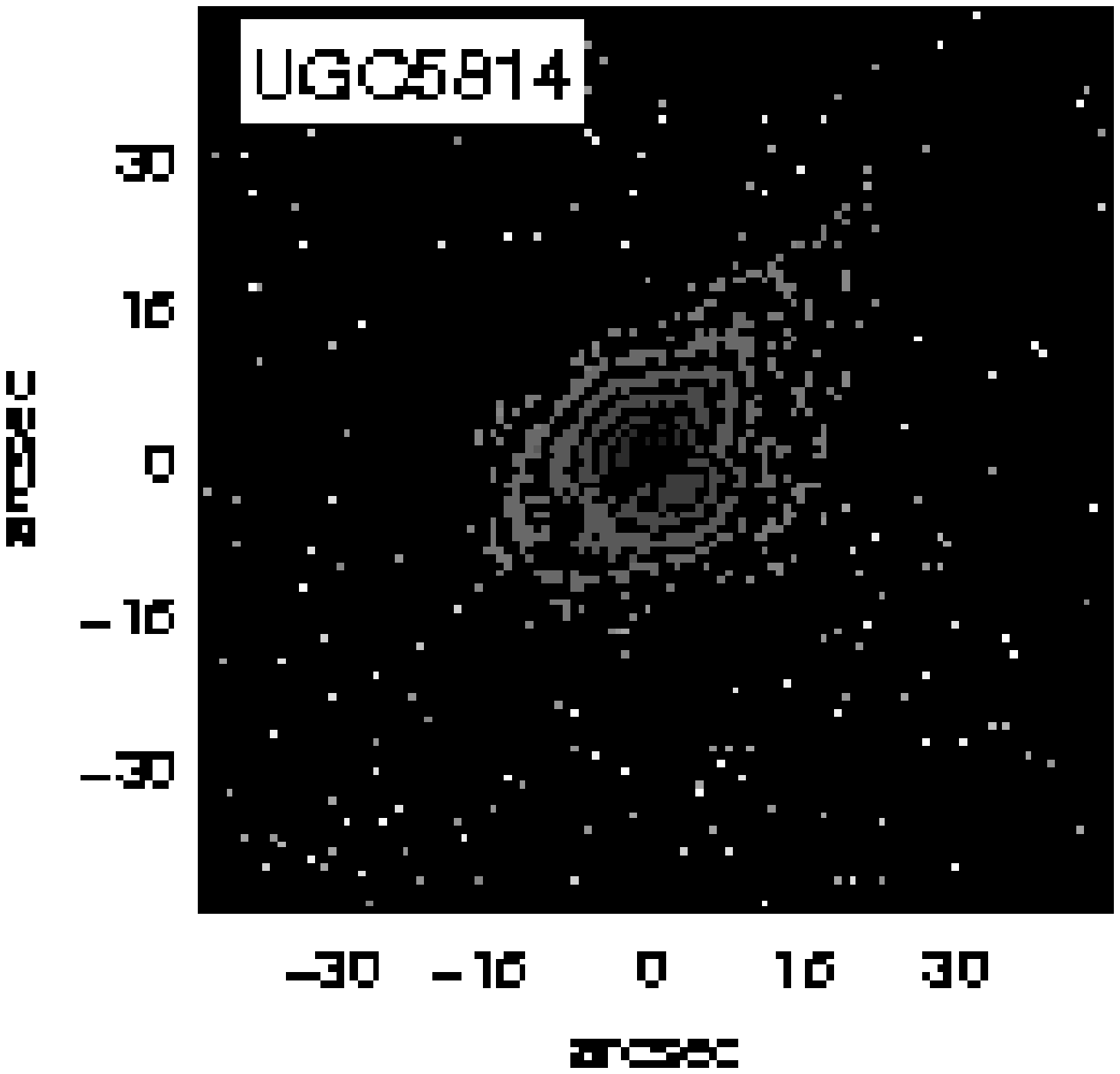} & \includegraphics[]{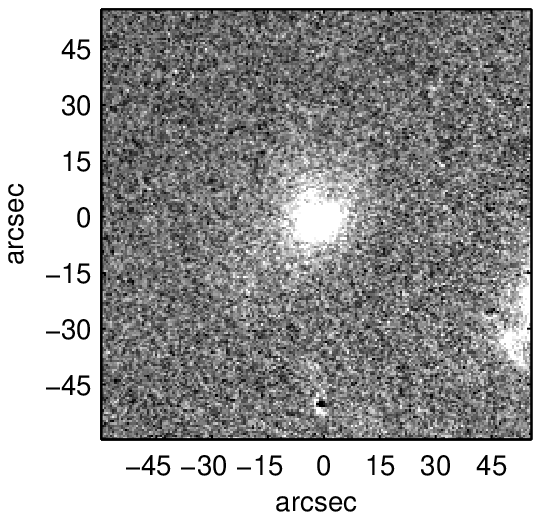}  & \includegraphics[]{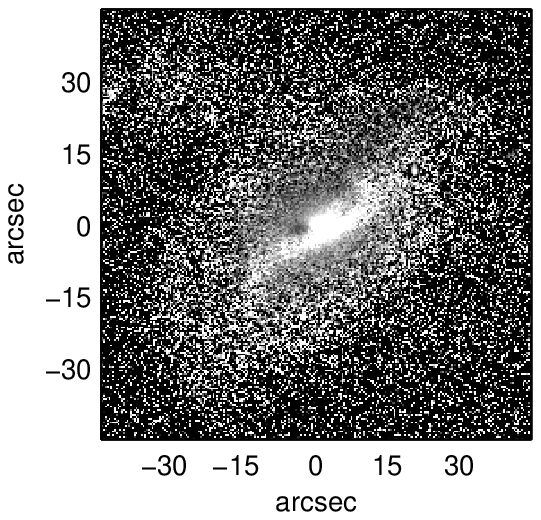}\\
   \vspace{-3mm}
\includegraphics[width=6cm]{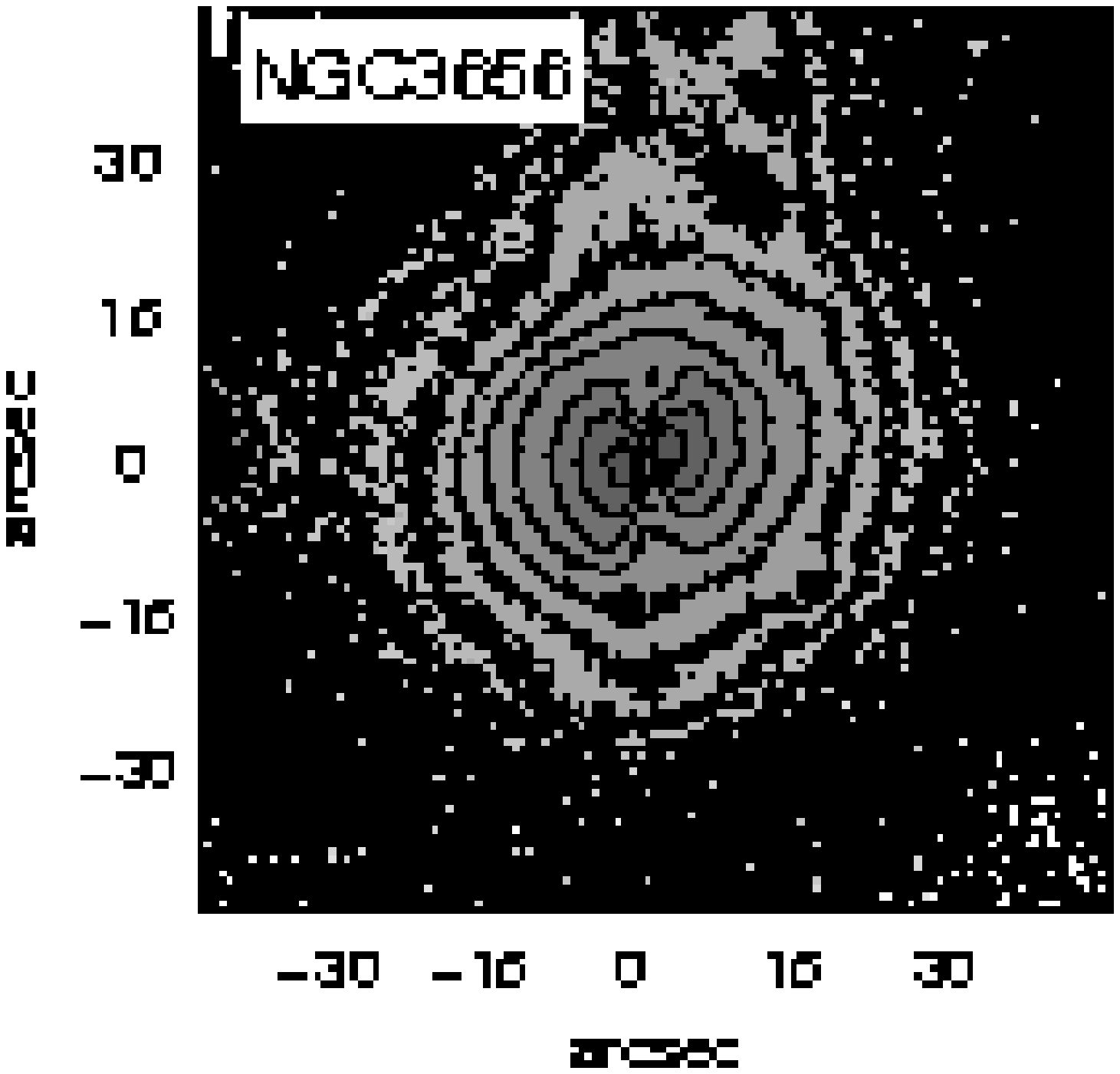} & \includegraphics[]{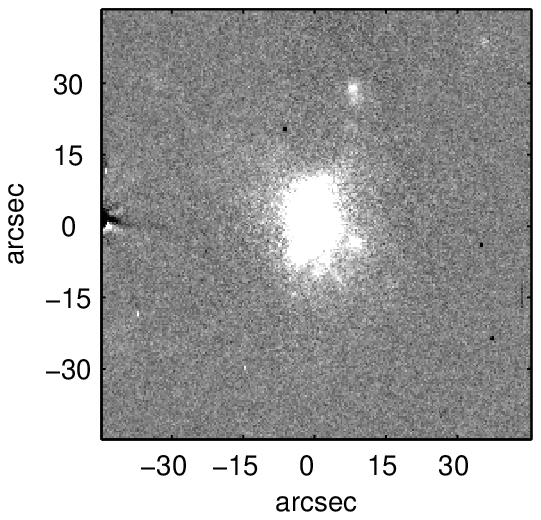} & \includegraphics[]{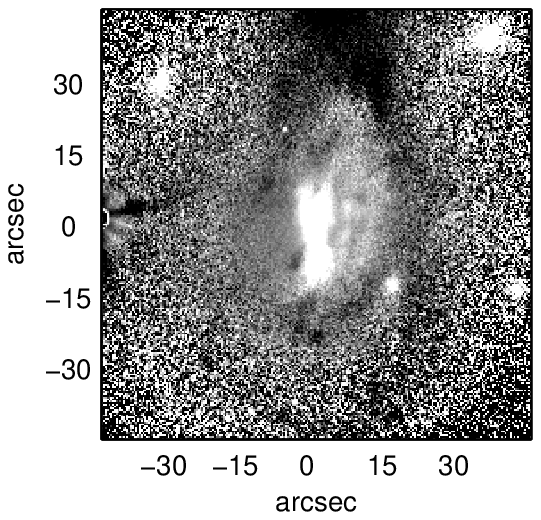}\\
  \vspace{-3mm}
\includegraphics[width=6cm]{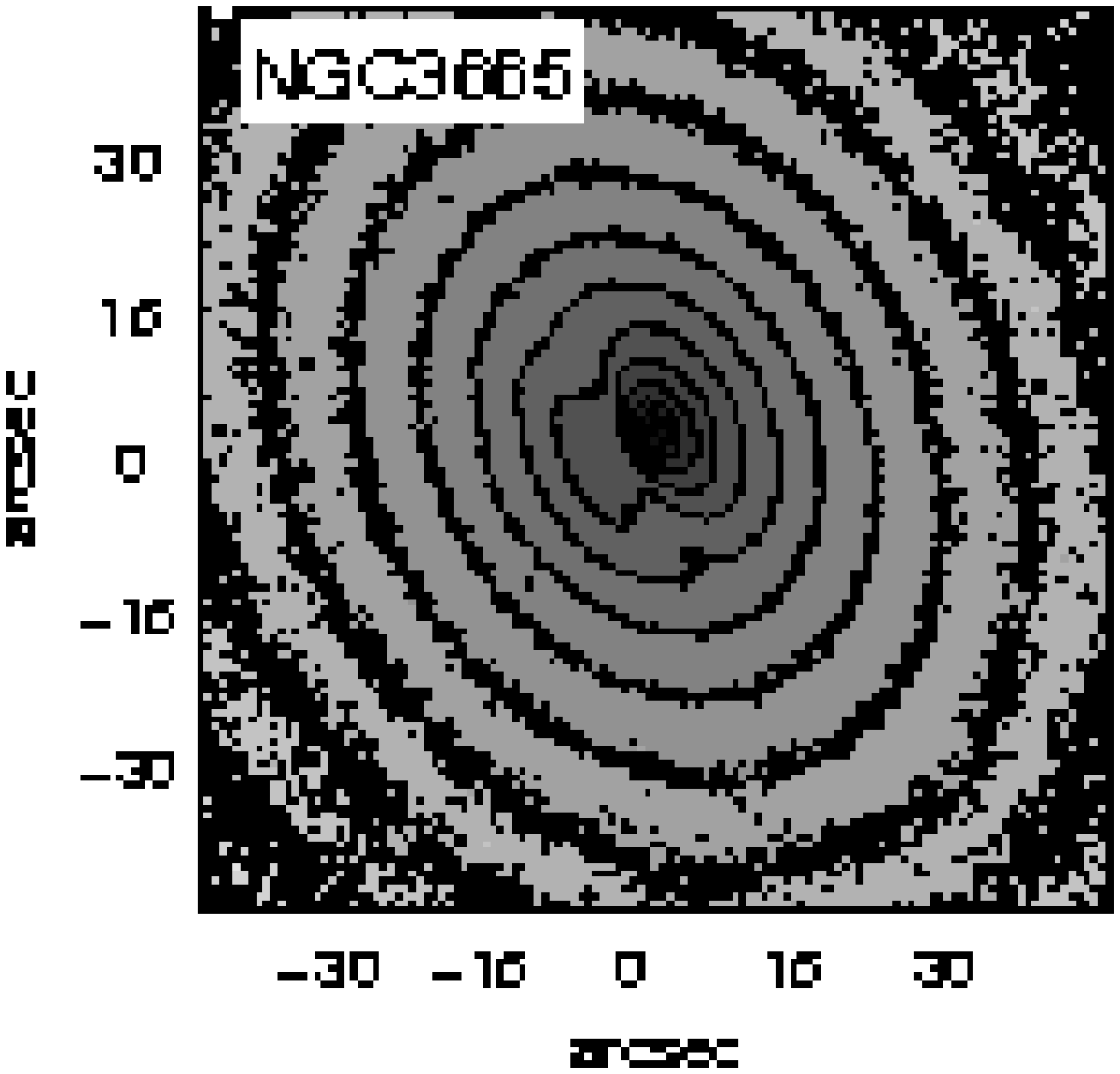} & \includegraphics[]{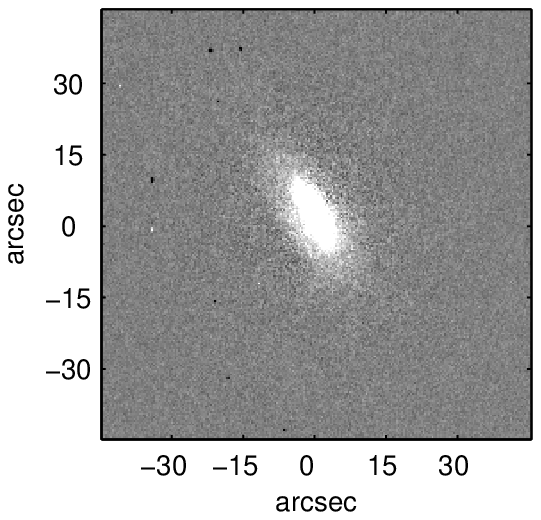}&\includegraphics[]{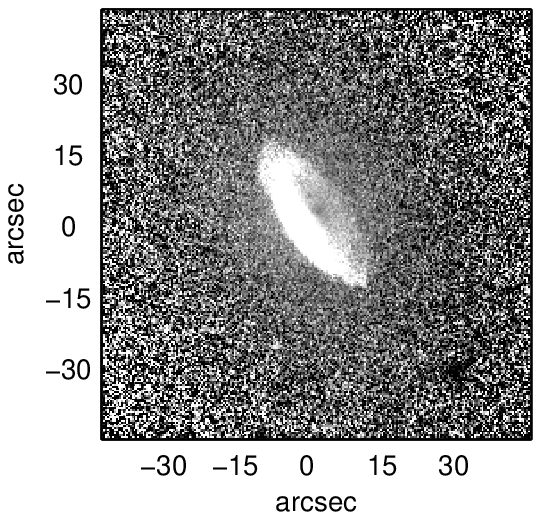}
\end{tabular}
\end{center}
\caption{R-band contour maps, continuum-subtracted H$\alpha$+[NII] images and B-R colour-index maps.}
 \label{fig:BImaps4}
\end{figure*}

\begin{figure*}
\begin{center}
\begin{tabular}{ccc}
 \includegraphics[width=6cm]{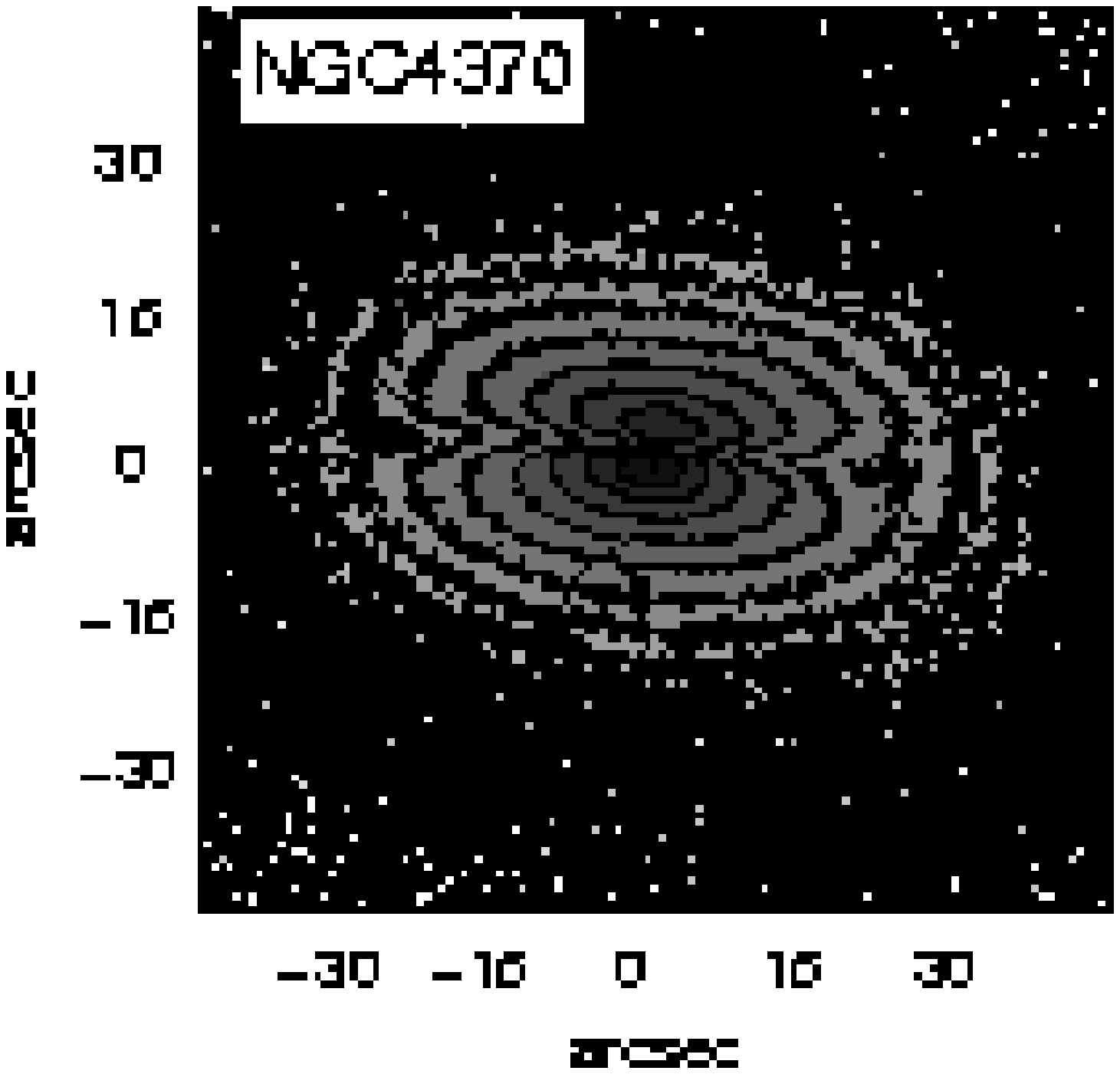} &  \includegraphics[]{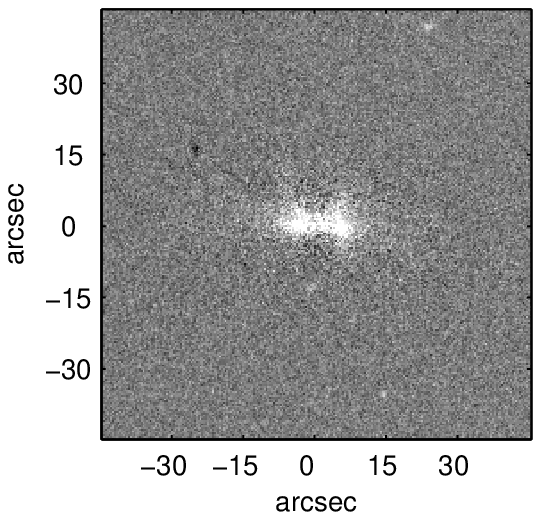} &\includegraphics[]{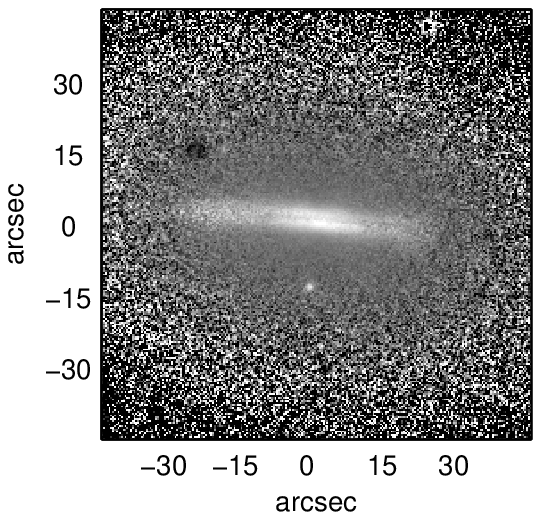}\\
  \vspace{-3mm}
 \includegraphics[width=6cm]{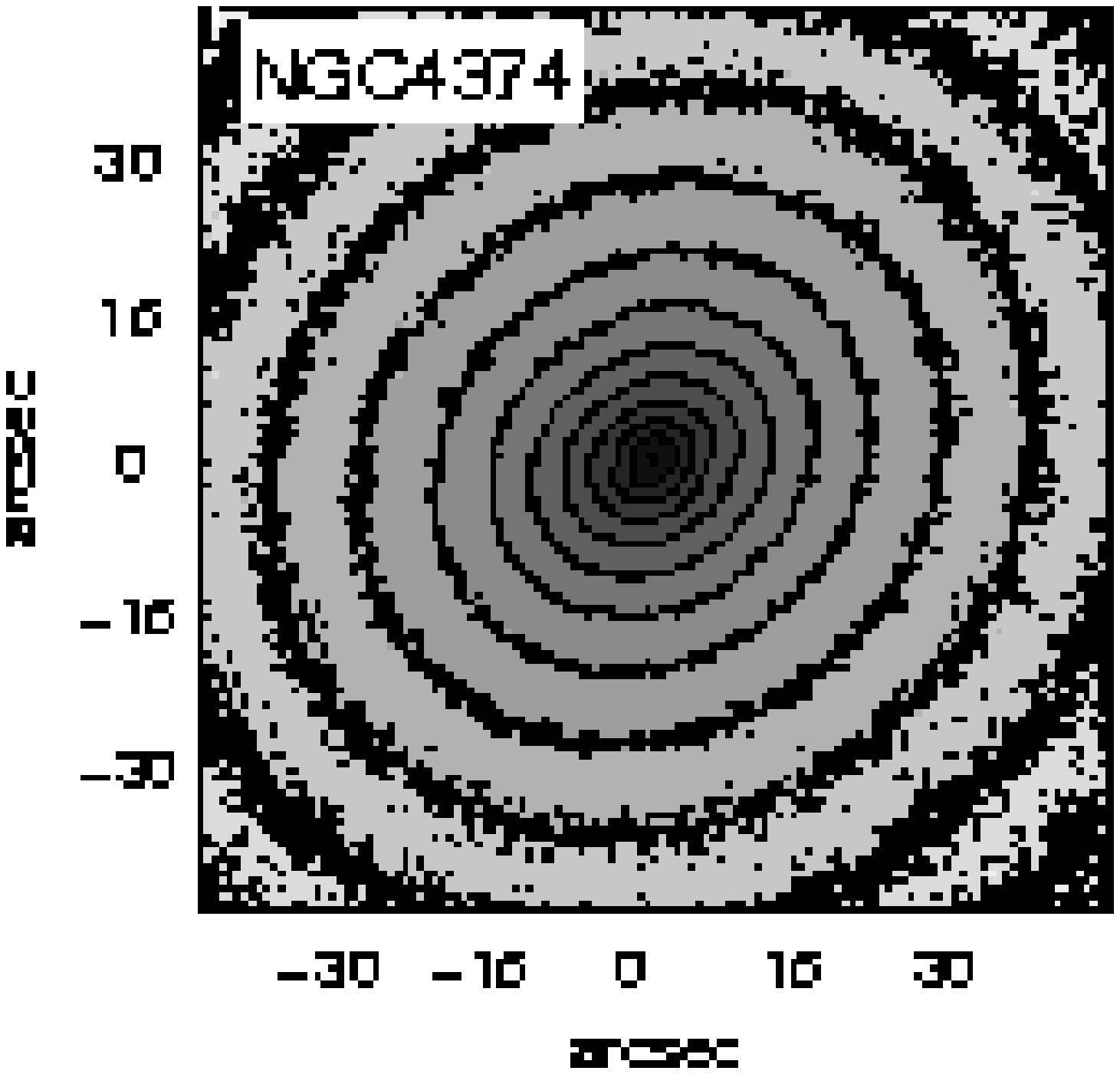} & \includegraphics[]{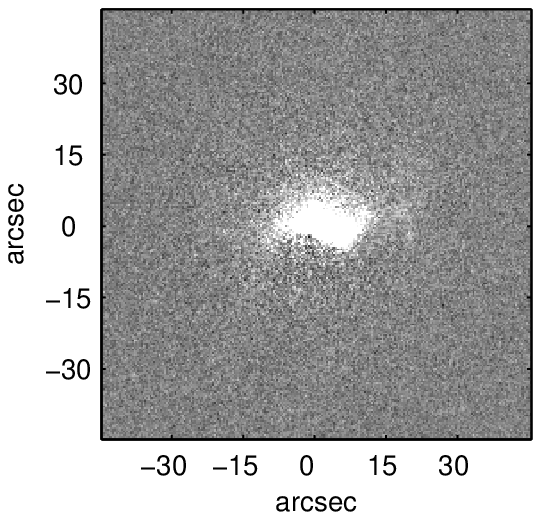}  & \includegraphics[]{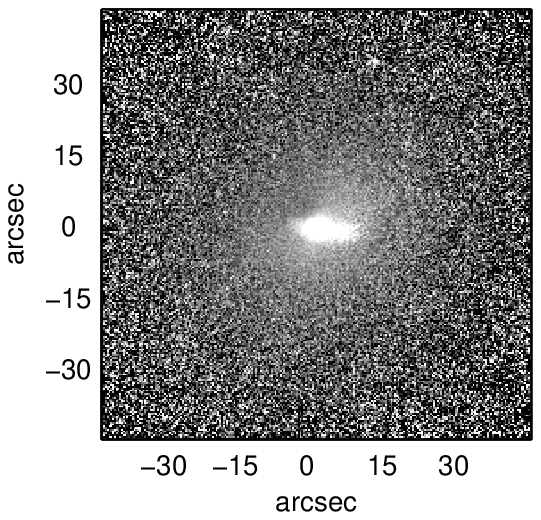}\\
  \vspace{-3mm}
\includegraphics[width=6cm]{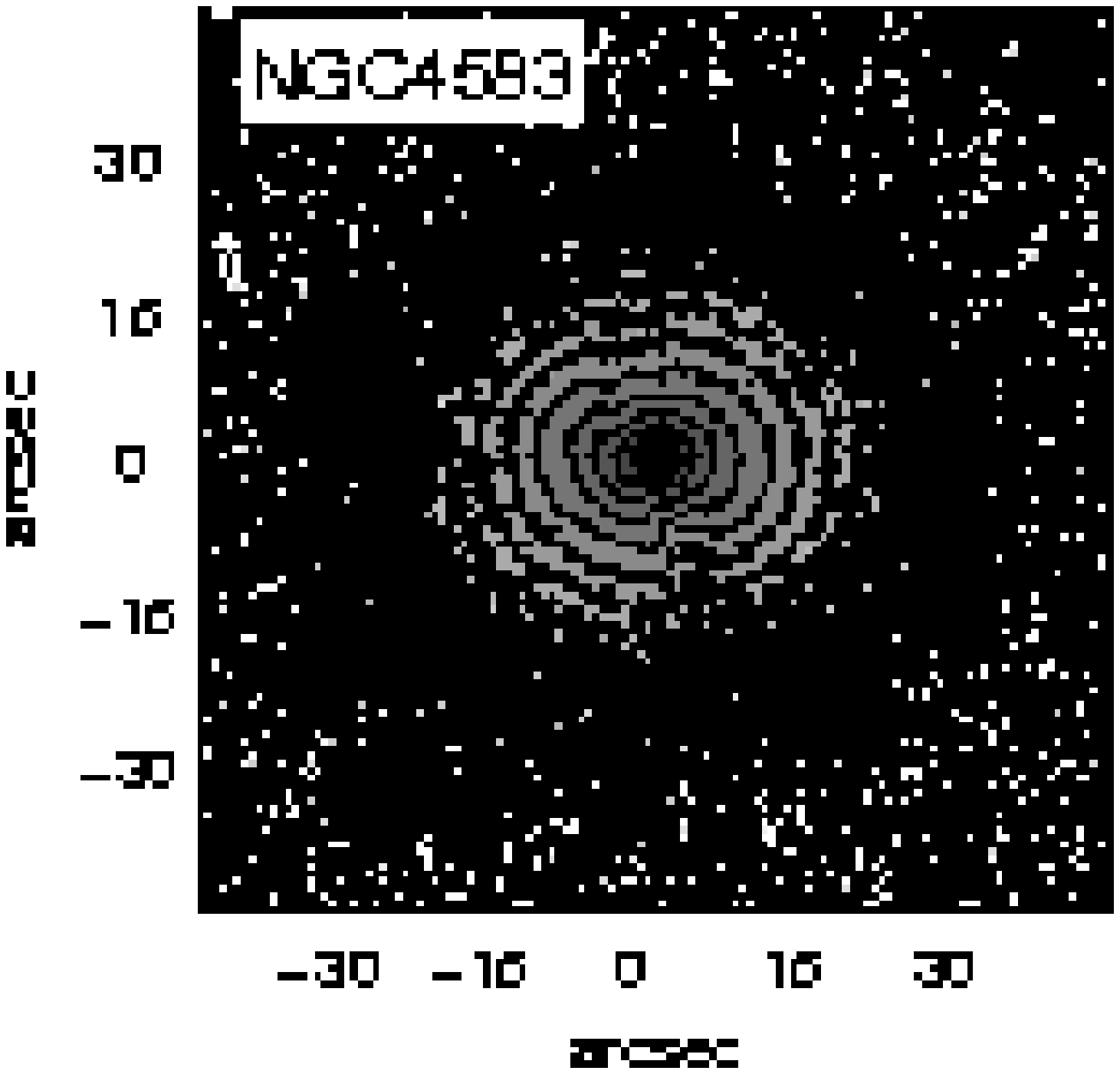} & \includegraphics[]{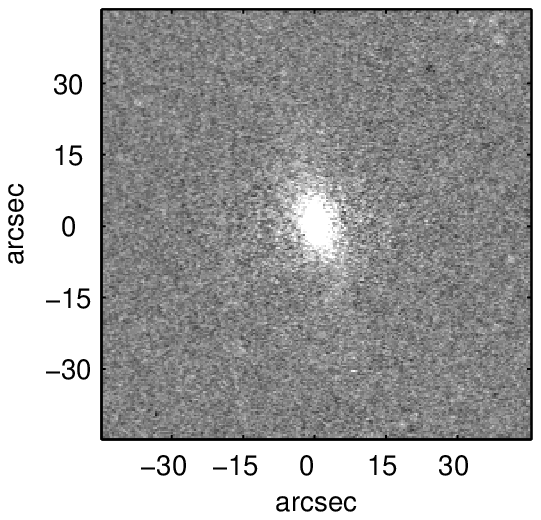} & \includegraphics[]{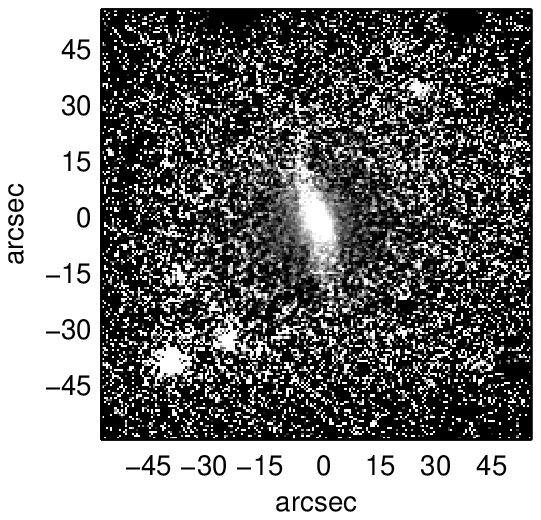}\\
 \vspace{-3mm}
\includegraphics[width=6cm]{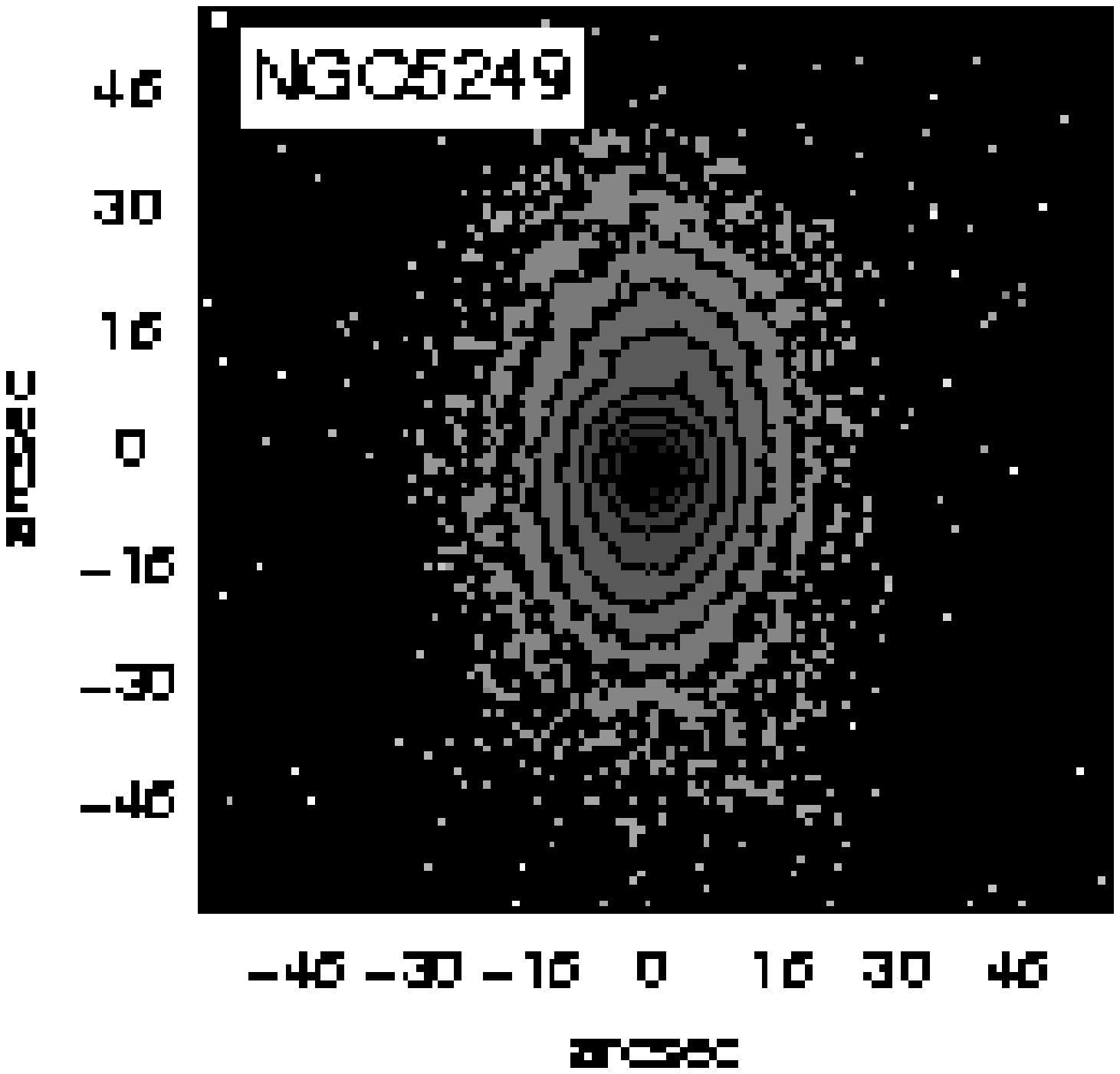} & \includegraphics[]{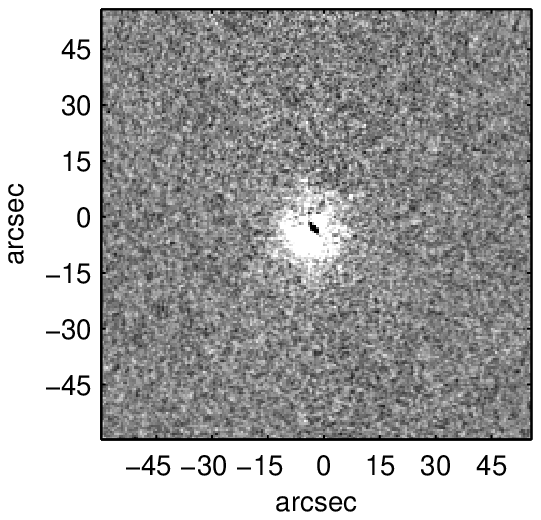} &\includegraphics[]{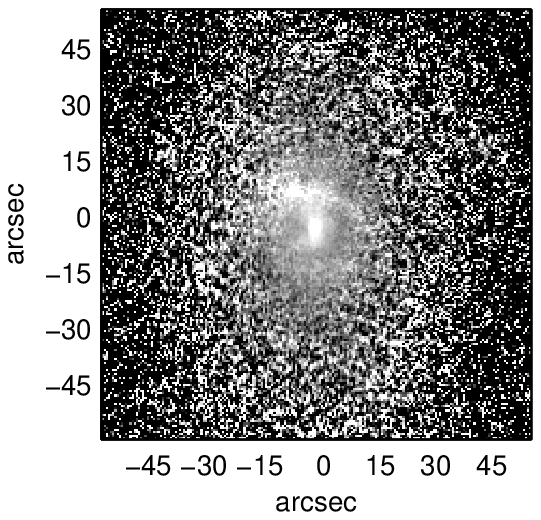}
\end{tabular}
\end{center}
\caption{R-band contour maps, continuum-subtracted H$\alpha$+[NII] images and B-R colour-index maps.}
 \label{fig:BImaps5}
\end{figure*}

\begin{figure*}
\begin{center}
\begin{tabular}{ccc}
 \includegraphics[width=6cm]{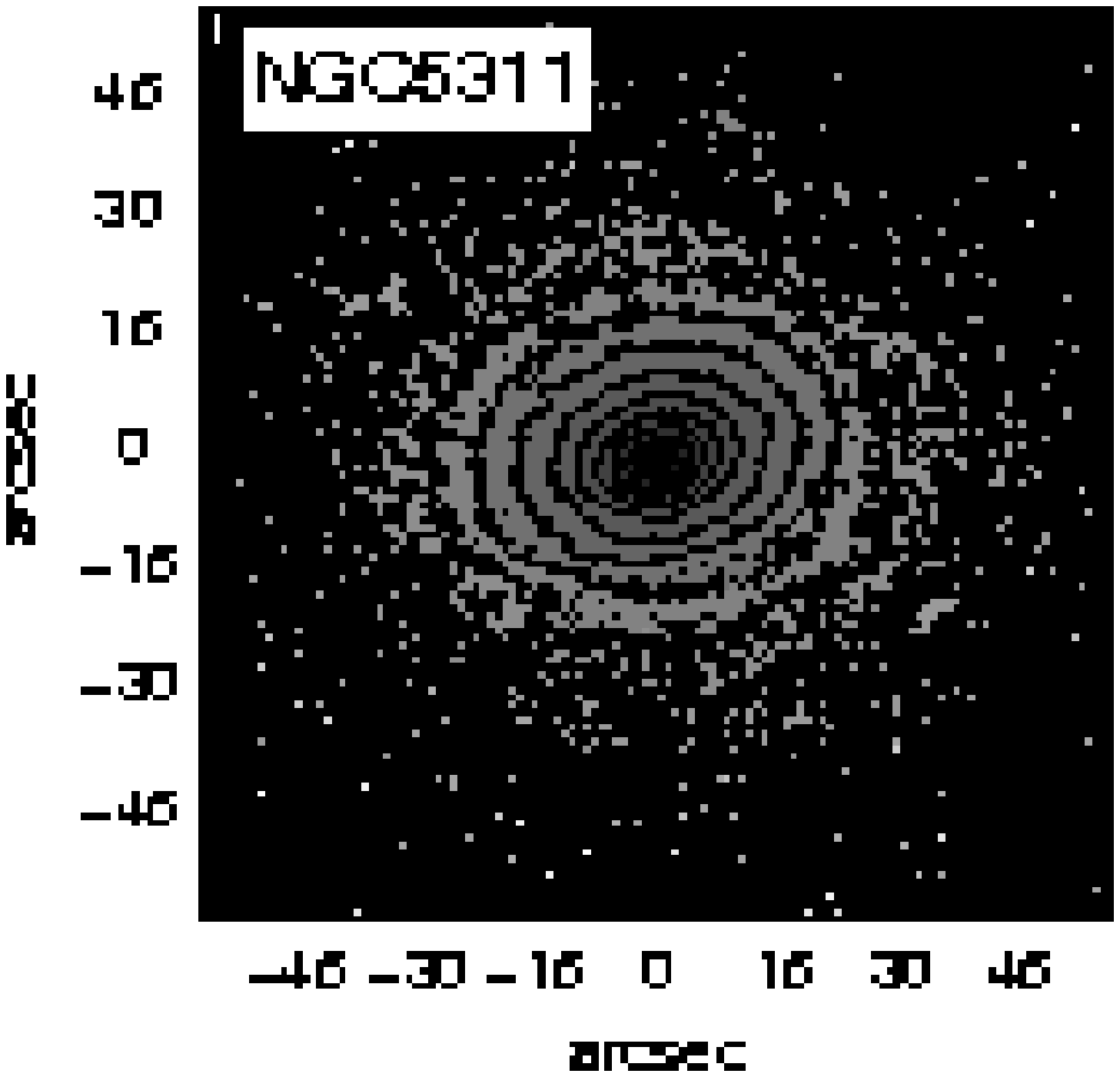} &  \includegraphics[]{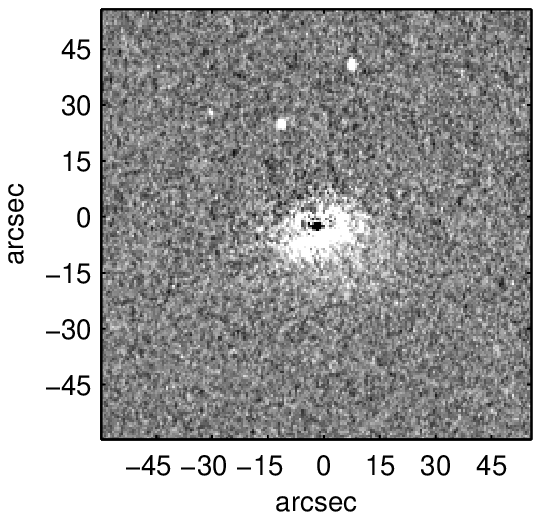} &\includegraphics[]{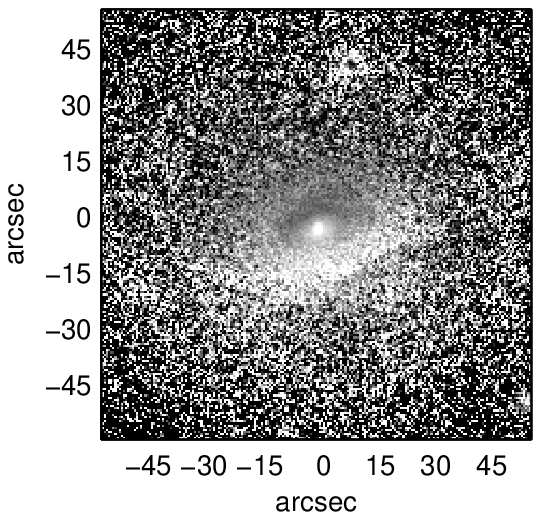}\\ 
 \vspace{-3mm}
 \includegraphics[width=6cm]{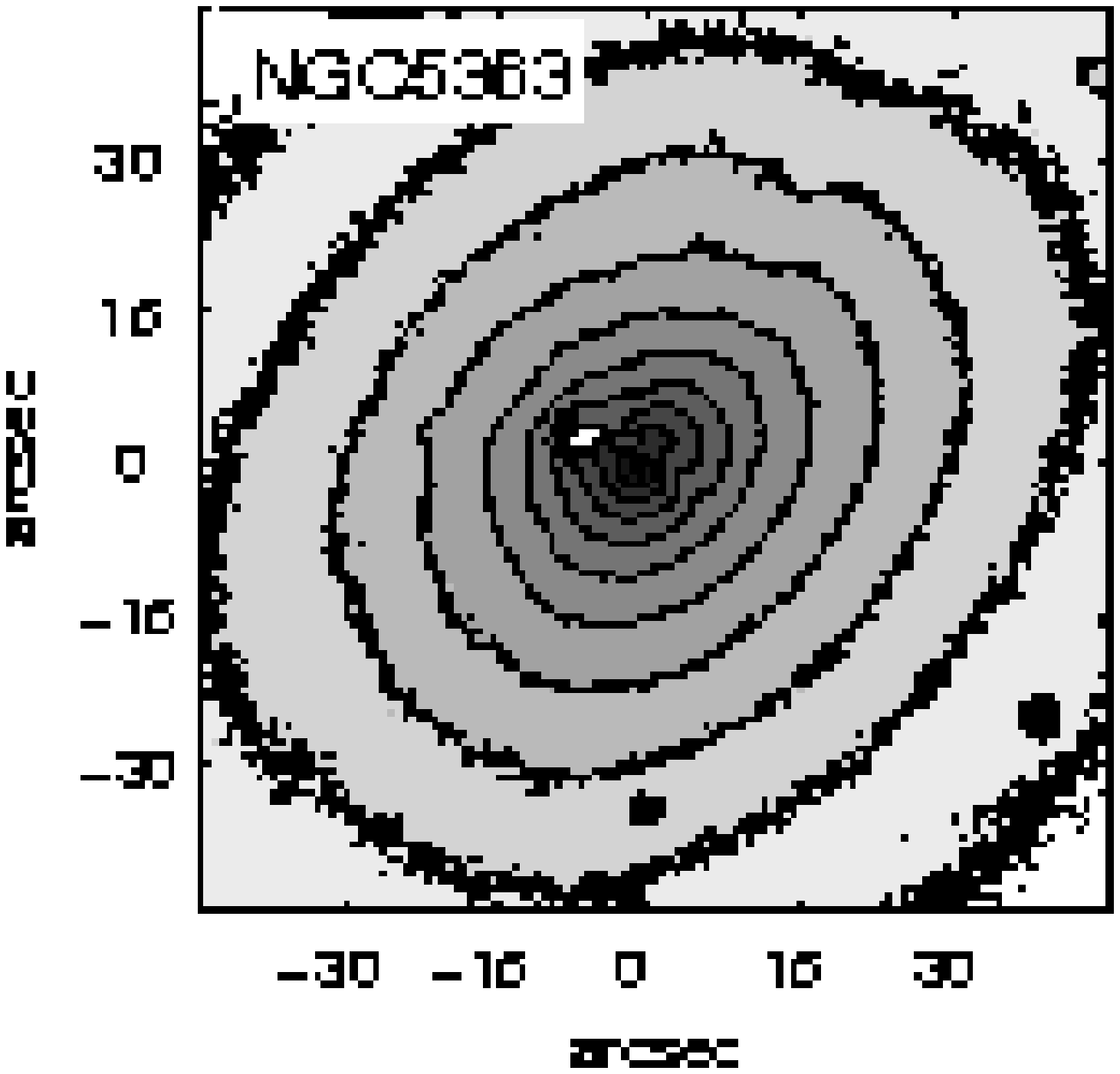} & \includegraphics[]{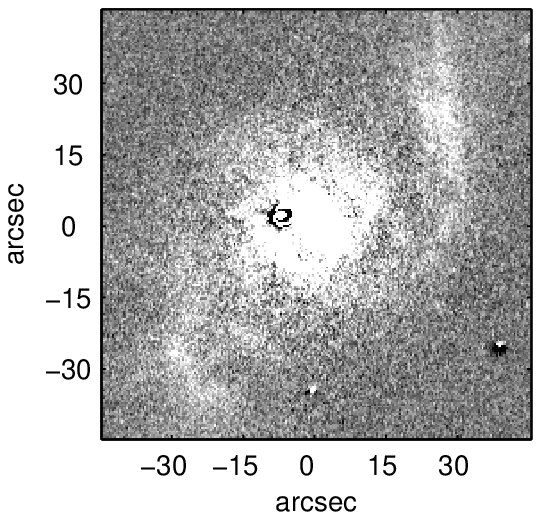}  & \includegraphics[]{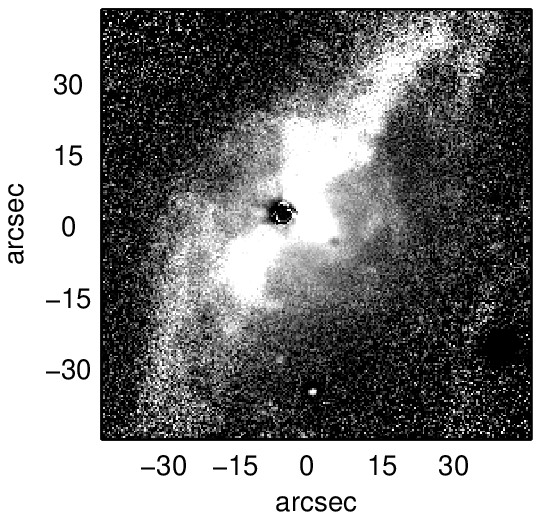}\\
  \vspace{-3mm}
\includegraphics[width=6cm]{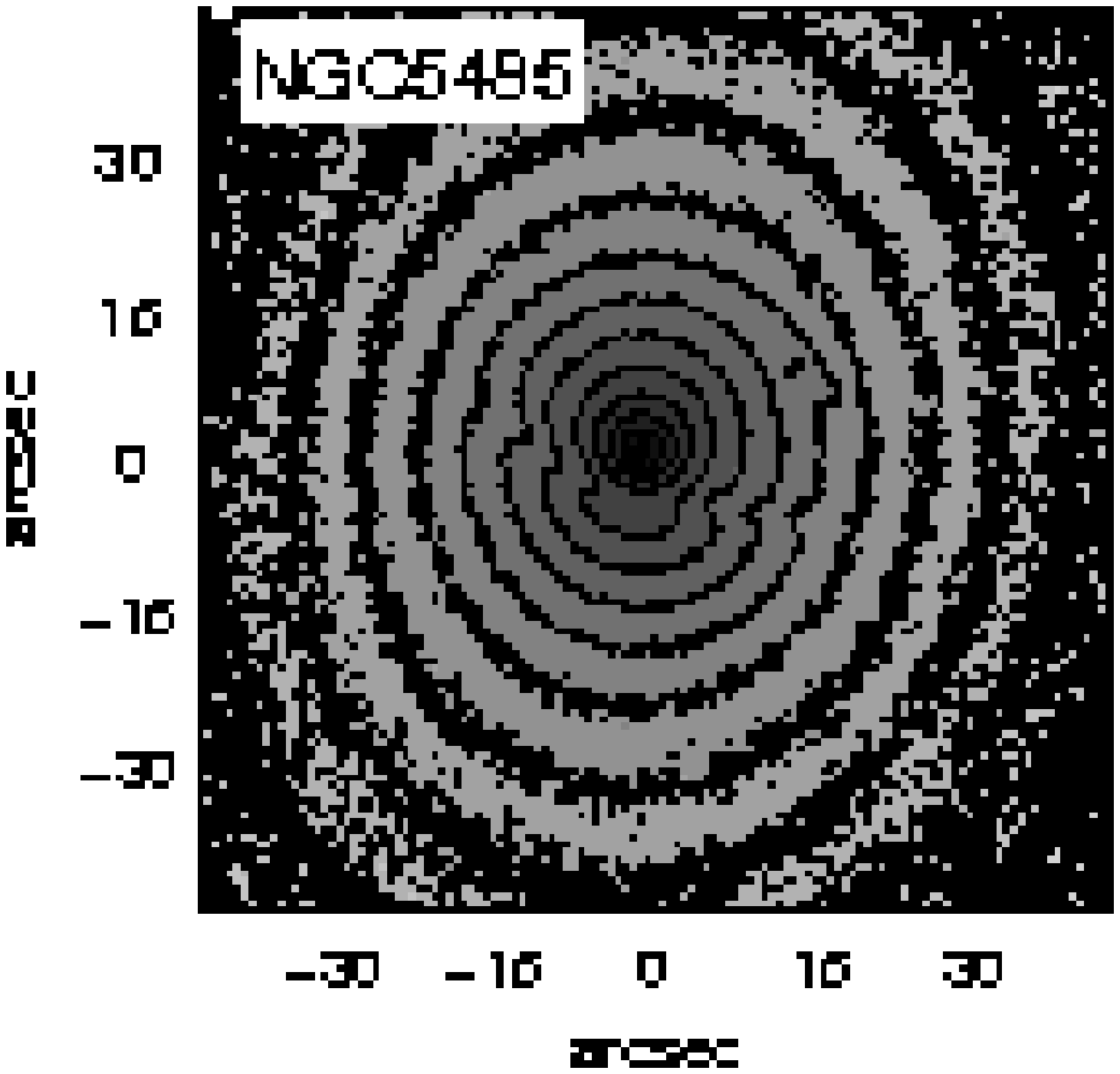} & \includegraphics[]{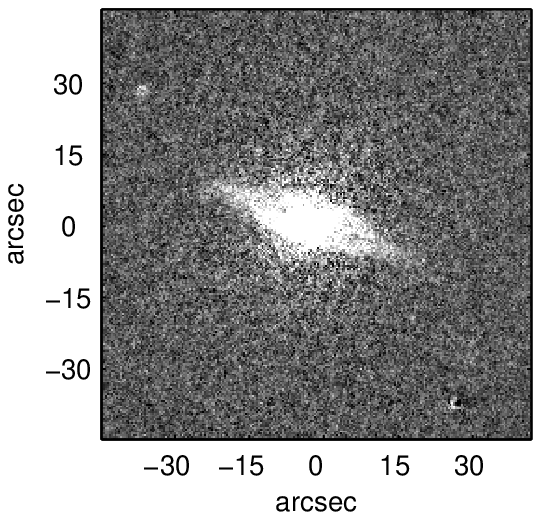} & \includegraphics[]{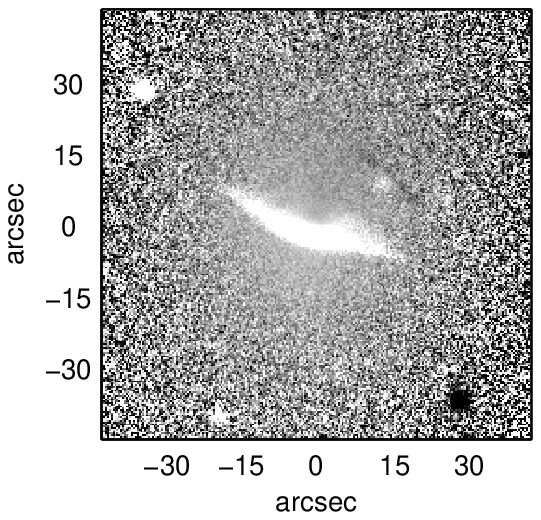}\\
 \vspace{-3mm}
\includegraphics[width=6cm]{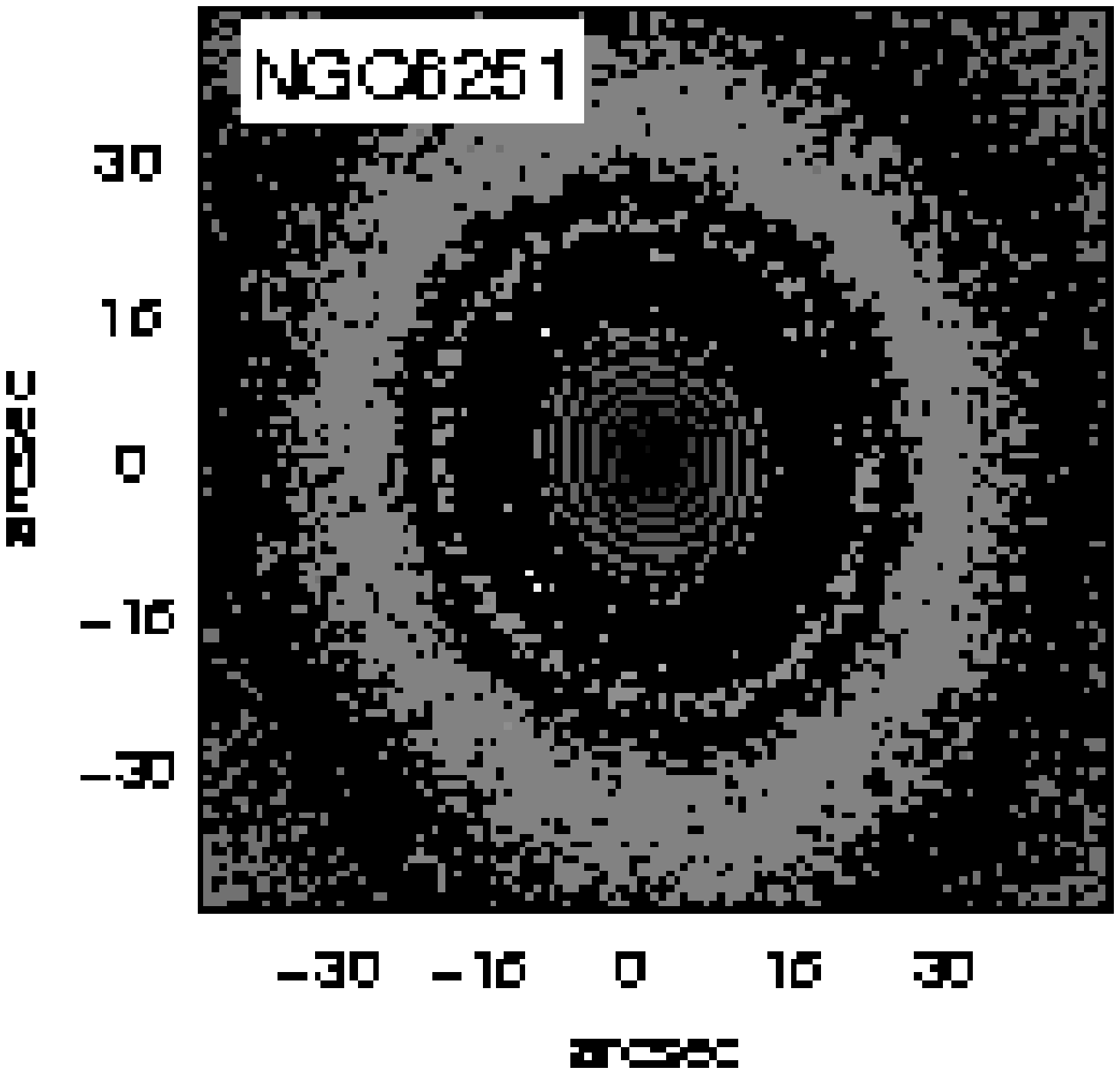} & \includegraphics[]{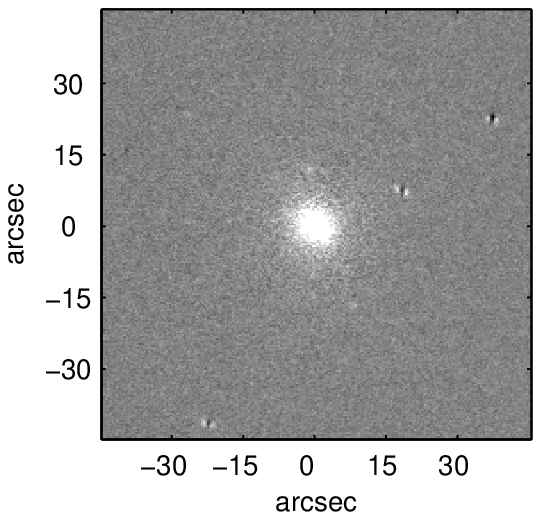} &\includegraphics[]{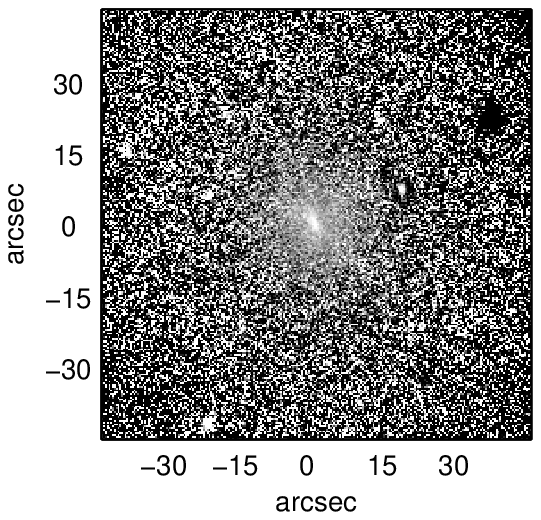}
\end{tabular}
\end{center}
\caption{R-band contour maps, continuum-subtracted H$\alpha$+[NII] images and B-R colour-index maps.}
 \label{fig:BImaps6}
\end{figure*}

\begin{figure*}
\begin{center}
\begin{tabular}{ccc}
 \includegraphics[width=6cm]{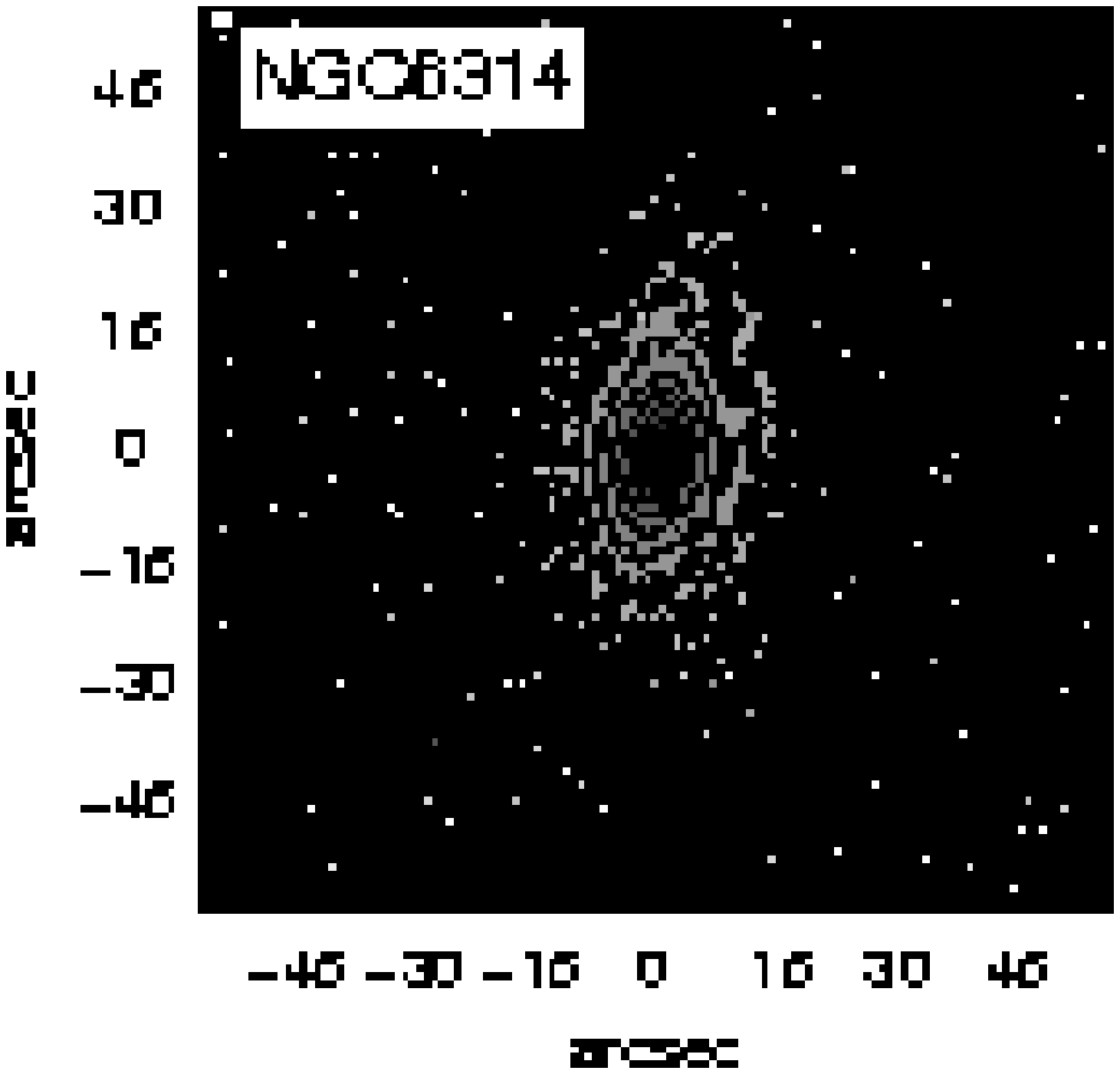} &  \includegraphics[]{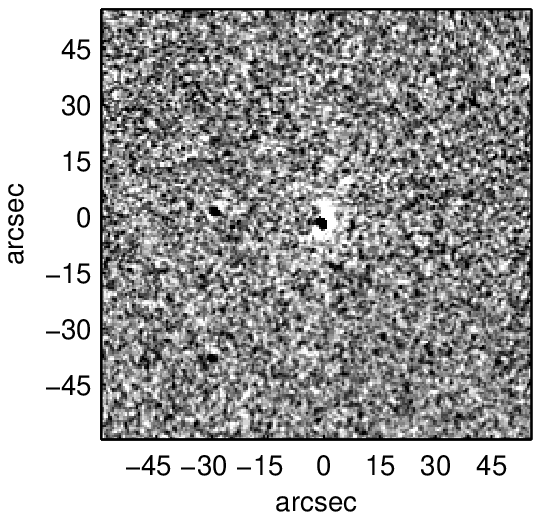} &\includegraphics[]{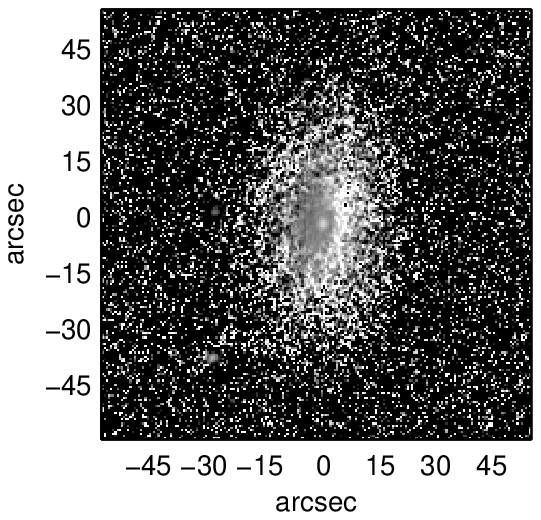}\\
  \vspace{-3mm}
 \includegraphics[width=6cm]{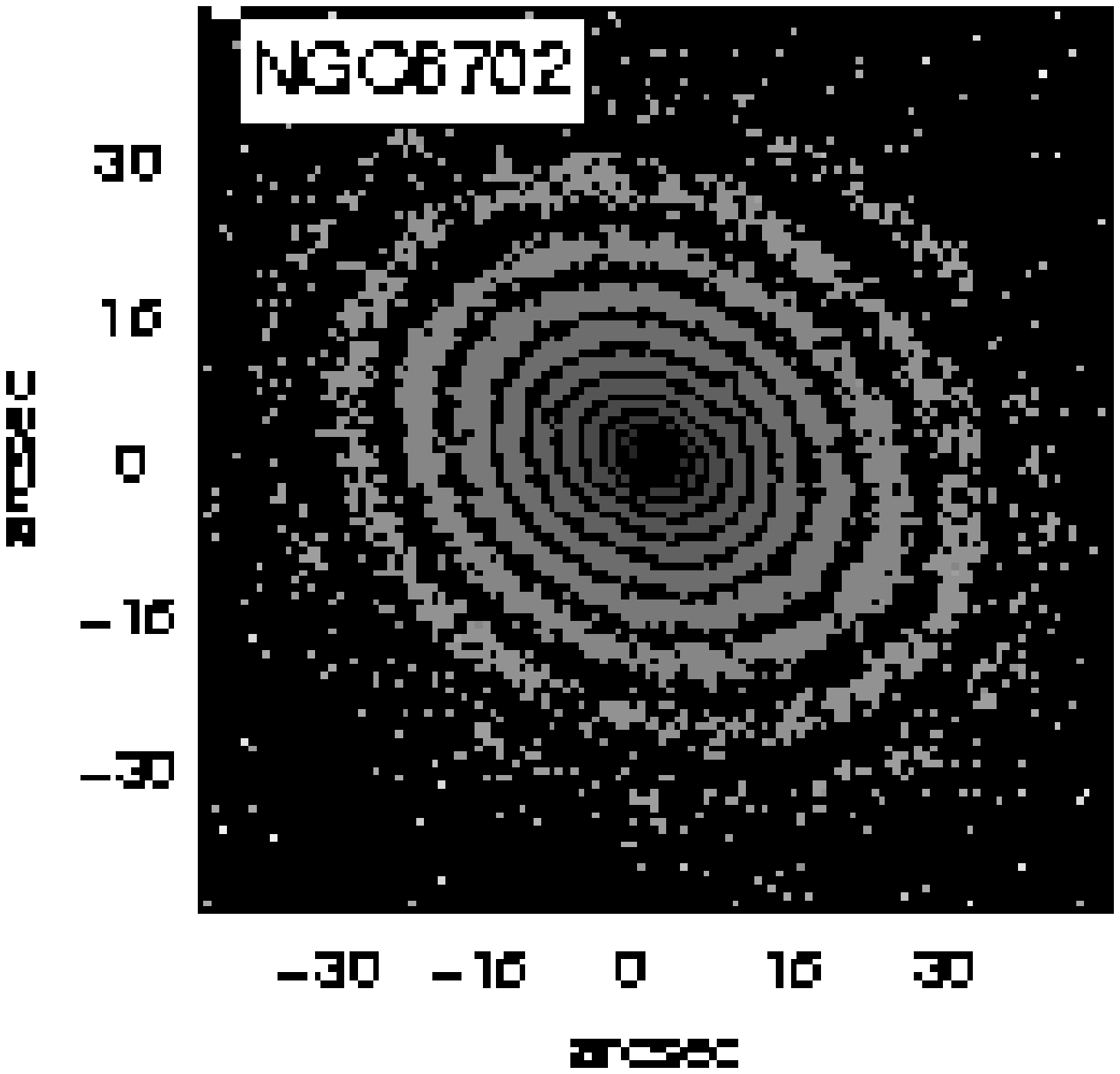} & \includegraphics[]{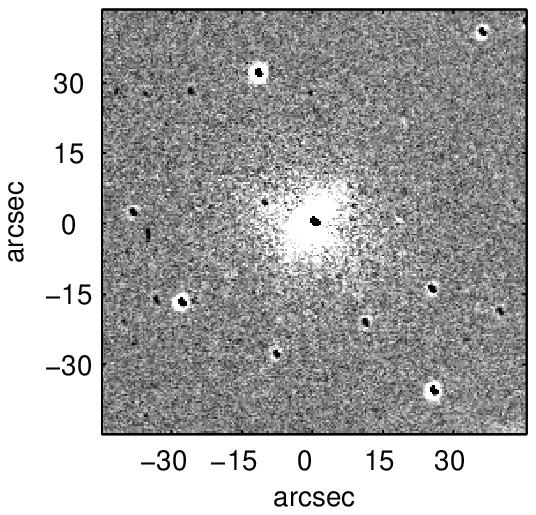}  & \includegraphics[]{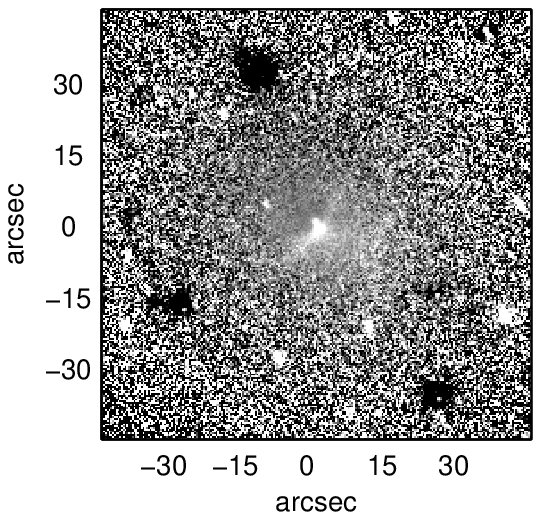}\\
  \vspace{-3mm}
\includegraphics[width=6cm]{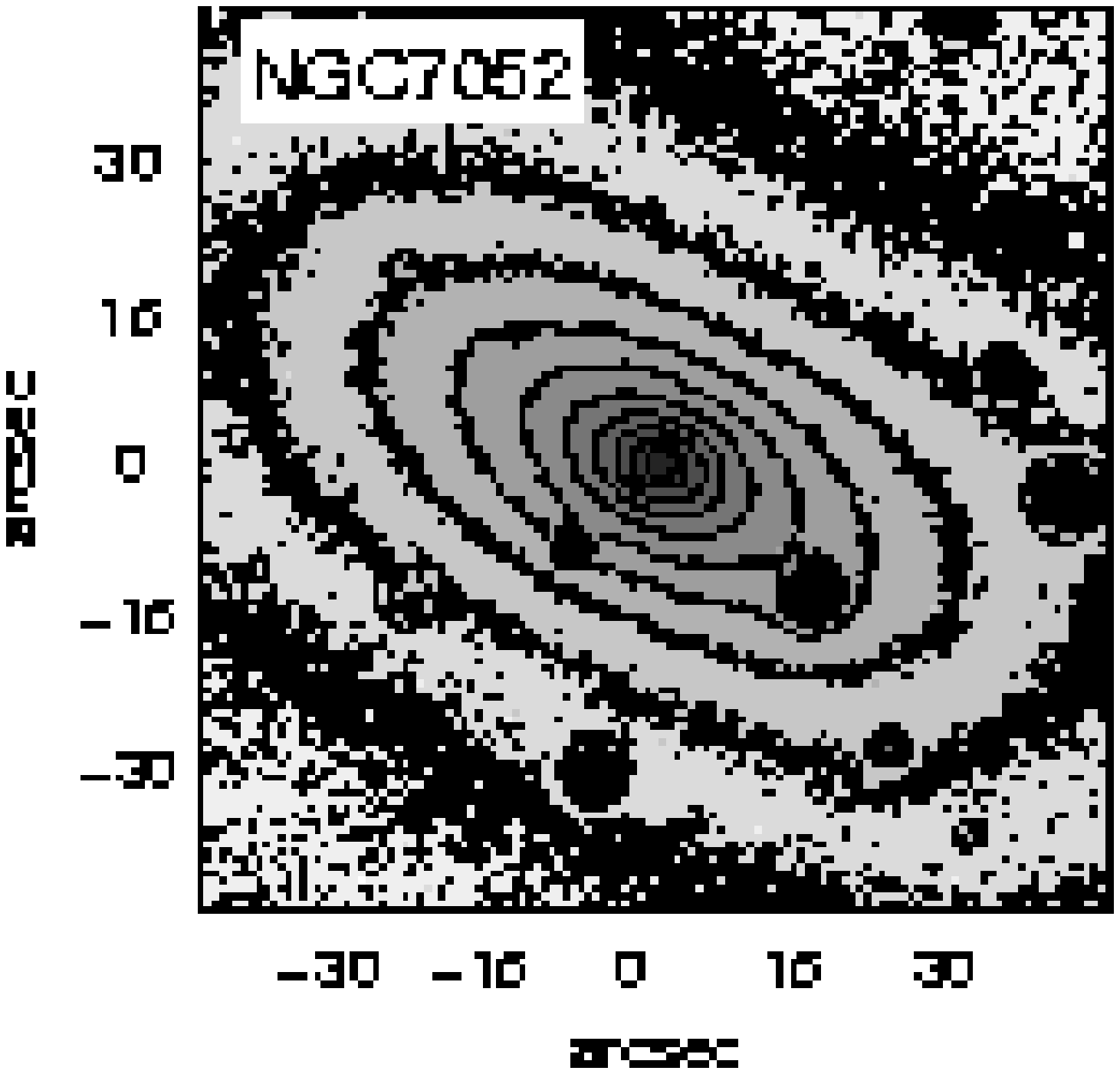} & \includegraphics[]{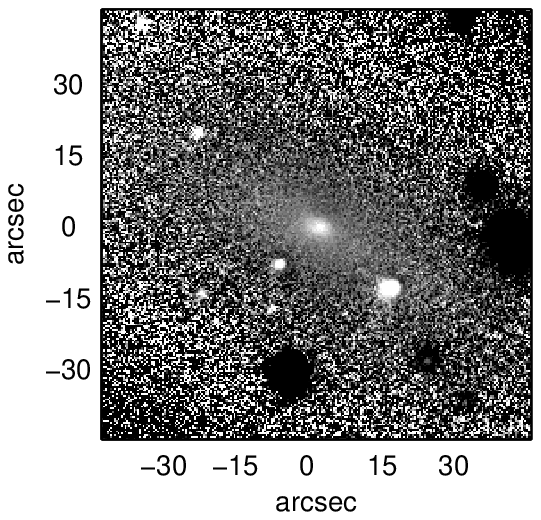} & \includegraphics[]{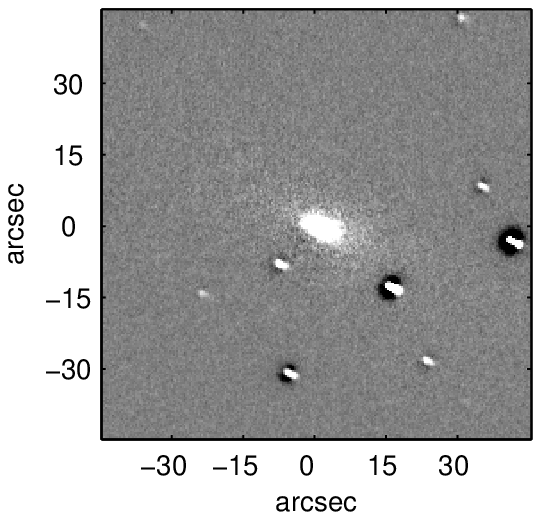}\\
 \vspace{-3mm}
\includegraphics[width=6cm]{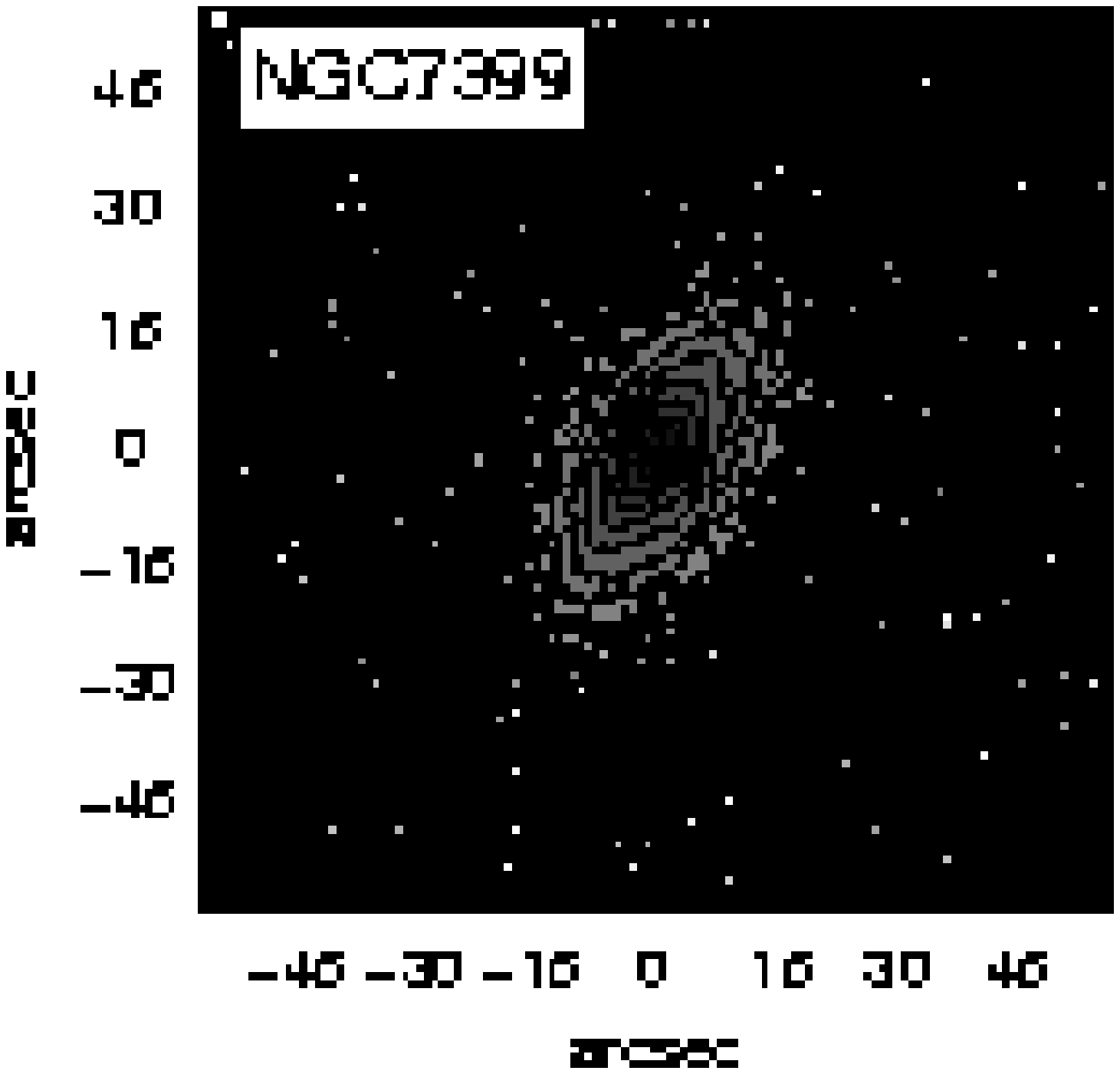} & \includegraphics[]{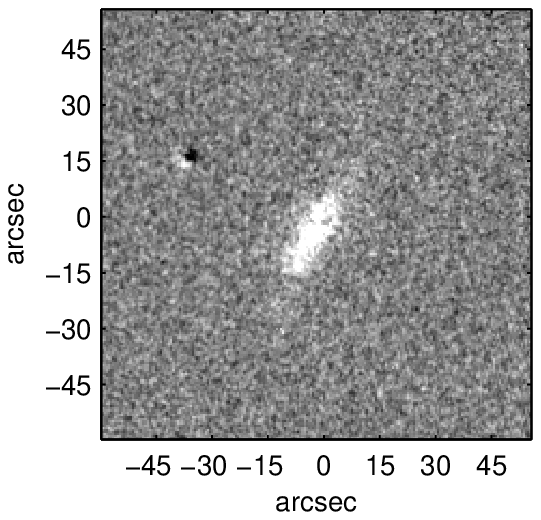} &\includegraphics[]{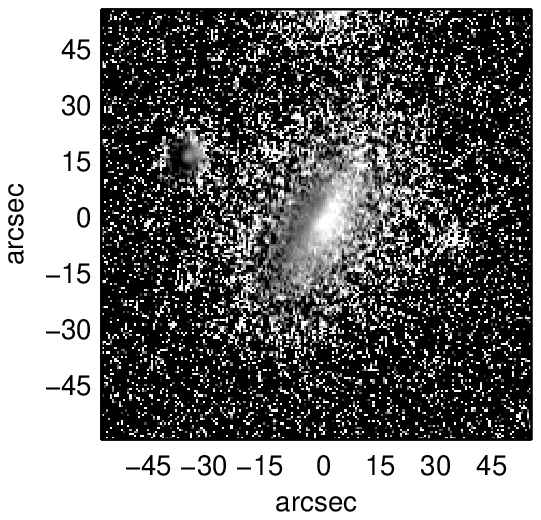}
\end{tabular}
\end{center}
\caption{R-band contour maps, continuum-subtracted H$\alpha$+[NII] images and B-R colour-index maps.}
 \label{fig:BImaps7}
\end{figure*}

\begin{figure*}
\begin{center}
\begin{tabular}{ccc}
 \includegraphics[width=6cm]{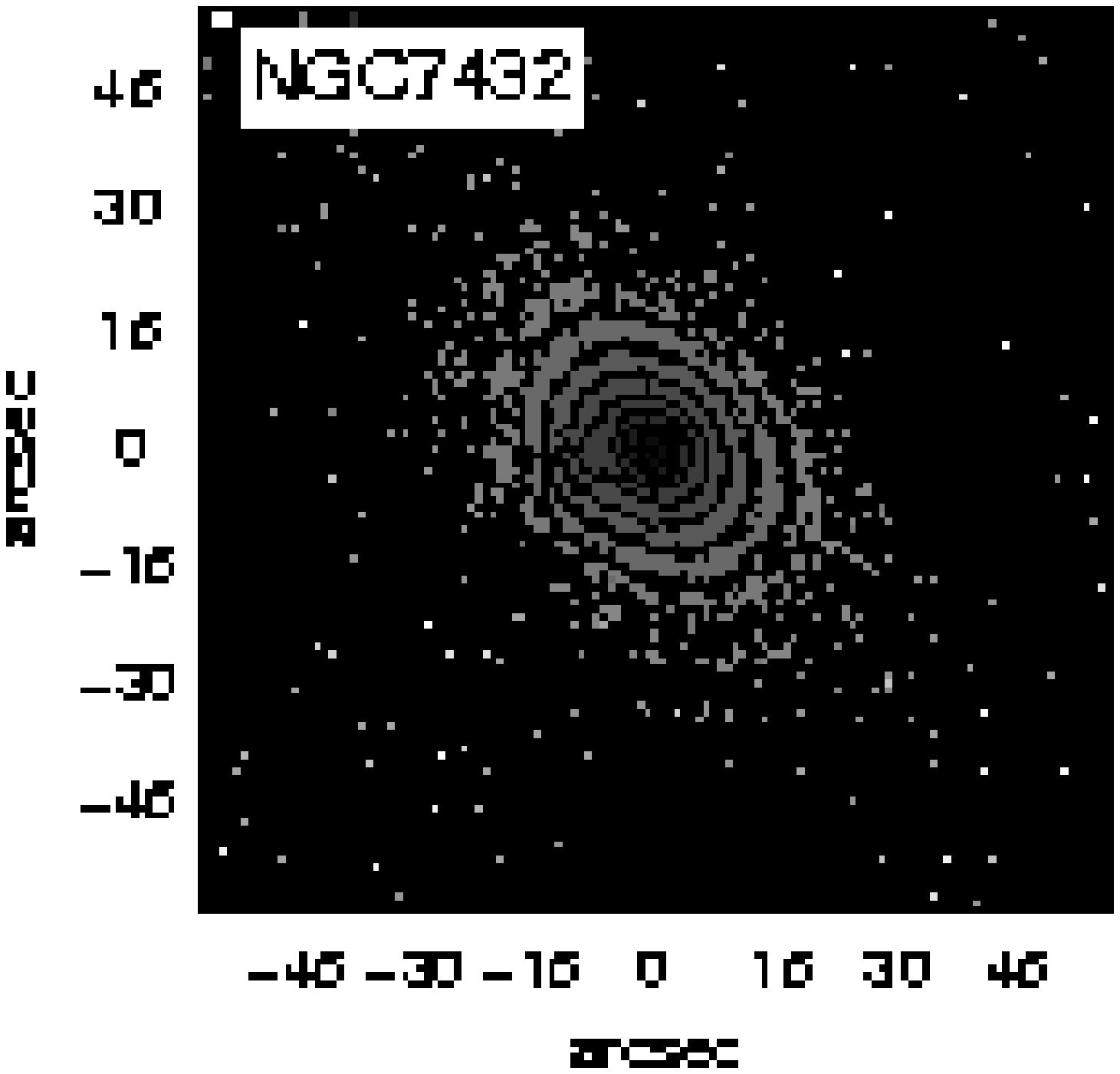} &  \includegraphics[]{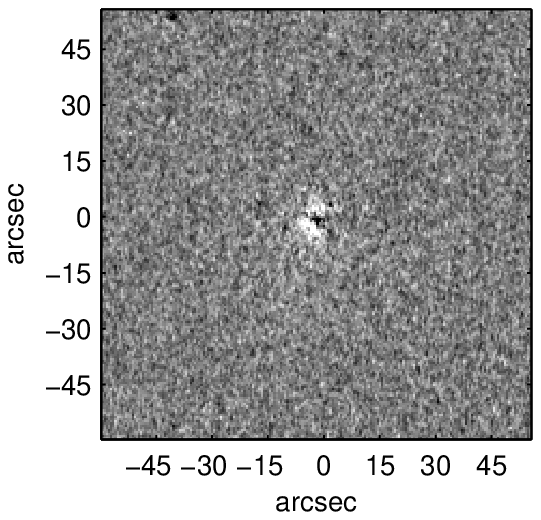} &\includegraphics[]{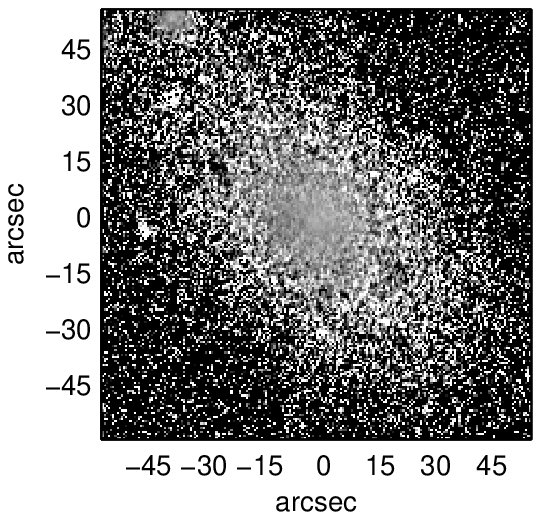}\\
 \vspace{-3mm}
 \includegraphics[width=6cm]{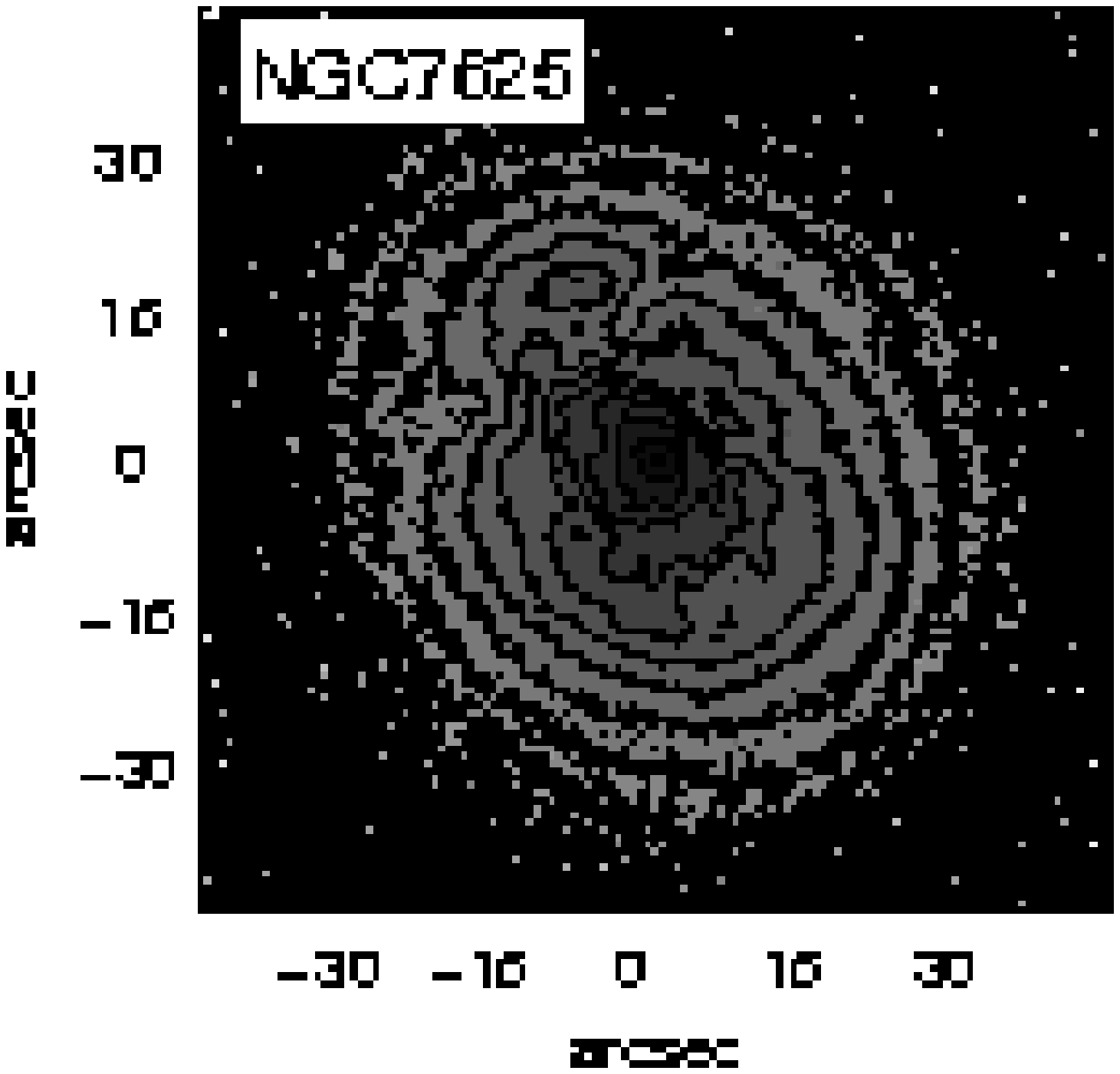} & \includegraphics[]{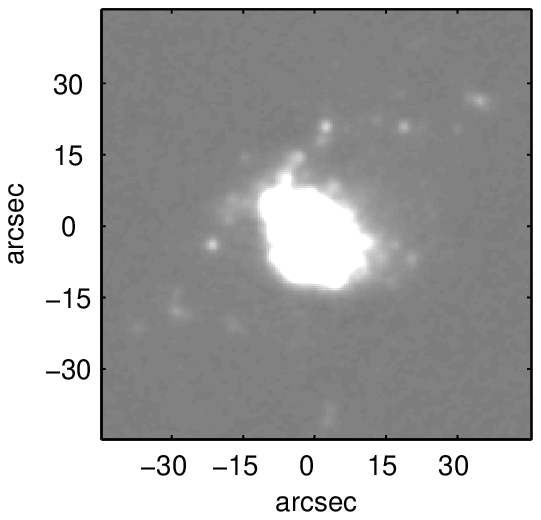}  & \includegraphics[]{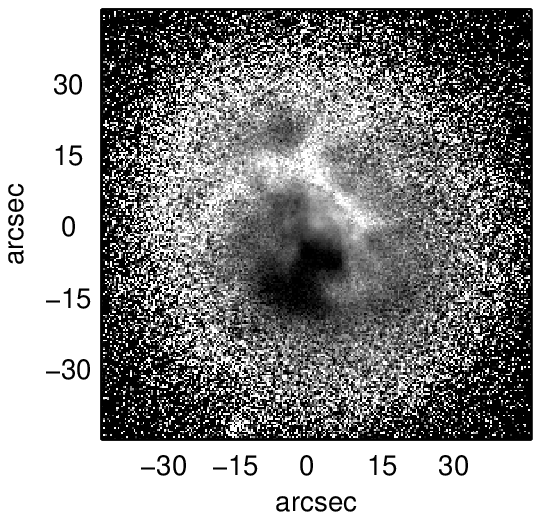}
\end{tabular}
\end{center}
\caption{R-band contour maps, continuum-subtracted H$\alpha$+[NII] images and B-R colour-index maps.}
 \label{fig:BImaps8}
\end{figure*}

Below we briefly describe each of our sample galaxies. In addition we present in Figures \ref{fig:BImaps1}-\ref{fig:BImaps8} a R-band contour map image, a continuum-subtracted  H$\alpha$+[NII] narrow-band image and a B-R colour image for each galaxy. The images correspond to the inner regions of the galaxies and are typically $\sim 45 \arcsec \times 45 \arcsec$.
\newline
\textbf{IC 1575.} This elliptical galaxy has an apparent DL along the minor axis and stellar shells in the direction of the major axis. 
It was included in a study of the optical and morphology of elliptical DL galaxies (M\"{o}llenhoff et al.\ 1992).
While the minor-axis DL may be an inclined dust disk, the colour image also reveals the presence of an off-center DL which is not aligned with either the minor or major axes but is at $\sim 30 ^{\circ}$. As shown by the continuum-subtracted H$\alpha$+[NII] image, the extended IG region follows the central dusty disk. \\
\textbf{NGC662.} This galaxy is classified S peculiar in the RC3 catalog. The colour and H$\alpha$+[NII] images show patches of dust which coincide with the line emission. Very weak radio emission was detected by Gregorini et al.\ (1989). This galaxy was included in a search for molecular gas in Es with DLs where $1.5 \times 10^9 M_\odot$ of cold molecular hydrogen gas were detected (Wang et al.\ 1992).\\
\textbf{ESO 477-7.} This galaxy is classified S0 in RC3. The colour image shows an inner dust disk surrounded by an asymmetric and more diffuse dust structure. The continuum-subtracted H$\alpha$+[NII] image reveals an inclined extended disk $\sim$orthogonal to the stellar isophotes.\\
\textbf{NGC708.} This E2 galaxy is the brightest in the Abell 262 cluster. The object shows two orthogonal gaseous components with similar masses and a total mass of $3.2\times 10^4 M_\odot$ (Plana \& Boulesteix 1997). The continuum-subtracted H$\alpha$+[NII] image shows an inclined disk which follows the dust component.\\
\textbf{ESO 197-10.} This E galaxy has a strong minor-axis DL which is likely a dust disk observed edge-on. The continuum-subtracted $\mbox{H} \alpha$+[NII] image shows an extended IG disk with a diffuse faint structure that follows the DL.\\
\textbf{ESO 355-8.} The colour image of this S0 galaxy shows a strong intermediate DL which may represent an edge-on disk lying in the same plane as the IG disk.\\
\textbf{NGC 1199.} This E galaxy is the brightest in a small group of galaxies. Its colour image reveals a barely resolved inner structure of a circumnuclear dust disk surrounded by a ring, where both components are aligned with the major axis.  The continuum-subtracted H$\alpha$+[NII] image shows an extended disk-like structure which follows the stellar isophotes.\\
\textbf{NGC 1297.} This galaxy shows a faint outer envelope which classifies it as an S0 in RC3. The derived colour map show a complex dust structure consisting of an inner circumnuclear disk with several concentric arcs along the minor axis. The continuum-subtracted H$\alpha$+[NII] image shows the IG as an inclined extended disk along the minor axis.\\
\textbf{ESO 118-19.} This S0 galaxy has a strong DL along its minor axis. The continuum-subtracted H$\alpha$+[NII] image reveals an inclined IG disk which follows the dust structure.\\ 
\textbf{NGC 2534.} This E1 galaxy was part of a study of molecular gas in E galaxies with DLs by Wang et al.\ (1992), where $10^9\times M_\odot$ of cold molecular hydrogen gas were detected.
The continuum-subtracted H$\alpha$+[NII] image reveals a gaseous disk inclined with respect to the galactic plane, with its long axis along the observed minor axis of the galaxy. The apparent lack of emission at the center of the galaxy in the continuum-subtracted image is likely the result of a problematic PSF matching of the $\mbox{H} \alpha$+[NII] and the R-band images. The colour image reveals a circumnuclear disk surrounded by a dusty arc. The arc might be part of an inclined dust ring/disk which follows the IG distribution to its detection limit.\\
\textbf{UGC4449.}  This galaxy shows an apparent DL along the minor axis of its inner body. Extended outer shells are present in the direction of the minor axis. The continuum-subtracted H$\alpha$+[NII] image does not show an extended structure and we did not detect significant line emission at the center of the galaxy.\\
\textbf{NGC2968.}  The colour image shows a strong DL along the minor axis of the galaxy, which may account for the apparent rotation of the inner isophotes with respect to the outer isophotes. While Gregorini et al.\ (1989) reported that the DL is confined to the inner part of the galaxy, the colour image clearly shows it following the luminous body almost to the detection limit. 
The continuum-subtracted H$\alpha$+[NII] image does not show an extended structure and we did not detect significant line emission in the center of the galaxy. \\ 
\textbf{Mrk33.} This elliptical blue compact galaxy is known to contain dust and a conspicuous central star formation region (Davidge 1989). The colour image shows that the galaxy is extremely blue at its center and increasingly redder outwards. The off-center dust patches reported by M\"{o}llenhoff et al.\ (1992) are not seen in our images. H$\alpha$ imaging by M\'{e}ndez \& Esteban (2000) showed the presence of at least three star-forming knots in the center of the galaxy. These are not resolved in the disk shown in our continuum-subtracted H$\alpha$+[NII] image. Summers, Stevens \& Strickland (2001) suggested that the colour gradient and the distinct discontinuities in the stellar luminosity function of Mrk33 are evidence for multiple episodes of intense star-formation in the history of the galaxy. \\
\textbf{UGC5814.} This E galaxy has a major-axis DL. The R-band reveals distorted outer isophotes and outer shells with a minor axis perpendicular to that of the stellar body. The continuum-subtracted H$\alpha$+[NII] image reveals an IG disk which follows the dust component.\\
\textbf{NGC 3656.} This peculiar galaxy shows nearly circular shells. The colour map reveals a minor DL with filaments extending towards the shells. The continuum-subtracted H$\alpha$+[NII] image reveals an inclined gaseous disk with filaments and clumps, which coincide with the dust morphology and are known sites of on-going star formation with a total rate of $\sim 0.1 M_\odot$ yr$^{-1}$ (M\"{o}llenhoff et al.\ 1992; Balcells 1997).
HI VLA observations reveal an edge-on, warped, minor axis HI disk with a mass of $\sim 2 \times 10^9 M_{\odot}$ extending up to 7 kpc from the center (see also Wang et al.\ 1992). Balcells (1997) reported the discovery of two tidal tails, with a dwarf galaxy lying in one of the tails, and suggested that a disk-disk major merger scenario is more likely to account for the observed morphology than a minor merger. HI gas is also found outside the optical images in the two tails and in the shell (Balcells et al.\ 2001). \\
\textbf{NGC 3665.} The colour image of this E galaxy reveals a $\sim$kpc wide dust ring along its major axis. 
The apparent colour variation along the dust ring is probably an inclination effect where the more distant part of the ring obscures less starlight and therefore appears bluer. The IG structure revealed by the continuum-subtracted $\mbox{H} \alpha$+[NII] image clearly shows an extended disk which becomes more luminous towards its center and which follows precisely the dust morphology. Radio observations reveal a continuum radio source with an elongated structure with the major axis nearly perpendicular to the DL (Kotanyi 1979).\\
\textbf{NGC 4370.} This galaxy is known to host of a multiphase ISM (see Huchtmeier \& Richter 1988; Bettoni, Galletta \& Garc\'{i}a-Burillo 2003; Caldwell, Rose \& Concannon 2003; Patil et al.\ 2009). The presence of a prominent DL along its major axis hinders the determination of its underlying morphological classification; it is classified S0 in Binggeli, Sandage \& Tammann (1985) and Sa in the RC3 catalog.  The colour image reveals a major axis DL of a few kpc. The DL may be a disk observed edge-on, although the apparent bluer regions in its central parts are more suggestive of a ring-like structure. The continuum-subtracted $\mbox{H} \alpha$+[NII] image reveals a weak clumpy disk, possibly due to the heavily obscuring edge-on disk/ring. It is possible that the gaseous structure extends out to the limit of the DL, as observed in other galaxies, but its luminosity is too faint to detect due to the extinction. \\ 
\textbf{NGC 4374.} This E1 galaxy is a strong radio continuum source. The colour image reveals a red core with a slightly bluer surrounding region, which implies a circumnuclear dust disk denser towards the center. The continuum-subtracted $\mbox{H} \alpha$+[NII] image reveals a distorted IG structure, possibly a warped gaseous disk, which extends to the DL limit. The images imply that the DL and the gaseous disk do not lie on either the apparent long or short axes. HST images show two DLs within the central 5'' of the galaxy oriented roughly parallel to a filamentary H$\alpha$+[NII] emission of similar size (Bower et al.\ 1997). In addition, several studies reported no detection of HI or CO emission (Huchtmeier 1994, Knapp \& Ruppen 1996, Combes et al.\ 2007). \\
\textbf{NGC 4583.} Although not apparent in the colour image, this galaxy has a minor axis DL which is revealed when fitting ellipse models to the broad-band images and subtracting these from the actual images. The continuum-subtracted $\mbox{H} \alpha$+[NII] image reveals an IG disk-like structure which follows the DL morphology.\\
\textbf{NGC5249.} The colour image of this galaxy shows an inner dust structure of a central disk surrounded by an arc, which might be a part of an inclined ring. The continuum-subtracted H$\alpha$+[NII] image reveals an inner IG disk which coincides with the dust disk.\\
\textbf{NGC5311.} This galaxy is classified S0 in the RC3 catalog. The colour image clearly reveals a circumnuclear dust disk surrounded by a dust ring. The continuum-subtracted H$\alpha$+[NII] image shows traces of central line emission, but the weak emission is below our detection limit, thus we cannot confirm the presence of an inner IG disk. \\
\textbf{NGC 5485.} This galaxy is classified S0 in the HyperLEDA catalog. The colour image reveals a minor axis DL which might be a dust disk/ring observed edge-on and is better shown by the model-subtracted image. The continuum-subtracted $\mbox{H} \alpha$+[NII] image reveals an inclined IG disk which follows the DL.\\
\textbf{NGC 6251.} This giant E2 hosts a powerful radio source. While our colour image reveals a core slightly redder than its surroundings, implying the presence of dust, HST observations clearly reveal the presence of a well-defined nuclear disk of gas and dust (see Ferrarese \& Ford 1999). Our dust mass estimation, based on the HST images that have better spatial resolution, is in agreement with the result by Ferrarese \& Ford (1999).\\
\textbf{NGC6314.} The colour image of this galaxy shows a strong DL which might be part of an ring in addition to a compact inner dust disk. As implied by the continuum-subtracted H$\alpha$+[NII] image we could not detect significant amounts of IG in this galaxy.\\
\textbf{NGC 6702.} This is an E3 galaxy. The continuum-subtracted $\mbox{H} \alpha$+[NII] image shows an inclined disk with its major axis along the galaxy minor axis. The colour image reveals a thin DL extending towards the IG disk limit.\\
\textbf{NGC 7052.} This galaxy is classified as an E4 radio galaxy featuring a core and jet, but no lobes (Morganti et al.\ 1987), and its dusty disk physical properties were discussed in the literature (Nieto et al.\ 1990; de Juan, Colina \& Golombek 1996). The colour image reveals a circumnuclear DL along the galaxy major axis. The continuum-subtracted $\mbox{H} \alpha$+[NII] image shows an IG disk following the dust structure. The detailed structure of the nuclear gas and dust disk as observed by HST are also discussed in van der Marel \& van den Bosch (1998; see their Fig. 1).  \\
\textbf{NGC7399.} The colour and continuum-subtracted H$\alpha$+[NII] images show an inclined disk of dust and IG along the major axis of the galaxy.\\
\textbf{NGC7432.} Although small amounts of dust were found in the inner part of this galaxy by Patil et al.\ (2007) and Gregorini et al.\ (1989), we could not detect significant amounts of dust or IG in our images.\\
\textbf{NGC 7625.} This galaxy is considered by some authors to belong to the polar-ring galaxy type (Whitmore et al.\ 1990) and its morphology is considered to be an extended blue compact galaxy (Cair\'{o}s et al.\ 2001). The colour image shows a complex structure with multiple DLs and that the galaxy is bluer in its northern part. The continuum-subtracted $\mbox{H} \alpha$+[NII] image shows that the line emission is concentrated in a central disk surrounded by several HII regions (see Cair\'{o}s et al.\ 2001; Martinez-Delgado et al.\ 2007). The HII clumps are mostly spread along two antennae-like structure, which can be interpreted as signs of recent interaction. The HII filamentary structure clearly coincides with the DLs.\\
\newpage

\label{lastpage}
\end{document}